\documentclass[prx,twocolumn,superscriptaddress,showpacs,nofootinbib,longbibliography]{revtex4-2}
\usepackage{graphicx}
\usepackage[normalem]{ulem}
\usepackage{braket}
\usepackage{bm}
\usepackage{epsfig}
\usepackage{amssymb,amsmath,amsfonts}
\usepackage{wasysym}
\usepackage{latexsym}
\usepackage{mathrsfs}
\usepackage{xcolor}
\usepackage{pifont}
\usepackage{booktabs}
\usepackage{chngcntr}
\usepackage{dsfont}
\definecolor{je}{rgb}{0.858, 0.188, 0.478}
\definecolor{so}{rgb}{0.19,0.32,1.52}
\usepackage{verbatim}
\usepackage{chemformula}
\usepackage[caption=false]{subfig}
\usepackage{braket}
\usepackage[colorlinks,bookmarks=true,citecolor=blue,linkcolor=red,urlcolor=blue]{hyperref}
\newcommand{\cmark}{\ding{51}}%
\newcommand{\xmark}{\ding{55}}

\newcommand{\K}{{\bf K}}

\newcommand{\I}{{\bf I}}

\newcommand{\br}{{\bf r}}

\newcommand{\G}{{\bf G}}

\newcommand{\R}{{\bf R}}

\newcommand{\be}{\begin{eqnarray}}
  \newcommand{\ee}{\end{eqnarray}}

\renewcommand{\k}{{\mathbf k}}
\newcommand{\q}{{\mathbf q}}

\renewcommand{\bm}{\mathbf}

\begin{document}

\title{Kekul\'e spiral order at all nonzero integer fillings in twisted bilayer graphene}
\author{Y.H. Kwan} 
\affiliation{Rudolf  Peierls  Centre  for  Theoretical  Physics, University of Oxford, Oxford  OX1  3PU,  United Kingdom}
\author{G. Wagner} 
\affiliation{Rudolf  Peierls  Centre  for  Theoretical  Physics, University of Oxford, Oxford  OX1  3PU,  United Kingdom}\author{T. Soejima}
\affiliation{Department of Physics, University of California, Berkeley, California 94720, USA}
\author{M.P. Zaletel}
\affiliation{Department of Physics, University of California, Berkeley, California 94720, USA}
\affiliation{Materials Sciences Division, Lawrence Berkeley National Laboratory, Berkeley, California 94720, USA}
\author{S.H. Simon} 
\affiliation{Rudolf  Peierls  Centre  for  Theoretical  Physics, University of Oxford, Oxford  OX1  3PU,  United Kingdom}
\author{S.A. Parameswaran} 
\affiliation{Rudolf  Peierls  Centre  for  Theoretical  Physics, University of Oxford, Oxford  OX1  3PU,  United Kingdom}
\author{N. Bultinck} 
\affiliation{Rudolf  Peierls  Centre  for  Theoretical  Physics, University of Oxford, Oxford  OX1  3PU,  United Kingdom}
\affiliation{Department of Physics, Ghent University, 9000 Ghent, Belgium}

\begin{abstract}
We study magic angle graphene in the presence of both strain and particle-hole symmetry breaking due to non-local inter-layer tunneling. We perform a self-consistent Hartree-Fock study that incorporates these effects alongside realistic interaction and substrate potentials, and explore a comprehensive set of competing orders including those that break translational symmetry at arbitrary wavevectors.
We find that at all non-zero integer fillings very small strains, comparable to those measured in scanning tunneling experiments, stabilize a fundamentally new type of time-reversal symmetric and spatially non-uniform order. This order, which we dub the `incommensurate Kekul\'e spiral' (IKS) order, spontaneously breaks both the emergent valley-charge conservation and moir\'e translation symmetries, but preserves a modified translation symmetry $\hat{T}'$ --- which simultaneously shifts the spatial coordinates and rotates the $U(1)$ angle which characterizes the spontaneous inter-valley coherence. We discuss the phenomenological and microscopic properties of this order. We argue that our findings are consistent with all experimental observations reported so far, suggesting a unified explanation of the global phase diagram in terms of the IKS order.
\end{abstract}

\vskip2pc
\maketitle

\section{Introduction}
\label{sec:introduction}
The discovery of superconductivity proximate to correlated insulating behaviour in a variety of graphene moir\'e heterostructures \cite{Cao2018mott,Cao2018sc,Yankowitz2019,Efetov2019} has triggered intensive efforts to explore the phase structure of these highly tunable two-dimensional materials. In the best-studied example, twisted bilayer graphene (TBG) tuned to a `magic angle' of approximately 1\textdegree, the enhancement of correlations is associated to the formation of extremely narrow bands due to the reconstruction of the electronic  dispersion by the moir\'e superlattice. As noted by Bistritzer and MacDonald (BM) \cite{Bistritzer2011}, this can be elegantly captured within a continuum model \cite{CastroNeto2007,CastroNeto2012,Shallcross2010} where Dirac cones contributed by isolated graphene layers are coupled by interlayer tunneling modulated at the moir\'e scale. The BM model serves as a starting point for most theoretical studies of TBG. On combining the degeneracies corresponding to spin and microscopic valley indices with the Dirac structure enforced by  exact and approximate symmetries of the moir\'e band structure, the model reveals that the striking  effects reported in experiments occur when the Fermi energy is tuned to lie within an octet of  nearly-flat bands that straddle charge neutrality. The phase structure of TBG then turns on the question of how electron correlations and other perturbations such as strain and substrate effects lift the approximate degeneracy within this subspace to select between a variety of competing ground states.

At first sight, the phase diagram of TBG resembles that of the cuprate high-temperature superconductors, with electrostatic gating playing the role of chemical doping. This prompted initial attempts to model correlation effects within a single-band Hubbard model for electronic states localized to a triangular moir\'e superlattice. Although this approach  has proven fruitful in studying moir\'e heterostructures of MoS$_2$ and other transition-metal dichalcogenides \cite{WuLovorn2018}, its applicability to TBG is limited by the `fragile topology' of the BM bands \cite{Zou2018,Song2019}. The latter requires that the simplest tight-binding model that faithfully captures the symmetries of TBG involves a pair of crystallographically-inequivalent Wannier orbitals centered on sites of a honeycomb lattice, but with their charge densities peaked on a triangular lattice formed by the centers of its hexagons \cite{Zou2018,KangVafekPRX,KoshinoFuPRX}. A corollary of this Wannier representation is that it implies a higher degree of itineracy than can be captured via a minimal honeycomb lattice Hubbard model with on-site repulsion and nearest-neighbor hopping. 

The utility of a Hubbard description is further challenged by the early experimental observation of a quantized anomalous Hall (QAH) resistance in TBG samples aligned with a hexagonal boron nitride (hBN) substrate, at an electron density corresponding to filling 7 of the 8 bands \cite{Serlin2019}. [In the  convention where $\nu=0$ represents the filling at neutrality, this corresponds to $
\nu=+3$.] Since this occurs in the absence of an external magnetic field, it indicates the spontaneous breaking of time reversal symmetry (TRS). One explanation of this phenomenon invokes a compelling analogy to the {\it other} paradigmatic setting for strong correlations: the  celebrated Landau levels (LLs) of an electron in a magnetic field. The TBG flat bands are endowed with non-trivial topology encoded in the winding of their Bloch functions across the moir\'e Brillouin zone (mBZ); the inclusion of substrate potential triggers the opening of gaps between the moir\'e Dirac points (by breaking one of the protecting symmetries), and assigns nonzero Chern numbers to the bands, making them topologically equivalent to LLs. The absence of explicit TRS breaking is reflected in the assignment of equal and opposite Chern numbers to different valleys, which are exchanged by TRS. At $\nu=3$, electrons in the two remaining unfilled bands spontaneously polarize into one of the two valleys. This allows them to minimize their interaction energy by virtue of Pauli exclusion, leading to an insulator that {\it spontaneously} breaks TRS, with a  quantized Hall resistivity of $\rho_{xy}=h/e^2$ protected by the charge gap. {Recently, a similar QAH state was also observed at $\nu = 1$ \cite{BernevigEfetov2020}}. While the formation of such orbital Chern insulator states can in principle be captured within a Hubbard description~\cite{VafekQMC}, its close parallels to quantum Hall ferromagnetism~\cite{Girvin,SondhiSkyrmion,MoonYang} (QHFM) has motivated a distinct perspective, where TBG is viewed as a generalized multicomponent quantum Hall system. This naturally explains both the observed QAH response as well the propensity for insulating states at commensurate filling, and motivates a sigma model description based on a hierarchy of perturbations around a `hidden' limit with $U(4)\times U(4)$ symmetry \cite{Bultinck2020,SkyrmionSC,KangVafekPRL}. The QHFM picture receives further experimental support by the observed stabilization of  QAH insulators\footnote{Strictly speaking, these are not anomalous since they only emerge in a finite magnetic field. However the small fields required (significantly smaller than $\sim10$\,T corresponding to one flux per unit cell) suggest that these Chern states remain competitive at zero field. }
with Chern numbers $C = \pm 3, \pm 2, \pm 1$ at $\nu = \pm 1, \pm 2, \pm 3$  on applying a small out-of-plane magnetic field, even in the absence of substrate alignment \cite{BernevigYazdani,AndreiChern,EfetovChern,BernevigEfetov2020,AndreaChern}. However, the TBG bands nevertheless retain features absent in LLs. For instance, their dispersion (though small) remains nonzero, and is enhanced when particle-hole symmetry breaking effects are incorporated --- especially in the electron-doped regime ---  or upon inclusion of strain. Such effects  are likely important in giving an accurate description of experimental samples. As a case in point, even at commensurate fillings some experiments\footnote{Note that while Ref.~\cite{Sharpe2019} reported an anomalous Hall effect at $\nu=+3$, the Hall conductance was not quantized, which is consistent with their samples hosting a gapless state at this filling.} report gapless states or insulators with Chern numbers distinct from those of the noninteracting bands~\cite{Sharpe2019,BernevigEfetov2020,Pierce2021}. This suggests that departures from the flat band/QHFM limit are non-negligible and that the competition between itineracy and localization characteristic of Hubbard physics remains relevant to TBG.

Given its enticing position at the intersection of two dominant themes of strong correlations, it is natural to conjecture that orders that are `natural' from both perspectives could be particularly robust candidate ground states in TBG. One example (and our focus here) is furnished by states with broken translation symmetry, which emerge in relatively well-understood limits of both the Hubbard and quantum Hall settings. The formation of charge and/or spin stripe order is believed to be a near-universal consequence of hole-doping the cuprates away from commensurate filling: while purely on-site Hubbard repulsion favors phase separation, the inevitably present further-neighbour interactions frustrate this in favor of spatially-ordered phases~\cite{ZaanenStripes,PoilblancRice,InnuiLittlewood,GiamarchiLhuillier, EMERY1993597, KEFLiquidCrystal, KivelsonEmeryTranquada, RevModPhys.75.1201, RevModPhys.87.457, Zheng1155, DevereauxStripes,MACHIDA1989,Machida1990}. For similar energetic reasons a variety of stripe and bubble phases are known to be competitive ground states in high Landau levels~\cite{KFS,FKS, MoessnerChalker, FoglerReview}: phase separation is driven by exchange physics and frustrated by Hartree contributions. As noted above, any Hubbard description of TBG must involve substantial further-neighbour interactions. Meanwhile, corrections to the flat band limit --- particularly from strain and particle-hole symmetry breaking --- can significantly modify the Hartree potential, penalizing full occupation of the mBZ. Since exchange interactions still favor insulating behaviour, one resolution is to reconstruct the bands via finite-wavevector ordering. Thus, from both points of view, it appears that conditions in TBG might favor translational symmetry breaking states over their competitors. Despite this, relatively little work has focused on this possibility, with rare exceptions \cite{YouVishwanath,IsobeFu,KangVafekPRL,Christos_2020,Kang2020,VafekQMC,Pierce2021,Lin2019,Lin2020}, of which we highlight a few. Refs.~\cite{Kang2020,VafekQMC} identified a unidirectional charge density wave order that doubles the moir\'e unit cell in an interaction-only model with spin and valley degeneracy suppressed, but this does not appear to be energetically competitive with translationally-invariant states in more realistic situations. More recently, a different period-doubling stripe order stabilized by Hartree effects was invoked to explain the occurrence of commensurate-filling insulators whose Chern numbers depart from those expected from the naive QHFM picture \cite{Pierce2021} (although we suggest an alternative and possibly more natural translation-breaking order at these fillings below). 
Ref.~\cite{Christos_2020} focuses on the Dirac cones and classifies the possible flavor-breaking orders that can be connected to superconductivity via Wess-Zumino-Witten terms. Among the candidate normal-state orders are moir\'e density waves which couple the different mini-valleys of TBG (see also Ref.~\cite{christos2021TTG} for similar discussion in twisted trilayer graphene) and hence break translation symmetry, but no microscopic studies have yet been performed to determine their energetic competitiveness. The nature of translation symmetry-breaking differs from that in the state described in the subsequent sections, which retains the modified translation symmetry $\hat{T}'$. Hence it is not associated with charge/spin modulations between moir\'e AA regions, unlike the moir\'e density waves.
Ref.~\cite{Lado2021} studies a valley spiral state in magnetically encapsulated TBG. This is a similar state to our proposed one, albeit the physical origin and the parameter space within which that state exists are quite different\footnote{The spiral state proposed in Ref.~\cite{Lado2021} relies on large Rashba and Zeeman terms and occurs for twist angles close to $2^\circ$.}. For completeness, we note that translational symmetry breaking has recently been observed in closely related twisted monolayer-bilayer graphene moir\'e heterostructures~\cite{AndreaCDWmonobi}, and was proposed theoretically to explain insulating states observed in twisted bilayer WSe$_2$ \cite{BiFuExcitonDW}. But despite these previous works, to date there has been no systematic analysis of translational symmetry breaking in realistic TBG systems, and so the extent to which such symmetry breaking is a common phenomenon across the wide range of parameters relevant to experimental samples remains unclear.

In this work, we explore translational symmetry breaking order at commensurate integer fillings in TBG. Our analysis incorporates three experimentally important deviations from the Bistritzer-Macdonald limit --- particle-hole symmetry breaking from non-local tunneling perturbations, a substrate potential, and uniaxial strain\footnote{Specifically, heterostrain in which one layer is strained relative to the other.} --- and also studies different interaction strengths and twist angles. In the balance of this introduction, we provide a digest of our main results, which also serves to signpost the organization of the remainder of this paper.

\begin{figure*}
    \centering
    \includegraphics[width=1\linewidth]{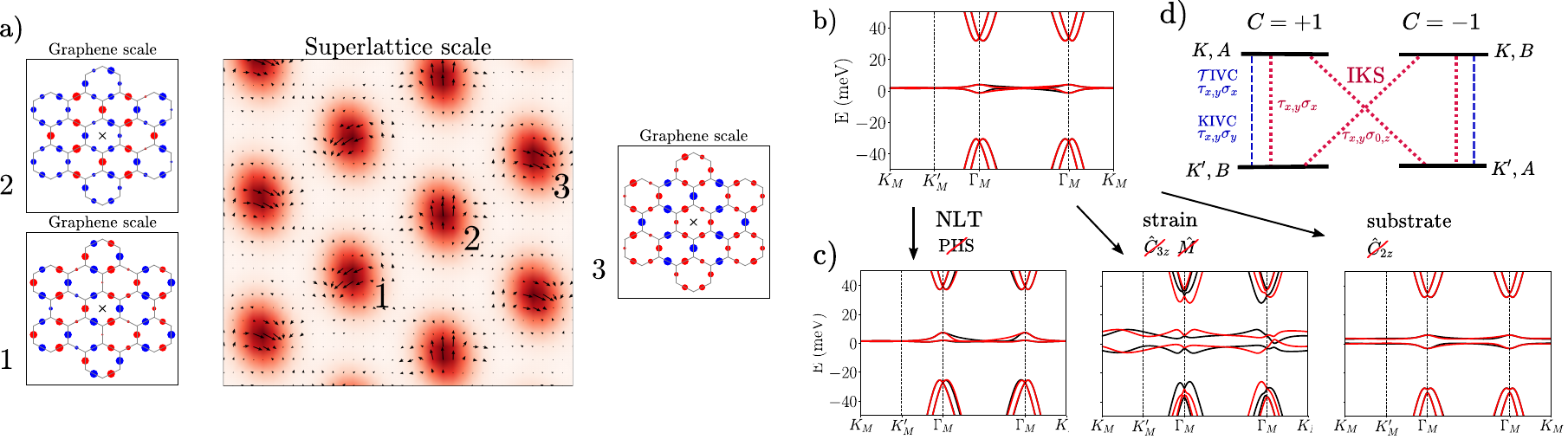}
    \caption{a) Real-space picture of a IKS state with $\q = \G_1/3$. The colorplot on the superlattice scale shows the charge density, with darks spots corresponding to AA regions. Black arrows represent the complex IVC order parameter $\sim \langle \tau_x\sigma_x\rangle + i\langle\tau_y \sigma_x\rangle$. For each of the three inequivalent AA regions, the expectation value of $c^\dagger_A c_B + c^\dagger_B c_A$ on the microscopic graphene bonds is shown. Blue (red) dots correspond to positive (negative) expectation values, and the center of the AA region is marked with a black cross. The different inequivalent AA regions have different approximate $\sqrt{3}\times\sqrt{3}$ Kekul\'e-like patterns on the graphene scale. b) Red (black) line shows BM band structure along a cut in the mBZ for valley $K$ ($K'$). c) The presence of non-local tunneling, strain and substrate potential breaks various symmetries and affects the dispersion. d) If these single-particle perturbations are weak, the interacting model has an approximate $U(4)_{C=1}\times U(4)_{C=-1}$ symmetry. Dashed/dotted lines indicate the channel of inter-valley coherence that generically occurs for $U(1)_V$-breaking phases. $\tau_{x,y}$ denotes any valley-off-diagonal components.}
    \label{fig:IVCSrealspace}
\end{figure*}

\subsection{Summary of Results}

    \begin{figure*}
    \includegraphics[ width=1\linewidth]{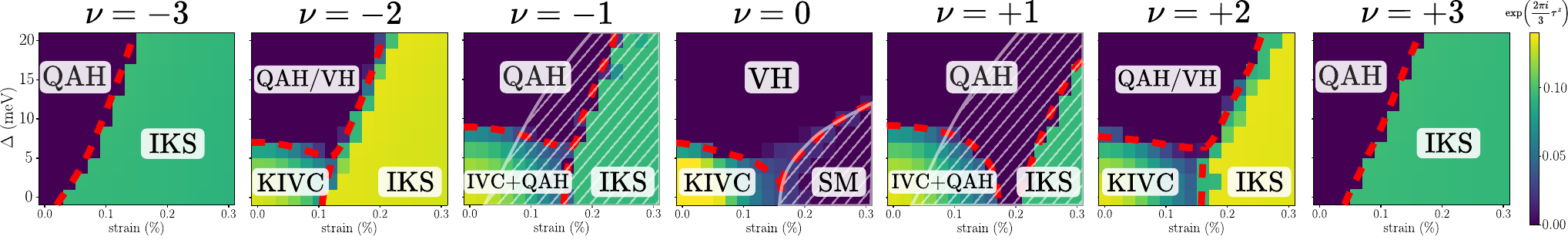}
    \caption{Summary of phases found in self-consistent Hartree-Fock at all integer fillings $\nu$ for different heterostrains and substrate potentials $\Delta$, with non-local tunneling included. Color plot diagnoses IVC order (dark blue indicates unbroken $U(1)_V$ symmetry). Dashed red lines indicate approximate phase boundaries, and hatched areas denote absence of a charge gap. Properties of each phase are tabulated in Table~\ref{tab:phases}. The phase diagrams here are presented in greater detail in Figs.~\ref{fig:phase_nu2} and \ref{fig:phase_nu3}.  {[IKS: incommensurate Kekul\'e spiral, QAH: quantized anomalous Hall state, KIVC: Kramers intervalley coherent state, VH: valley Hall state, IVC: intervalley coherence, SM: semi-metal]}}
    \label{fig:all_phases}
    \end{figure*}

Our central finding, obtained from extensive Hartree-Fock simulations (discussed in Sec~\ref{sec:models}), is that for modest (as little as $0.2\%$) uniaxial strain and largely independent of all other parameters,  the ground state of TBG at {\it all non-zero integer fillings} $\nu=\pm 1, \pm2,\pm3$ is a time-reversal invariant state that breaks superlattice translational symmetry\footnote{Note that strictly speaking, superlattice translation symmetry is present only for commensurate twist angles. We return to discuss this point in more detail in Sec. \ref{sec:HF}.} by modulating intervalley coherence at an incommensurate wavevector  $\q$. Since the intervalley coherence corresponds to a Kekul\'e pattern at the microscopic graphene scale, one can view this order as a Kekul\'e pattern that rotates at the moir\'e scale with period $2\pi/|\q|$. We therefore dub this the `incommensurate Kekul\'e spiral' (IKS) order. The IKS state is insulating at $\nu=\pm2,\pm3$, but does not show a charge gap at $\nu=\pm 1$. In contrast to its ubiquity at nonzero integer fillings, the IKS order is absent at charge neutrality, where we instead find the ground state for comparable strain to be a nematic semimetal, as identified in previous work \cite{Parker2020}. An overview of all the phases found in Hartree-Fock is given in Fig.~\ref{fig:all_phases}.

Modulations in the valley coherence are fundamental to the IKS state, which hence relies on the interplay between the moir\'e pattern and graphene-scale physics. This makes its properties and phenomenology distinct both from the previously-studied period-2 stripe states in TBG \cite{Pierce2021} and from various stripe orders in other correlated systems. It also differs in a few important ways from other proposed states with spatially-modulated valley coherence in either TBG, twisted bilayer WSe$_2$ or twisted monolayer-bilayer graphene \cite{YouVishwanath,IsobeFu,BiFuExcitonDW,YahuiYankowitz}. First, IKS order generally occurs at an incommensurate wavevector, unlike the moir\'e density waves in Ref.~\cite{Christos_2020}. Second, it does not rely on the presence of higher-order van Hove singularities or Fermi-surface nesting and thus has valley coherence over almost the entire mBZ. Third, the IKS state apparently requires a small, but non-zero, amount of strain to be stabilized against competing $\q=0$ orders. And finally, the IKS state is time-reversal symmetric, unlike the state discussed for twisted monolayer-bilayer graphene in Ref. \cite{YahuiYankowitz}. 

Although the IKS order parameter breaks the moir\'e translation $\hat{T}_{\mathbf{a}_i}$ and  $U(1)_V$ valley symmetries, it preserves $\hat{{T}}'_{\mathbf{a}_i} = \hat{T}_{\mathbf{a}_i}e^{i\q\cdot\mathbf{a}_i\tau_z/2}$ (where $\bm{a}_i$ are moir\'e lattice vectors and $\tau_\mu$ denote a set of Pauli matrices acting in valley space).
Performing a valley-dependent gauge transformation therefore yields eigenstates that satisfy a generalized Bloch theorem. This transformation, which amounts to shifting the dispersion in the valleys by $\pm \q/2$ in the mBZ, allows us to label electronic states in the IKS state by an analog of the crystal momentum associated with $\hat{{T}}'_{\mathbf{a}_i}$ (rather than $\hat{{T}}_{\mathbf{a}_i}$), without folding the mBZ. This shifted-mBZ perspective yields a simple yet quantitatively accurate \textit{ansatz} for the HF projector in the IKS state, which in turn allows us to link the origin of the order to features of the interaction-renormalized BM bands. We can also use the preservation of  $\hat{{T}}'_{\mathbf{a}_i}$ to derive a modified Lieb-Schulz-Mattis theorem that forces gapped IKS order to only appear at integer fillings unless it triggers subsidiary topological or symmetry-breaking order. This explains why, despite involving a {\it modulation} that is incommensurate with the moir\'e lattice, the IKS insulator is tied to {\it densities} that are commensurate with it. 

As we show below, the precise magnitude and direction of the spiral wavevector $\q$ are controlled by details of the dispersion of the interaction-renormalized bands in the mBZ. The former is roughly $1/3$ of the width of the mBZ while the latter is approximately aligned along a moir\'e crystallographic axis, but the energy cost for varying away from these values (the `stiffness' of the IKS order) is quite small. In a sense one can view this softness as being partly responsible for the robustness of the IKS state, since it allows the ordered state to respond to variations in external parameters such as substrate potential, twist angle, or interaction strength by adjusting $\q$ while keeping its other properties essentially unchanged.

The above results are obtained in Sec.~\ref{sec:SSstripe} by focusing initially on fillings $\nu =\pm 2$  where the IKS order is especially simple to describe. We broaden the scope of our analysis in Sec.~\ref{sec:FMstripes} to also consider $\nu=\pm 1$ and $\nu= \pm 3$, which differ primarily in the spin structure and the stability of IKS order against competing phases. Interestingly, the time-reversal symmetric IKS order provides a way to obtain insulating states with zero Chern number at the odd integer fillings. Such states are difficult to obtain within the QHFM formalism, and were observed experimentally in Refs. \cite{Yankowitz2019,BernevigEfetov2020}. In Sec.~\ref{sec:relation} we show that the $\nu=-3$ state provides a `basis spiral' that serves as a building block for IKS states at other fillings, in a manner that we make quantitatively precise.

We derive a Landau-Ginzburg theory of IKS order in Sec.~\ref{sec:LG}, and use this to link the  circular (elliptical) nature  of the spiral order in valley coherence to the absence (presence) of subsidiary charge density modulations. We also consider the response of the IKS state to quenched disorder (Sec.~\ref{sec:quenched}) and thermal fluctuations (Sec.~\ref{sec:thermodyn}). A key point is that although disorder on the microscopic graphene scale can couple to the Kekul\'e distortion as a random field, this is suppressed in powers of the ratio of the graphene and moir\'e lattice constants, scaling as $\theta^4$ for small twist angles $\theta$. Consequently, the dominant disorder fluctuations are those that occur on the moir\'e scale. This justifies our assignment of a definite Kekul\'e order to each AA-stacking region that defines a  superlattice `site'. Thermal effects are more delicate, owing to a rich set of symmetries broken by the IKS state: namely, valley-$U(1)$, superlattice translation, and three-fold rotation. Since the superlattice translation is broken solely by valley-$U(1)$ charged operators, long range order in both is absent at any temperature $T>0$, and is replaced by algebraic correlations, which in turn become exponential decay above a Berezinski-Kosterlitz-Thouless transition temperature $T_{{\text{BKT}}}$ at which vortices in the valley order proliferate. In contrast, the rotational symmetry breaking persists as true long-range $T>0$ order, so that (ignoring explicit symmetry breaking from strain) the finite-temperature IKS state has a nematic order up to a finite Ising transition at $T_N$. Depending on the ratio of $T_N$ to $T_{{\text{BKT}}}$, we can have a variety of different thermal melting scenarios based on which order is lost first, though the precise details are subtle and may depend on the ability of the superlattice to pin $\q$ to a commensurate value (which is likely weak). The relevant scales for $T_N$ and $T_{\text{BKT}}$ are similar and  are set by the IKS stiffness, and are $\sim 7~\text{K}$, which is comparable with the experimental temperature scales at which gapped insulators are observed.

We emphasize (Sec.~\ref{sec:experiments}) that our results  closely match  current experiments: most notably, through the absence of spin polarization in the $\nu=\pm 2$ IKS, the relatively greater robustness (as measured by the charge gap) of insulating states on the electron-doped side ($\nu>0$), and the ability of the IKS to `reset' the Chern number to zero at odd integer fillings. The IKS order is a nematic at  finite temperature. Thus, we expect the orientational symmetry breaking effects of strain to be heavily enhanced in the IKS state, making it a natural proximate order to the reported nematic superconductors near $\nu=-2, -3$. A subset of experiments find correlated insulators at all integer fillings except at neutrality where they see evidence of a gapless state, and $\nu=\pm1$ where weak insulating peaks have been observed~\cite{Efetov2019}. This is readily reconciled with  our results --- since (see Fig.~\ref{fig:all_phases}) we find IKS  order for all integer $\nu$ {\it except} $\nu=0$ --- if we assume that the relevant experimental samples are subject to a small amount of heterostrain. (This is a relatively mild assumption given the weak strain needed to stabilize the IKS state and the $\nu=0$ nematic semimetal.) Direct verification of IKS order is a more challenging goal. Owing to the unusual nature of IKS states, the translational symmetry breaking is invisible to valley-diagonal observables, and a Landau-Ginzburg analysis reveals that the circular spiral order does not trigger a parasitic charge-density wave order. Nevertheless, since intervalley coherence necessarily triggers spatial order on the graphene scale, the associated Kekul\'e distortion should be apparent, e.g. in the locally-AA regions of the superlattice, but will be modulated at the moir\'e scale. This order can in principle be detected via scanning probes, though the sensitivity required may be difficult to achieve in the very near term.  We close with  a discussion of future directions motivated by this work, in Sec.~\ref{sec:discussion}.  We provide details of numerical simulations and additional analysis in five technical appendices.

\begin{table*}
\centering
\newcommand{\colskip}{\hskip 0.15in}
\renewcommand{\arraystretch}{1.15}
\begin{tabular}{
l @{\hskip 0.3in} 
c @{\colskip} 
c @{\colskip} 
c @{\colskip} 
c @{\colskip} 
c @{\colskip} 
c @{\colskip} 
c @{\colskip} 
c @{\colskip} 
c }\toprule[1.3pt]\addlinespace[0.3em]
Phase & 
$|\nu|$ & 
spin pol. & 
valley pol. & 
$U(1)_V$ &
$\hat{\mathcal{T}}=\tau_x\hat{\mathcal{K}}$ & 
$\hat{\mathcal{T}}'=\tau_y\hat{\mathcal{K}}$ & 
$\hat{T}_{\bm{a}_i}$ & 
$\hat{T}_{\bm{a}_i}e^{ i \bm{q}\cdot \bm{a}_i\tau_z/2}$ & 
$|C|$   
\\ \midrule
IKS& 1 & * & 0 & \xmark & \cmark & \xmark & \xmark & \cmark & 0 \\
& 2 & 0 & 0 & \xmark & \cmark & \xmark & \xmark & \cmark & 0 \\
& 3 & * & 0 & \xmark & \cmark & \xmark & \xmark & \cmark & 0 \\
QAH & 1 & 1 & 1 & \cmark &  \xmark & \xmark & \cmark &  & 1 \\
 & 2 & 0 & 2 & \cmark & \xmark & \xmark & \cmark &  & 2 \\
 & 3 & 1 & 1 & \cmark & \xmark & \xmark & \cmark &  & 1 \\
 KIVC & 0 & 0 & 0 & \xmark & \xmark & \cmark & \cmark &  & 0 \\  
 & 2 & * & 0 & \xmark & \xmark & \cmark & \cmark &  & 0 \\  
VH  & 0 & 0 & 0 & \cmark & \cmark & \cmark & \cmark &  & 0 \\ 
    & 2 & * & 0 & \cmark & \cmark & \cmark & \cmark &  & 0 \\
QAH+IVC  & 1 & 1 & 1 & \xmark & \xmark & \xmark & \cmark &  & 1 \\
SM  & 0 & 0 & 0 & \cmark & \cmark & \cmark & \cmark &  & 0 \\\bottomrule[1.3pt]
\end{tabular}
\caption{Symmetries and order parameters of HF phases at integer fillings $\nu$. Spin (valley) polarization indicates the number imbalance of $\uparrow$ vs $\downarrow$ ($K$ vs $K'$) electrons per moir\'e unit cell. The spin quantization axis is arbitrary due to $SU(2)_S$-symmetry, and an asterisk indicates a degenerate manifold obtained by performing a valley-dependent spin-rotation. $\tau_\mu$ are Pauli matrices in valley space, and $\hat{\mathcal{K}}$ is complex conjugation. IKS phases break a subset of moir\'e translations $\hat{T}_{\bm{a}_i}$, but preserve the combined valley-rotation + translation symmetry $\hat{T}_{\bm{a}_i}e^{ i \bm{q}\cdot \bm{a}_i\tau_z/2}$, where $\bm{q}$ is the IKS wavevector.} 
\label{tab:phases}
\end{table*}

\section{Model and Numerical Techniques}
\label{sec:models}

In this section, we discuss the interacting Bistritzer-MacDonald (BM) model in the presence of strain, substrate potential and non-local tunneling (NLT), and describe our HF calculations. Further details can be found in the Appendices.

We begin with the standard single-particle BM model~\cite{Bistritzer2011} describing the band structure of two graphene layers $l=1,2$ stacked with a relative twist $\theta$ near the magic angle. For concreteness, we orient the coordinate system such that the untwisted monolayer Dirac points lie at $\bm{k}=\pm K_\mathrm{D}\hat{x}$. The interlayer coupling, which is modulated by the moir\'e pattern, is parameterized by sublattice-dependent hopping constants $w_{\mathrm{AA}}=82.5\,\text{meV}$ and $w_{\mathrm{AB}}=110\,\text{meV}$. The presence of different coupling constants arises from corrugation effects~\cite{Nam2017,Carr2019} that increase the interlayer spacing in AA stacking regions compared to AB (Bernal) regions. 

Throughout this work, we fix $\theta=1.08^\circ$, unless stated otherwise. Including both spins and valleys (with corresponding Pauli matrices $s_\mu$ and $\tau_\mu$ respectively), the BM Hamiltonian has 8 central bands near the neutrality point with narrow bandwidth $\sim5\,\mathrm{meV}$ and large gaps $\sim30\,\mathrm{meV}$ to the remote bands (Fig.~\ref{fig:IVCSrealspace}b). The central band wavefunctions are concentrated at the AA-stacked regions ~(Fig.~\ref{fig:IVCSrealspace}a), which form the moir\'e lattice. The model possesses (spinless) TRS $\hat{\mathcal{T}}=\tau_x\hat{\mathcal{K}}$ (where $\hat{\mathcal{K}}$ denotes complex conjugation) and enjoys emergent $D_6$ and valley-charge conservation ($U(1)_V$) symmetries, and an approximate particle-hole symmetry (PHS)~\cite{Zou2018,Hejazi2019,Song2019}. A related antiunitary symmetry $\hat{\mathcal{T}}'=\tau_y\hat{\mathcal{K}}$ can be defined, which is a signature of the Kramers intervalley coherent (KIVC) phase~\cite{Bultinck2020} to be reviewed below. The relevant spatial symmetries of the single-valley BM Hamiltonian are $\hat{C}_{2z}\hat{\mathcal{T}},\hat{C}_{3z}$, and $\hat{M}$, where the latter corresponds to an in-plane two-fold rotation around the $x$-axis which interchanges the two layers. Spin-orbit coupling is neglected, resulting in a total $U(2)_K\times U(2)_{K'}$ flavor symmetry.

\subsection{Chern basis and effect of a substrate potential}
The central bands bear a remarkable resemblance to zero Landau levels in opposite fields (an analogy which is sharpened in the chiral limit $w_{AA}=0$~\cite{Tarnopolsky2019}). For a given spin/valley, we can take advantage of the weak dispersion to rotate the pair of central BM bands into a $C=\pm1$ Chern basis by diagonalizing the sublattice operator $\sigma_z$ \cite{Bultinck2020}. Each band carries substantial sublattice polarization (tending to $\pm 1$ in the chiral limit), and hence we use $\sigma$ to also refer to this basis. The Chern number of each of the degenerate bands is tied to its valley according to $C=\sigma_z\tau_z$~\cite{Bultinck2020,Zou2018,Liu2019pseudo}. [Note that the `Chern basis' defined by a definite value of $C$ does not coincide with the eigenbasis of the single-particle dispersion in the absence of a substrate potential.]

Alignment of TBG with the hBN substrate directly couples to the Chern basis via a sublattice mass $\sim\sigma_z$ with strength $\Delta\simeq 10-20\,\mathrm{meV}$~\cite{Jung2015,Bultinck2019,Zhang2019} (we ignore the additional moir\'e potential coming from the mismatch between the graphene and hBN lattice constants, although it has recently been argued to be important for explaining some of the experimental features \cite{MaoSenthil,ShiZhuMacDonald}), and violates $\hat{C}_{2z}$ and $\hat{M}$. The breaking of $\hat{C}_{2z}\hat{\mathcal{T}}$ gaps the Dirac points~(Fig.~\ref{fig:IVCSrealspace}c), resulting in the formation of Chern bands. Polarization into a subset of these Chern bands (akin to quantum Hall ferromagnetism) is believed to explain the observation of the anomalous Hall (AH) effect at $\nu=+3$ in aligned samples~\cite{Sharpe2019,Serlin2019}. The substrate-reconstructed central bands are also used as a starting point for constructing more exotic correlated states~\cite{Abouelkomsan2020,Ledwith2020,Repellin2019,Kwan2021,Kwan2020fqhe,Stefanidis2020}.

\subsection{Strain effects}
Uniaxial strain of strength $\epsilon=0.1-0.7\%$ is observed in many TBG samples using STM/STS~\cite{Kerelsky2019,Choi2019,Xie2019stm}. At charge neutrality, this small strain is believed to be an important driving force behind the weakening of symmetry-broken insulators found in numerics at zero strain in favour of semimetallic phases~\cite{Parker2020,Liu2021,Kang2020}. In the context of van der Waals homobilayers, it is useful to distinguish between homostrain, where strain is applied identically to both layers, and heterostrain, where the layers are strained independently. Since homostrain, to first order, does not account for the experimentally observed distortion of the moir\'e lattice, and has a substantially smaller impact on the electronic structure~\cite{Huder2018}, we focus on heterostrain~\cite{Bi2019,Parker2020}, which is also believed to be experimentally relevant. The moir\'e lattice vectors $\bm{a}_{1,2}$ are deformed depending on the value of the strain ratio $\epsilon$ and strain angle $\varphi$ with respect to the $x$-axis. The orthogonal direction is also stretched/compressed due to the Poisson ratio $\simeq 0.16$ \cite{Poisson}. To first order in $\epsilon$ and $\theta$, the twist angle is unaffected. The BM model is modified by taking into account the deformed superlattice basis vectors, as well as adding an effective layer-dependent vector potential $\bm{A}_l$ (similar to the orbital effect of an in-plane magnetic field~\cite{Kwan2020parallel}).
Strain preserves $\hat{C}_{2z}$ but breaks $\hat{C}_{3z}$ and $\hat{M}$. Hence the Dirac points remain intact, but are unpinned from the $K_M$ and $K'_M$ points and migrate towards the mBZ center \cite{Bi2019}. The Dirac points also separate in energy leading to Fermi pockets at charge neutrality (CN), and the overall bandwidth of the central bands increases dramatically~(see Fig.~\ref{fig:IVCSrealspace}c).

\subsection{Non-local tunneling and breaking of particle-hole symmetry}
The standard BM Hamiltonian obeys PHS very well---the only violations come from small twists in the Dirac cone kinetic terms which are suppressed in $\theta$~\cite{Song2019,Hejazi2019}. However, many experiments show pronounced electron-hole asymmetry~\cite{Cao2018sc,Lu2019,Sharpe2019,Serlin2019,Yankowitz2019,Cao2020nematicity,Pierce2021}, with stronger superconductors on the hole side and more robust insulators on the electron side. We model this PHS-breaking by augmenting the BM model with a non-local interlayer tunneling term~\cite{Fang2019,Carr2019,Xie2020,WaletGuinea}. Consistent with density functional theory calculations, the effect of this term is to make the conduction bands more dispersive than the valence bands~\cite{Fang2019,Carr2019} (Fig.~\ref{fig:IVCSrealspace}c). We use the form of NLT motivated in Ref.~\cite{Fang2019} and choose values $\lambda_{2} =2 \lambda_{1}= 0.18\, \mathrm{eV}$\AA, $\lambda_{3} = 0$ (see App.~\ref{secapp:continuummodel} for definitions and a discussion on combining NLT and strain).

\subsection{Hartree-Fock procedure}
We perform self-consistent HF calculations on the single-particle Hamiltonian with dual-gate screened Coulomb interactions $V(q)=\frac{e^2}{2\epsilon_0\epsilon_r q}\tanh{qd}$, where the screening length $d=25\,\textrm{nm}$ and relative permittivity $\epsilon_r=10$. We neglect terms which scatter electrons between the valleys, as they are suppressed for small $\theta$. Along with electron-phonon scattering, such `intervalley-Hund's couplings' would weakly break the $U(2)_K\times U(2)_{K'}$ symmetry~\cite{Bultinck2020}. In order to avoid double-counting interaction effects, we subtract off a density matrix corresponding to decoupled graphene layers at charge neutrality~\cite{XieSub,Bultinck2020}. The results shown the main text were obtained by considering only the central bands as active, with the remote valence/conduction bands frozen to be filled/empty. However we have checked that increasing the number of active bands does not lead to a qualitative change in the results, and mainly leads to a decrease in the bandgap and a slight shift in the phase boundaries. For more details on the effects of adding more active bands, we refer to App. \ref{secapp:additional}.

In our numerical simulations, we consider completely general moir\'e translation symmetry breaking Slater determinants with single-particle density matrices of the general form
\begin{equation}
\langle \hat c^\dagger_{\mathbf{k}\tau as}\hat c^{\phantom{\dagger}}_{\mathbf{k}'\tau'a's'}\rangle=P_{\mathbf{k}\tau as;\mathbf{k}'\tau'a's'},
\end{equation}
satisfying $\text{Tr}\,P=(\nu+4)N_1N_2$, where $N=N_1N_2$ is the number of moir\'e unit cells, and $a,a'$ are BM band indices. We use periodic boundary conditions in both directions which leads to discrete values for the allowed momenta, viz.~$\mathbf{k},\mathbf{k}'=\frac{n_1}{N_1}\mathbf{G}_1+\frac{n_2}{N_2}\mathbf{G}_2$, with $\mathbf{G}_i$ the moir\'e reciprocal lattice vectors. 

For most calculations we enforce co-linearity of the spins and accelerate convergence using the optimal damping algorithm \cite{ODA,Cances2000}. All the states found in our HF simulations (discussed in detail in the following sections) contain at most a single wavevector modulation, meaning that $P_{\mathbf{k}\tau as;\mathbf{k}'\tau'a's'}\neq0$ only for $\mathbf{k'}=\mathbf{k},\mathbf{k}+\mathbf{q},\mathbf{k}-\mathbf{q}$. As detailed below, whenever translational symmetry breaking occurs we find that the finite-$\mathbf{q}$ component of the projector $P$ is entirely off-diagonal in valley space. 

\section{Spin-unpolarized Kekul\'e spirals at $\nu=\pm 2$}
\label{sec:SSstripe}

    \begin{figure*}
    \includegraphics[ width=1\linewidth]{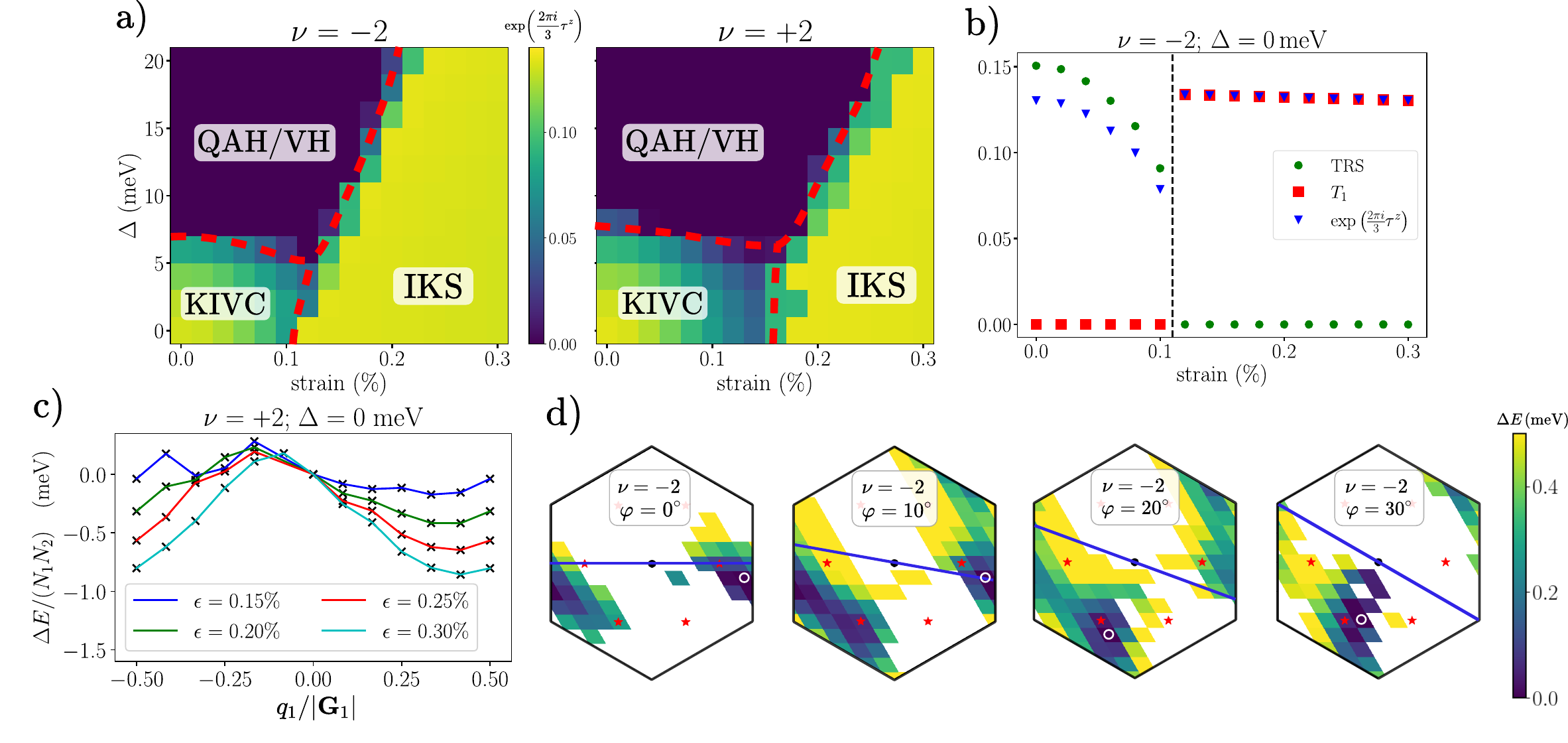}
    \caption{
    a) $\nu=\pm2$ strain-substrate HF phase diagram (enlarged view of panels from Fig.~\ref{fig:all_phases}). Strain is along the $x$-axis ($\varphi=0^\circ$), and translational symmetry-breaking is restricted to $\bm{q}=\bm{G}_1/3$. Color plot diagnoses IVC order by taking the Frobenius norm of the difference of the density matrix after acting with $e^{\frac{2\pi i}{3}\tau_z}$. Dashed red lines indicate approximate phase boundaries. Particle-hole breaking is introduced via non-local tunnelling (NLT). System size is $12\times12$, and only central bands are active. b) Phase transition between KIVC and IKS states along the strain axis of a) at $\nu=-2$. In the IKS phase, combined $U_V(1)$ rotation by $2\pi/3$ and real-space translation by $\bm{a}_1$ is a symmetry. c) IKS energy relative to the lowest translation-symmetric state for different strains and spiral wavevectors $q$ along $\bm{G}_1$. NLT is not included. d) Relative energies of IKS states for different strain angles $\varphi$ (blue axis) and enforced symmetry-breaking wavevectors $\bm{q}$ (see Eq.~\ref{eq:IVCSmatrixelements}). Red stars denote $\bm{q}$'s corresponding to period-tripling along the moir\'e axes, white circles denote minimum energy wavevectors $\bm{q}_0$. Non-IKS states that converged to higher energies were discarded.  Strain is $0.2\%$, and NLT is not included. Data points have been transformed to fit on a hexagonal mBZ.}
    \label{fig:phase_nu2}
    \end{figure*}

We now explore how the trio of realistic modifications to the BM model introduced above --- namely, substrate effects, strain, and particle-hole symmetry breaking --- stabilize  phases that compete with those previously proposed for the idealized situation where these modifications are absent~\cite{Xie2020,XieSub,Bultinck2020,CeaGuinea,Zhang2020HF,KangVafekPRL,Kang2020,Liu2021,TBGIV,TBGVI,SoejimaDMRG,PotaszMacDonaldED,liu2021theories,shavit2021theory,zhang2021correlated}. In this section, we first focus on fillings $\nu = \pm 2$, as the phenomenology of the translational-symmetry-breaking states is especially clear here. 

\subsection{Numerical Hartree-Fock results}
\label{sec:HF}

Fig.~\ref{fig:phase_nu2}a presents the HF phase diagrams at $\nu=\pm2$ in the presence of NLT as a function of both strain and substrate potential. The color scale diagnoses the magnitude of $U(1)_V$-breaking. Without symmetry-breaking perturbations, the lowest energy state is the $\hat{\mathcal{T}}'$-symmetric KIVC state~\cite{Bultinck2020} (see also Ref. \cite{TBGIV}). It consists of a filled intervalley coherent (IVC) band in each Chern sector, and can be succinctly described by ordering of $\tau_{x,y}\sigma_y$ where $\tau_{x,y}$ denotes any off-diagonal component in valley space~(Fig.~\ref{fig:IVCSrealspace}d). The absence of coherence between opposite Chern sectors sidesteps the energy penalty induced by vortices in the order parameter that would be topologically required for other IVC candidates~\cite{Bultinck2019}. At $U(2)_K\times U(2)_{K'}$ level, there is a manifold of degenerate states with different spin polarizations (with maximum $2\mu_B$ spin moment per moir\'e cell), but intervalley-Hund's perturbations will lift this degeneracy. 

At a finite substrate potential strength, the optimal state becomes a sublattice-polarized $U(1)_V$-symmetric ferromagnet, which can either be the QAH state $\sim\sigma_z\tau_z$ or the valley Hall state (VH) $\sim\sigma_z$. These are exactly degenerate at HF level since the VH state is obtained by applying $\hat{\mathcal{T}}$ on one spin component of the QAH state. 

Along the strain axis, we find a first-order transition to a novel phase, which we dub the incommensurate Kekul\'e spiral (IKS) state, at an experimentally relevant strain ratio of $\epsilon \sim 0.1 - 0.2 \%$ (Fig.~\ref{fig:phase_nu2}b). The main characteristic of the IKS state is the breaking of moir\'e translation symmetry at a single wavevector $\bm{q}$. The translation-breaking occurs entirely in the intervalley channel, and is clearly identified in HF by the non-vanishing of the following density matrix elements in the sublattice-polarized basis
\begin{equation}\label{eq:IVCSmatrixelements}
\langle \hat c^\dagger_{\bm{k}+\bm{q},\tau=+,\sigma}\hat c^{\phantom{\dagger}}_{\bm{k},\tau'=-,\sigma'}\rangle\sim
f_{\sigma,\sigma'}(\bm{k})
\end{equation}
where spin labels have been omitted. Importantly, the IVC occurs at a single $\bm{q}$, leading to a \emph{circular inter-valley coherent spiral} of definite handedness, as there is no symmetry relating the spiral we find in HF to the analogous spiral at $-\bm{q}$ (Fig.~\ref{fig:phase_nu2}c). The IKS state also preserves TRS $\hat{\mathcal{T}}$, and has zero total spin and valley polarization (Table~\ref{tab:phases}). Since the spins within each valley are also unpolarized, inclusion of intervalley-Hund's coupling does not lead to qualitative changes.

The IKS order persists for fairly large substrate potential strengths. This is expected since the intervalley coherence is flexible enough to polarize onto one sublattice, as evidenced from the fairly constant magnitude of IVC throughout the phase. On the other hand, the KIVC is progressively weakened  under increasing sublattice potential, and gives way to $U(1)_V$-preserving ferromagnets since its mechanism relies on inter-sublattice coherence~\cite{Bultinck2020}. The strong PHS-breaking effect of NLT manifests in the shifted phase boundaries between $\nu=-2$ and $+2$. Furthermore, the zero-substrate band gaps (of order $10\,\mathrm{meV}$) of both the KIVC and IKS phases are larger on the electron side than the hole side by $10-30\%$, which is consistent with the experimental trend of more robust insulators at positive fillings.

The numerical phase diagram in Fig.~\ref{fig:phase_nu2}a was constructed by restricting to a strain angle $\varphi =0$ and period-tripling order along $\bm{G}_1= |\bm{G}_1|\hat{x}$. However, when we relax this requirement, we find that the IKS phase actually consists of a family of spirals which differ only in their ordering wavevector $\bm{q}$ and are close in energy. Fig.~\ref{fig:phase_nu2}c plots the IKS energy relative to best translation-symmetric state, as a function of $q_1$ ($q_2$ is fixed to $0$, so translation symmetry is maintained along $\bm{a}_2$). The ideal wavevector $\bm{q}_0$ is slightly greater than $1/3$ of the mBZ, and evolves weakly with strain magnitude. Hence, the spiral ordering generically occurs at an incommensurate $\bm{q}_0$ (see also Fig.~\ref{fig:phase_nu3}b). 

Fig.~\ref{fig:phase_nu2}d, where we fully relax the constraints on the wavevector $\bm{q}$ in our HF calculations, reveals that the dispersion about $\bm{q}_0$ is very soft in both directions. Note that we have checked that even without enforcing a particular $\bm{q}$, the HF still only converges to a single-$\bm{q}$ state. We find the energy density of the IKS state to have a term of the form $\frac{\rho_s}{2}(\bm{q}-\bm{q}_0)^2$, from which we estimate the wavevector stiffness to be $\rho_s\sim0.4\,\mathrm{meV}$, without strong spatial anisotropy. In Fig.~\ref{fig:phase_nu2}d we show that as the strain angle $\varphi$ rotates, $\bm{q}_0$ also changes, but appears to have roughly constant magnitude and predominantly lies near a moir\'e crystallographic axis.  Figs.~\ref{fig:phase_nu2}c,d are computed without NLT; including NLT does not affect the qualitative features of these plots.

Before concluding the discussion of our numerical HF results, we want to point out the following subtlety. In the absence of $\hat{C}_{3z}$ symmetry (which is broken by strain), the $\Gamma$-point of the single-valley BM model is no longer a high symmetry point. From this one might conclude that the choice of $\Gamma$ in one of the two valleys becomes arbitrary (the $\Gamma$-point in the other valley is still fixed by either $\hat{C}_{2z}$ or $\hat{\mathcal{T}}$). Making a different choice for $\Gamma$ does not go without consequences for the IKS state, as this changes the wavevector $\q$ at which the inter-valley coherence occurs. For commensurate twist angles, however, there is a preferred $\Gamma$-point in the mBZ even in the absence of $\hat{C}_{3z}$ --- namely, it is the point that should fold on top of the $\Gamma$-point of the mono-layer graphene BZ (which is fixed by $\hat{C}_{2z}$ or $\hat{\mathcal{T}}$). From this it is clear that the wavevector $\q$ is well-defined for commensurate twist angles, and that the corresponding superlattice translation symmetry is unambiguously broken. In our numerical simulations at incommensurate twist angles, we have always used the same choice for the mBZ $\Gamma$-point as in the commensurate twist angle case, such that the location of the mBZ $\Gamma$-point varies continuously as a function of $\theta$. However, for incommensurate twist angles a different choice for $\Gamma$ is possible in principle, and thus the wavevector $\q$ of the IKS state becomes `gauge dependent'. This is consistent with the fact that for incommensurate twist angles, there is strictly speaking no superlattice translation symmetry.

\subsection{Generalized Bloch and Lieb-Schulz-Mattis theorems for IKS states}
A defining property of the IKS state, which has circular IVC spiral order, is that it is invariant under the combination of a translation along superlattice vector $\mathbf{a}_i$ and a valley-$U(1)$ rotation which shifts the IVC angle by $\bm{a}_i \cdot \q$. Let us therefore define modified translation operators $\hat{T}'_{\bm{a}_i} \equiv \hat{T}_{\bm{a}_i}e^{i\bm{a}_i\cdot \q \,\tau_z/2}$. Because the IKS state preserves $\hat{T}'_{\bm{a}_i}$, a generalized Bloch theorem applies which states that the single-particle wavefunctions should satisfy
\begin{equation}\label{GBloch1}
    \psi_{\bm{\tilde{k}}}(\bm{r}+\bm{a}_i) = e^{i\bm{\tilde{k}}\cdot \bm{a}_i} e^{-i \bm{a}_i \cdot \q\, \tau_z/2}\psi_{\bm{\tilde{k}}}(\bm{r}).
\end{equation}
Here, $\bm{\tilde{k}}$ is a new `momentum' label restricted to the first mBZ, which differs from the conventional crystal momentum. In particular, $\bm{\tilde{k}}$ labels real, physical momenta $\bm{\tilde{k}} + \tau_z \q/2$ in the two valleys $\tau_z = \pm$. From Eq. \eqref{GBloch1}, it follows that we can write the single-particle wavefunctions as
\begin{equation}
    \psi_{\bm{\tilde{k}}}(\bm{r}) = e^{i\bm{r}\cdot(\bm{\tilde{k}}-\tau_z \q/2)}u_{\bm{\tilde{k}}}(\bm{r})\, ,
\end{equation}
where $u_{\bm{\tilde{k}}}(\bm{r})$ is the periodic part satisfying $u_{\bm{\tilde{k}}}(\bm{r}+\bm{a}_i) = u_{\bm{\tilde{k}}}(\bm{r})$. As a result, we can define a Hartree-Fock band structure in the mBZ for general IKS states, even if the order wavevector $\q$ is incommensurate with the moir\'e lattice. We note that a similar observation has previously been made for incommensurate circular spin spiral states \cite{Sandratskii,Savrasov}.

Another, but closely related, consequence of the $\hat{T}'_{\bm{a}_i}$ symmetry is that the IKS state can only have a non-zero energy gap (ignoring the Goldstone modes) at integer fillings --- unless it breaks additional symmetries or develops non-trivial topological order. To see why this is the case, first add a small perturbation of the form

\begin{eqnarray}
    \hat{V} & = & h \int \mathrm{d}^2\bm{r}\, \big[ \cos(\q\cdot\bm{r} + \alpha)\hat{\psi}^\dagger(\bm{r})\tau_x\sigma_x\hat{\psi}(\bm{r}) \nonumber \\ 
    & & + \sin(\q\cdot \bm{r} + \alpha) \hat{\psi}^\dagger(\bm{r})\tau_y\sigma_x\hat{\psi}(\bm{r})\big]
\end{eqnarray}
to the Hamiltonian. This perturbation preserves $\hat{T}'_{\bm{a}_i}$, but explicitly breaks the valley-$U(1)$ symmetry. As a result, the Goldstone mode of the IKS state acquires a small gap $\Delta_G \propto |h|$. Next, we invoke a generalized Lieb-Schulz-Mattis (LSM) theorem which states that the IKS state with gapped Goldstone modes can have a unique ground state on the cylinder geometry which is separated by a non-zero energy gap from all other states in the spectrum only if the charge per unit cell is integer. To show that such a generalized LSM theorem indeed holds, one can simply use the standard adiabatic flux-insertion argument put forward by Oshikawa \cite{Oshikawa1,Oshikawa2}. The redefinition of the translation symmetry operator by multiplying it with $e^{i\bm{a}_i\cdot \q \,\tau_z/2}$ does not change this argument, as the additional factor commutes with the electric-charge $U(1)$ symmetry\footnote{One subtlety, however, is that in the simplest version of the adiabatic flux-insertion setup, one wants to have the stripes run along the axis of the cylinder, which means that the system has to be compactified in the direction parallel to $\q$. Forcing states with incommensurate $\q$ on the cylinder in this way will cause them to adjust $\q$ to take on the closest commensurate value compatible with the periodic boundary conditions. However, because we can make the cylinder circumference arbitrarily big, and thus the shift in $\q$ arbitrarily small, this will not change whether or not the system has a gap.}. In general, one expects that the gapless states which occur at non-integer fillings (excluding topological order and additional symmetry breaking) will have a vanishing charge gap, meaning that it is possible to create well-separated particle-hole pairs with arbitrarily small energy.

    \begin{figure*}
    \includegraphics[ width=0.85\linewidth]{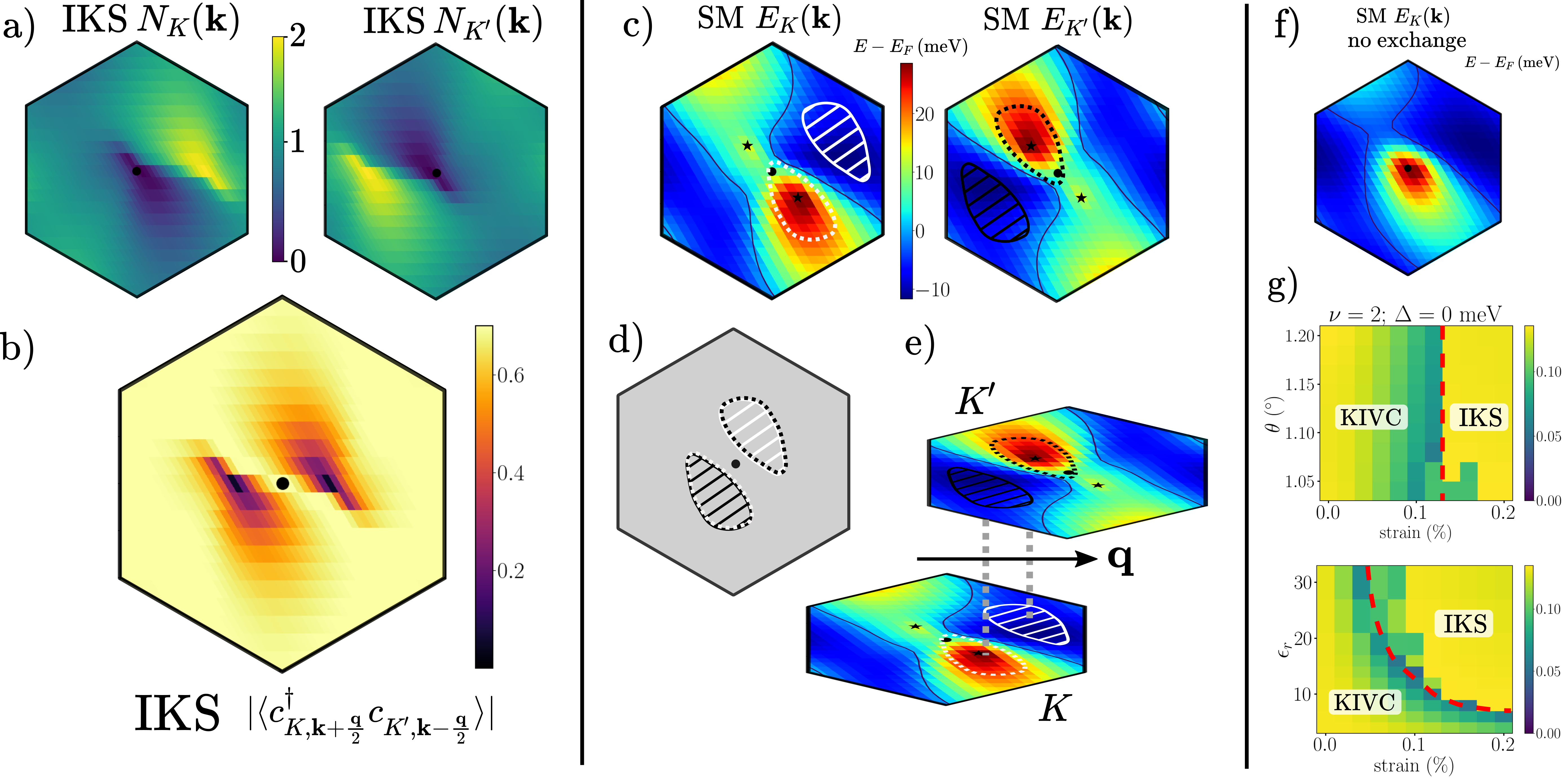}
    \caption{ mBZ-resolved properties of the IKS state at $\nu=-2$. 
    a) Valley populations of the IKS state.
    b) Frobenius norm of the IVC at wavevector $\bm{q}=\bm{G}_1/3$. System size is $48\times16$, strain is $0.2\%$, substrate is $\Delta=0\,\mathrm{meV}$, and NLT is not included.
    c) Dispersions of the lower band for the metastable symmetry-preserving self-consistent SM. Hatched (dotted) lobes, situated near low (high) energy regions, will be predominantly filled (empty) in the IKS state. Black lines indicate the Fermi surface, and black stars mark Dirac point locations. 
    d,e) Schematic construction of the IKS state---a relative momentum boost of the valleys by $\bm{q}$ allows the lobes to overlap each other. Regions not within the lobes participate strongly in IVC.
    f) HF spectrum for the same density matrix used in c) except that exchange has been neglected.
    g) $\nu=+2$ HF phase diagram in the strain-twist angle plane, and the strain-relative permittivity plane. System size is $12\times12$.}
    \label{fig:largespiral_nu-2}
    \end{figure*}

\subsection{Structure and energetics of the IKS state}
\label{sec:IKSproperties}

Microscopically, the $\hat{\mathcal{T}}$-invariant IVC order of the IKS state induces a Kekul\'e-like pattern on the graphene scale (Fig.~\ref{fig:IVCSrealspace}), with orientation determined by the local IVC angle $\theta_{\mathrm{IVC}}$. Since the symmetry-breaking occurs predominantly within the central bands (see Appendix~\ref{secapp:additional}, specifically Fig.~\ref{figapp:band_coherences}), the Kekul\'e or $\sqrt{3}\times\sqrt{3}$ pattern, which triples the graphene unit cell, is strongest in the AA regions where the flat band wavefunctions are spatially localized. However, the finite-$\bm{q}$ character of the IKS state means that the microscopic Kekul\'e-like patterns differ between different AA regions, as dictated by the combined moir\'e lattice translation and valley-$U(1)$ rotation symmetry $\hat{T}_{\bm{a}_i}'$ noted above ~(Fig.~\ref{fig:phase_nu2}b). For $\bm{q}=\bm{G}_1$/3, the system forms stripes along the $\bm{a}_2$ direction where the graphene-scale Kekul\'e pattern is the same. Since the translation-breaking order is purely IVC, with no $-\bm{q}$ or higher harmonic components, there is no additional charge reconstruction at the moir\'e scale (see Section~\ref{sec:LG}).

Further insight into the properties of the IKS state can be gained by analyzing its momentum-resolved single-particle density matrix in more detail. Figure~\ref{fig:largespiral_nu-2}b plots the strength of the IVC in momentum space, showing that it is close to the maximum value $1/\sqrt{2}$ throughout most of the mBZ. The exceptions are at two lobes in the mBZ, where the electron populations $N_{\tau}(\bm{k})$ in the two valleys Fig.~\ref{fig:largespiral_nu-2}a are close to 0 or 2. The total occupation at each $\bm{k}$ is 2 consistent with an insulating state. The strong momentum-dependence of the IKS state sets it apart from previously studied mean-field phases \cite{Bultinck2020}. From our numerics we find that the same coherence structure is repeated for both spin species. Therefore, henceforth we consider spin to simply be a spectator degree of freedom, an assumption which is further validated by the `basis spiral' analysis in Section~\ref{sec:FMstripes}.

The locations of IVC-depletion provide strong clues as to the mechanism underlying IKS formation. In Figure~\ref{fig:largespiral_nu-2}c, we calculate for each valley the HF spectrum of the lower band of the self-consistent symmetry-preserving semimetal (SM). This captures the major momentum-dependent effects that strain and interactions have on the band structure. The dispersions of the two valleys are related by TRS. All Dirac points lie above $E_F$ at $\nu=-2$. Near $\Gamma_M$, there is a region of very high energy (red) that coincides with one of the Dirac points. There is also a region of low energy (dark blue) lying in some other region of the mBZ. Because of TRS, the low/high energy lobes (indicated by hatched/dotted regions) in the two valleys are related by $\bm{k}\rightarrow-\bm{k}$.

We now sketch an intuitive picture for how these dispersion features influence the parameters of the IKS order. Figure~\ref{fig:largespiral_nu-2}d,e demonstrates that coupling the two valleys at a finite $\bm{q}$ can pairwise align a high energy lobe with a corresponding low energy lobe in the other valley. In these momentum regions, the system will choose to polarize into the energetically favorable valley (Fig.~\ref{fig:largespiral_nu-2}a). Elsewhere, substantial valley hybridization is induced. In this way, the IKS state is able to maximize IVC while respecting the prominent characteristics of the band dispersion. Each $\bm{\tilde{k}}$ is equally populated, allowing for an insulating state. Note that attempting to induce IVC at $\bm{q}=0$ instead runs into issues---a large portion ($\sim 4\times$lobe area) of the mBZ would be unable to participate in the IVC since the lobes have small overlap. Furthermore, the total electron occupations would vary as a function of $\bm{\tilde{k}}$, meaning the state cannot be insulating. 

\begin{figure}
    \includegraphics[ width=1\linewidth]{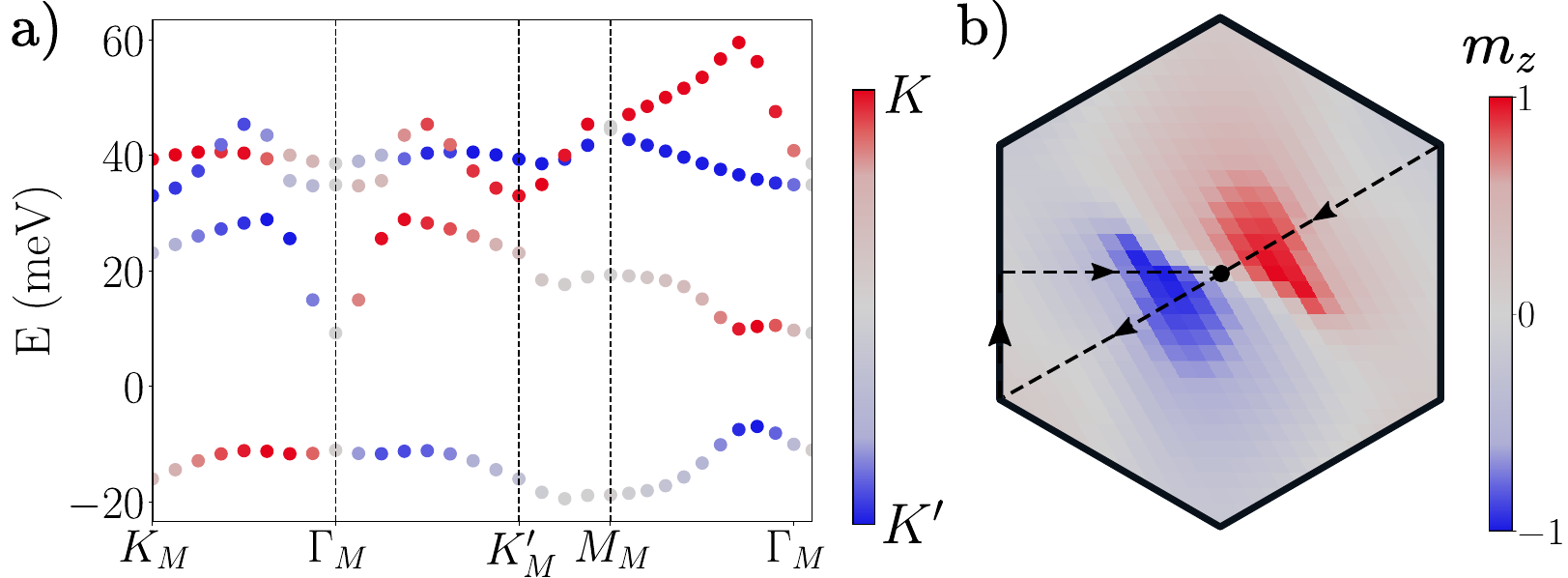}
    \caption{a) Band structure of the IKS state at $\nu=-2$, with color indicating the valley polarization of the HF orbitals. b) $z$-component of the vector $\bm{m}(\tilde{\bm{k}})$ in the parameterization of Eq.~\ref{eq:IKSansatz}. Dashed lines indicate mBZ path in a). In both plots, $\tilde{\bm{k}}$ refers to the valley-dependent `boosted' momentum $\tilde{\bm{k}}+\tau_z \bm{q}/2$.}
    \label{fig:IKS_bandstruct_mz}
\end{figure}

This perspective naturally explains the strong $\bm{k}$-dependence of IVC and the slow variation of the IKS energy with $\bm{q}$. The somewhat diffuse features of Fig.~\ref{fig:largespiral_nu-2}c mean that for nearby $\bm{q}$, the locations/shapes of the lobes only change slightly, leading to a small and roughly isotropic wavevector stiffness. A simple estimate for the ideal wavevector $\bm{q}_0$ can be made by connecting the minimum energy momentum in valley $K'$ with the maximum energy peak in valley $K$. The predicted $\bm{q}_0$ is broadly consistent with HF results of the IKS state for a range of strain angles $\varphi$ (for details see App. \ref{secapp:additional}, in particular Fig.~\ref{figapp:varphi_q0}).  
We emphasize that this scenario opens a gap at $E_F$ but, unlike most of the translation-invariant insulators, does not rely on gapping out the Dirac points, which remain high in energy above $E_F$. Instead the $\bm{k}$-dependent IVC hybridizes the two valleys at finite $\bm{q}$ and pulls the occupied band below the rest of the states (Fig.~\ref{fig:IKS_bandstruct_mz}a).

We can construct a simple \textit{ansatz} for the IKS projector in the absence of substrate alignment, which matches the HF numerics extremely well. With $\sigma$ denoting the sublattice-polarized basis, we define two mutually commuting $SU(2)$ Lie algebras 
$\boldsymbol{\gamma}=(\sigma_x,\tau_z\sigma_y,\tau_z\sigma_z)$
and $\boldsymbol{\eta}=(\tau_x\sigma_x,\tau_y\sigma_x,\tau_z)$, in terms of which $C=\gamma_z$ \cite{SkyrmionSC}. We partially fix the gauge by requiring that $\hat{C}_{2z}$ acts as $\tau_x\sigma_x$ and $\hat{\mathcal{T}}$ as $\tau_x \hat{\mathcal{K}}$---the remaining gauge freedom acts as $e^{i\varphi_{\bm{k}}\tau_z\sigma_z}$. The spin-singlet IKS state at $\nu=-2$ can then be parameterized by the projector
\begin{equation}\label{eq:IKSansatz}
    P_{\nu=-2}(\bm{\tilde{k}})=\frac{1}{4}(\mathds{1}+\bm{n_{\tilde{k}}}\cdot\boldsymbol{\gamma})(\mathds{1}+\bm{m_{\tilde{k}}}\cdot\boldsymbol{\eta})\,,
\end{equation}
where $\bm{\tilde{k}}$ labels the eigenvalues of $\hat{T}'_{\bm{a}_i}$ as discussed in the previous subsection, the gauge-variant $\bm{n_{\tilde{k}}}$ is entirely in the $x-y$ plane, and an identity matrix in spin space is implicit. Across most of the mBZ, the vector $\bm{m_{\tilde{k}}}$ lies in-plane with a constant angle that can be changed by a global $U(1)_V$-rotation. At the lobes, $\bm{m_{\tilde{k}}}$ orients towards the poles, reflecting the valley polarization in these momentum regions (recall that $\eta_z = \tau_z$) (Fig.~\ref{fig:IKS_bandstruct_mz}b). Since $\gamma_x,\gamma_y$ in Eq.~\eqref{eq:IKSansatz} anticommute with the Chern number $\gamma_z$, there is both inter-Chern and intra-Chern IVC in the IKS state with equal magnitude. This implies that the IVC significantly entangles bands with opposite Chern number, in contrast to the usual $U(4)\times U(4)$ ferromagnets found in previous mean-field studies \cite{Bultinck2020}. Importantly, this distinguishes the type of IVC order found here from the uniform $\mathcal{T}$IVC state of Ref.~\cite{Bultinck2020} (which also preserves TRS at even integer fillings). In terms of symmetries, the only differences between the $\mathcal{T}$IVC and IKS states is that the former is translationally invariant and cannot be made spin singlet by opposite spin rotations in the two valleys. Both states can be made $\hat{C}_{2z}\hat{\mathcal{T}}$-symmetric by a suitable global valley-$U(1)$ rotation. The IKS projector at $\nu=+2$ can be found by particle-hole conjugation.

The construction outlined above, involving the nesting of features of a parent symmetry-preserving band structure, is suggestive of a weak-coupling instability. However the $U(1)_V$-breaking coherence occurs nearly everywhere in the mBZ, instead of just the lobe boundaries. Furthermore, the Fermi surfaces of the SM or the non-interacting BM model generically bear little relation to the momentum structure in the IKS state. Indeed, both strain and interactions play a vital role in renormalizing the central bands and setting the stage for symmetry-breaking phases---the non-interacting BM bands have a total bandwidth $\sim5\,\text{meV}$, which broaden to $\sim15\,\text{meV}$ in the presence of strain (breaking $\hat{C}_{3z}$ and shifting the Dirac points up/down), and finally $\gtrsim50\,\text{meV}$ with the inclusion of interactions. Strengthening the Coulomb interaction (by reducing $\epsilon_r$) favours the strong-coupling states (Fig.~\ref{fig:largespiral_nu-2}g). Strain thus effectively tunes the system from strong coupling, where Chern-diagonal ferromagnetic states dominate, to \emph{intermediate} coupling where other phases (such as the IKS order) emerge that violate the $U(4)\times U(4)$ hierarchy. In the language of Ref. \cite{TBGIV}, strain does not significantly impact the quality of the `flat-metric condition' (see App.~\ref{secapp:additional}, in particular Fig.~\ref{figapp:FMC}), which is used to prove that the uniform $U(4)\times U(4)$ ferromagnet is an exact ground state of the pure interaction Hamiltonian at all integer $\nu$. Instead, strain substantially increases the dispersion, thereby undermining the validity of the perturbative analysis about the ferromagnetic states, and allowing for alternative states such as the IKS to come in. On the other hand, twist angle, which weakly influences the non-interacting central band dispersion, does not have a significant impact on the phase diagram (Fig.~\ref{fig:largespiral_nu-2}g).

This strong- to weak-coupling crossover can also be viewed through the lens of direct versus exchange energy~\cite{Guinea2018}. The intra-Chern states at small strain are stabilized by exchange, in analogy with quantum Hall ferromagnetism. At larger strains, including just the Hartree piece of the interaction already recreates the key features of the band renormalization, as shown in Fig.~\ref{fig:largespiral_nu-2}f. As verified in Section~\ref{sec:FMstripes}, the IKS state is more competitive farther away from charge neutrality, in harmony with the larger Hartree peak (dip) expected for increasing hole (electron) doping \cite{Pierce2021}. All particles will feel this increased Hartree potential, while exchange effects are only applicable between electrons of the same flavor. We caution though that this direct-exchange dichotomy is not so clear-cut in practice---separation of the IKS energy into its components reveals that both Hartree and Fock contributions change significantly with comparable magnitude as a function of $\bm{q}$. Also, exchange does significantly perturb the band structure of the self-consistent SM, and its inclusion is necessary to obtain reasonable predictions for $\bm{q}_0$. This implies that a proper treatment of both terms is required to adequately capture the physics of TBG for realistic parameters (further details in App.~\ref{secapp:additional}, in particular Fig.~\ref{figapp:q0_prediction} and~\ref{figapp:energy_breakdown}). 

    \begin{figure*}
    \includegraphics[ width=0.75\linewidth]{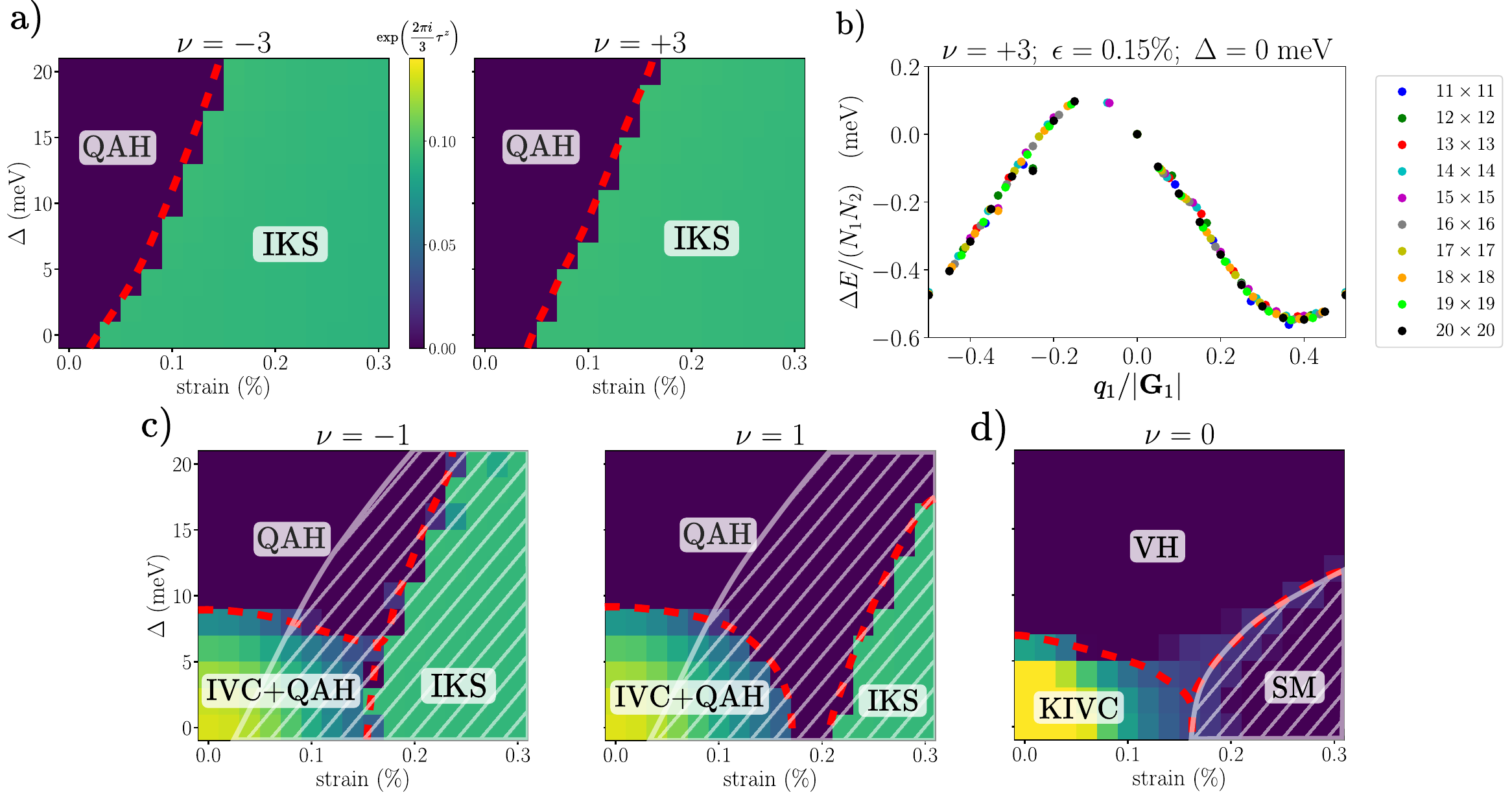}
    \caption{a) $\nu=\pm3$ strain-substrate HF phase diagram using same parameters as Fig.~\ref{fig:phase_nu2} (enlarged view of panels from Fig.~\ref{fig:all_phases}). b) Dispersion relation of the IKS state at filling $\nu=+3$ for different system sizes (with $\Delta=0$ meV, $\epsilon=0.15\%$ and no NLT). $\Delta E$ is the energy of the IKS solution at a given wavevector $q\hat x$ compared to the energy of the translationally invariant ($q=0$) QAH state.
    c,d) Same as a) except at $\nu=\pm1$ and $\nu=0$ ($\epsilon_r=15$), again reproducing Fig.~\ref{fig:all_phases}. Hatched regions denote absence of a charge gap.}
    \label{fig:phase_nu3}
    \end{figure*}

\section{Ferromagnetic Kekul\'e spirals at $\nu=\pm 1$ and $\nu=\pm3$}
\label{sec:FMstripes}

We now turn to other nonzero fillings, focusing on the same departures from the idealized BM model, fixing NLT as above but exploring the phase structure in the strain-substrate plane. As we show, IKS order appears to be a ubiquitous feature for relatively modest strain, and largely independently of substrate strength.

For $\nu=\pm 3$, a significantly smaller strain $(\epsilon \sim 0.02 - 0.05 \%)$ is able to destabilize the QAH state found upon adding interactions to the unperturbed BM Hamiltonian (Fig.~\ref{fig:phase_nu3}a). Depending on details of the system parameters, it is possible to nucleate an intervening spin-valley polarized nematic semimetal (see App. \ref{secapp:additional}, and in particular Fig.~\ref{figapp:nematic}). At zero substrate potential and strain, an alternative period-doubling stripe which preserves $\hat{C}_{2z}\hat{\mathcal{T}}$ and $U(1)_V$ has also been proposed~\cite{Kang2020}. With the parameters used in the main text, we find a direct transition from the QAH to an IKS state under increasing strain. The IKS state at $\nu = \pm 3$ has the same symmetry properties as the one at $\nu=\pm2$, except that it can carry a spin polarization at $U(2)_K\times U(2)_{K'}$ level (of maximum $1\mu_B$ per moir\'e cell). For (anti-)ferromagnetic Hund's coupling, the spin moments carried by the two valleys will (anti-)align. The dispersion curve in Fig.~\ref{fig:phase_nu3}b is also similar (compare Fig.~\ref{fig:phase_nu2}c), but it has a deeper minimum, and hence a larger IVC stiffness, due to the increased renormalization of the bands. 

Given the tiny strains required for the IKS to beat the QAH at $\nu=\pm3$, a natural question that arises is whether the IKS can in principle be the ground state at zero strain. We emphasize that there is no fundamental reason that prohibits this scenario from occurring; for strong enough perturbations about the $U(4)\times U(4)$ limit, the strong-coupling insulators can be superseded by states outside the QHFM paradigm. While our Hartree-Fock calculations suggest this is not the case for our choice of parameters, the fact that the IKS can be obtained self-consistently without strain (see App.~\ref{secapp:additional}, in particular Fig.~\ref{figapp:anneal} which shows that energy cost is less than 1~meV per unit cell) is a strong indication that the IKS remains a highly competitive state. Altering details of the model Hamiltonian or choosing different parameters (e.g. increasing the relative permittivity) may well tilt the balance in favor of IKS.

At $\nu=\pm1$, the weak substrate potential region of the $U(1)_V$-symmetric QAH phase has non-zero IVC order~\cite{Zhang2020HF,TBGIV,BernevigEfetov2020} (Fig.~\ref{fig:phase_nu3}c). The transition from the QAH and QAH+IVC states to the IKS state now occurs at a larger strain ($\epsilon \sim 0.15-0.2 \%$). The effect of intervalley-Hund's terms is similar to that at $\nu=\pm3$, i.e. a net spin polarization of (0) 1 for (anti-)ferromagnetic coupling. In contrast to the other integer fillings, at $\nu = \pm 1$, the IKS state never develops a charge gap. For completeness, in Fig.~\ref{fig:phase_nu3}d we also present the phase diagram at $\nu=0$, which shows no indications of translation symmetry breaking. The KIVC gives way to a symmetric SM at finite strain~\cite{Parker2020}, and a VH insulator at finite substrate.

All numerically obtained IKS states preserve spinless TRS $\mathcal{T}$, implying that the Chern number $C$ vanishes. This fact is remarkable for the odd fillings, since conventional spin-valley polarized ferromagnets can only accommodate phases with odd $C$. However, recent experiments have shown the existence of even-$C$ gapped phases extending down to zero magnetic field at odd fillings \cite{Pierce2021}. One possible route to achieving this is by folding the mBZ in half and forming period-2 charge order, as theoretically argued by some authors \cite{Pierce2021}. Each Chern band (a finite sublattice splitting was considered) splits into a $|C|=1$ and $0$ mini-band, and a variety of different $C$ states can be obtained by selectively polarizing these. Our work proposes a fundamentally distinct scenario, relying instead on IVC to produce the requisite $C=0$ bands, and on moir\'e translation breaking to minimize the energy. The IKS order is agnostic to the presence of substrate alignment, and is a natural robust insulating candidate for experiments where strain is often an external confounding factor. Furthermore, as explained in detail in the next section, translation-breaking phases with non-zero Chern number can be obtained by `stacking' a phase with IKS order with other translation-symmetric phases to achieve the requisite band filling. Characterizations of the moir\'e charge order or strain in the sample of Ref.~\cite{Pierce2021} would help determine which theoretical scenario is operative there.

\section{Relationship between IKS states at different fillings}
\label{sec:relation}
Since the IKS states have similar properties at all non-zero integer fillings, we expect them to be closely related. To make the connection explicit, we consider the $|\nu|=3$ IKS state as a `{basis} spiral'. We start with the $\nu=-3$ IKS state with spin-polarization enforced for simplicity. In order to construct a $\nu=-2$ IKS, we take two copies of the $\nu=-3$ basis spiral in order to obtain a spin-unpolarized IKS. The same construction is possible at positive filling by particle-hole conjugation. The notion of the $|\nu|=2$ stripe as two copies of the $|\nu|=3$ stripe is consistent with the relative scale of the $U(1)_V$-breaking order parameter in Fig.~\ref{fig:phase_nu2}a being $\sqrt{2}$ times that in Fig.~\ref{fig:phase_nu3}a.

For the $\nu=-1$ IKS state, we note that the translational symmetry breaking is entirely in one spin sector, whereas the other spin sector has the same symmetries as the VH state at $\nu=-2$. This motivates the following construction: We start with the spin-polarized VH state at $\nu=-2$ and add to it the $\nu=-3$ IKS state in the opposite spin sector. Consistent with this fact, the $U(1)_V$-breaking order parameter has the same magnitude in the IKS phases at $|\nu|=1$ and $3$. 

\begin{figure}
    \includegraphics[ width=1\linewidth]{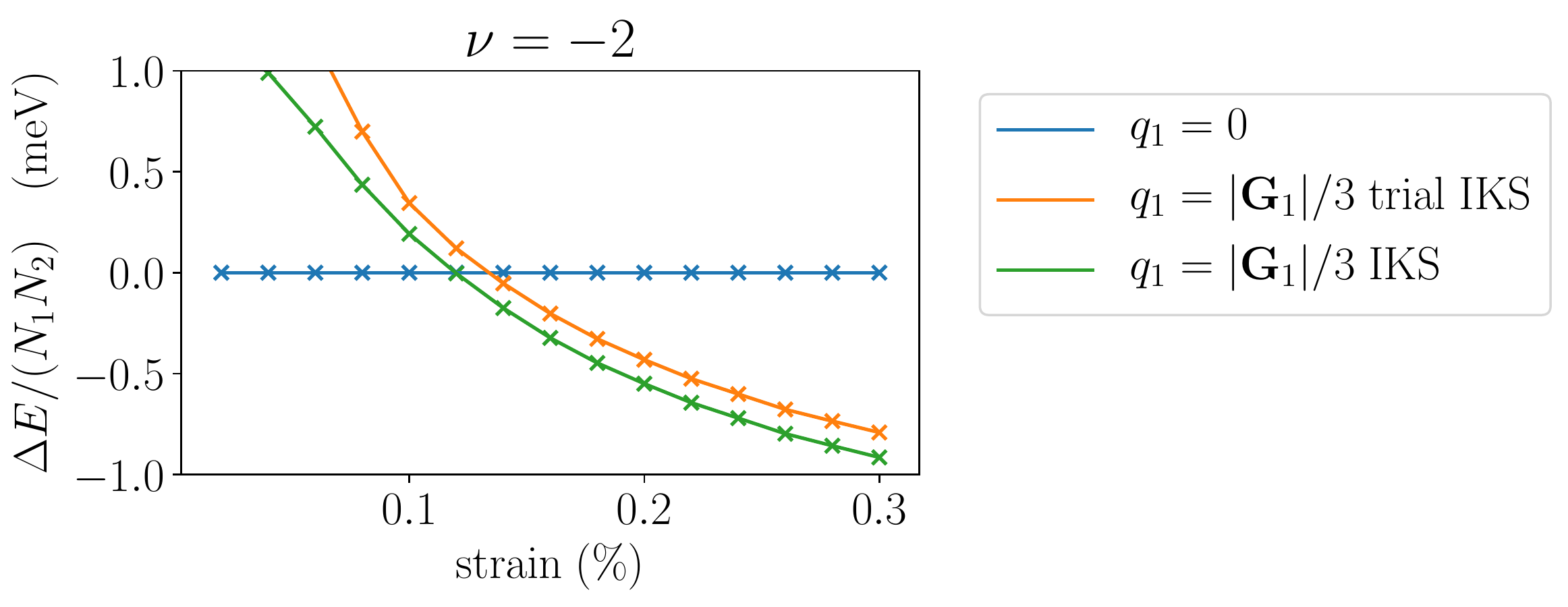}
    \caption{Comparison of the energies of the translationally invariant state, the IKS state and the IKS trial state at $\nu=-2$. The IKS trial state is obtained by taking two copies of the IKS HF solution at $\nu=-3$. System size is $12\times12$, $\Delta=0\ $ meV and NLT is included.}
    \label{fig:trial}
    \end{figure}

We show in Fig.~\ref{fig:trial} that the trial states for the $\nu=-2$ IKS state based on this construction have energies that are very close to the self-consistent HF solution at those fillings. We also find that HF simulations using these trial states as initial inputs converge very quickly to the self-consistent IKS ground state at that filling, demonstrating that the trial states have the correct correlations. For $\nu=-1$ the trial state energies are not as close to those of the self-consistent HF IKS state due to the fact the IKS state is not insulating at this filling (see App.~\ref{secapp:additional}).

\section{Landau-Ginzburg Analysis}
\label{sec:LG}

In this section, we explore the interplay between Kekul\'e spiral states and charge order, by generalizing the Landau-Ginzburg analysis of Ref.~\cite{ZacharKivelsonEmery}. This has two primary motivations. First,  charge order can be detected in a larger variety of experimental probes: for instance high-resolution scanning single-electron transistors (SETs) can detect charge modulation, as can scanning tunneling microscopy (STM) measurements (which can also directly access the moir\'e-scale-modulated Kekul\'e distortion characteristic of the IKS state.) Second, charge order is more readily pinned by external potentials, and so an IKS state with a subsidiary charge order is likely to respond more strongly to quenched disorder and also to experience stronger commensuration effects. 

Our Landau-Ginzburg construction involves the following order parameters\footnote{Note that there are other operators we could use instead of $I^{x/y}$. For example, $J^{x/y} = \frac{1}{N}\sum_{\k} \langle \hat{\psi}^\dagger_{\k+\q} \tau_{x/y}\hat{\psi}_{\k}\rangle$ and $K^{x/y} = \pm\frac{1}{N}\sum_{\k} \langle \hat{\psi}^\dagger_{\k+\q} \tau_{y/x}\sigma_z\hat{\psi}_{\k}\rangle$ have transform similarly to $\I$. However, using all these operators in the Landau-Ginzburg free energy would not change the conclusions of our analysis -- so for simplicity, and because $|\I| = \sqrt{\mathbf{J}^2+\mathbf{K}^2}$ in the IKS state, we consider only $\I$.}
\begin{eqnarray}
I^x_\q & = & \frac{1}{N}\sum_{\k} \langle \hat{\psi}^\dagger_{\k+\q} \tau_x\sigma_x\hat{\psi}_{\k}\rangle \, ,\\
I^y_\q & = & \frac{1}{N}\sum_{\k} \langle \hat{\psi}^\dagger_{\k+\q} \tau_y\sigma_x\hat{\psi}_{\k}\rangle \, ,\\
\rho_{\q} & = & \frac{1}{N}\sum_{\k} \langle \hat{\psi}^\dagger_{\k+\q}\hat{\psi}_{\k}\rangle\, ,
\end{eqnarray}
where $N$ is the number of moir\'e unit cells, and $\hat{\psi}^\dagger_{\k,\tau,\sigma}$ creates an electron in one of the active Chern bands with Chern number $C = \tau_z\sigma_z$. We have also partially fixed the gauge by requiring that $\hat{C}_{2z}\hat{\mathcal{T}}$ acts as $\sigma_x \mathcal{K}$ and $\hat{\mathcal{T}}$ as $\tau_x\mathcal{K}$. The notation $I$ refers to either `IVC' or `Isospin'.

Under the symmetries of strained TBG, the order parameters transform as
\begin{eqnarray}
{U}(1)_V:\quad& \I_\q & \rightarrow   R(\phi) \I_\q \nonumber\\
 & \rho_\q & \rightarrow  \rho_\q \\
  \hat{C}_{2z}:\quad & \I_\q & \rightarrow  \left(\begin{matrix}1  \\ & -1 \end{matrix}\right)\I_\q^*\nonumber \\
  & \rho_\q & \rightarrow \rho_\q^* \\
\hat{\mathcal{T}}:\quad & \I_\q & \rightarrow \I_\q \nonumber\\
 & \rho_\q & \rightarrow \rho_\q\, ,
\end{eqnarray}
where $\I_\q=(I^x_\q,I^y_\q)$, $R(\phi)$ is a $2\times 2$ rotation matrix, and we have used that $\I_{-\q} = \I_\q^*$ and $\rho_{-\q} = \rho_\q^*$. Note that both $\hat{C}_{2z}$ and $\hat{\mathcal{T}}$ act differently on the isospin vector $\I$ than on conventional spin-$1/2$ degrees of freedom. An interesting consequence of these unusual symmetry actions is that the bilinear $i\epsilon_{ij}I^i_\q I^{j*}_\q$ with $\epsilon_{ij}=-\epsilon_{ji}$ and $\epsilon_{xy}=1$, respects all symmetries of strained TBG, including superlattice translations.

As a disclaimer, we  point out that the Landau-Ginzburg free energy we  study below is not invariant under threefold rotations. One justification is that strain breaks $\hat{C}_{3z}$. However, since the strain is very small, a three-fold rotationally symmetric functional (plus small anisotropies) should actually still be an appropriate starting point. Instead, a better justification for a `uni-directional' free energy is that we are interested only in uni-directional physics here. In other words, we can think of our free energy functional as descending from a parent functional which is $\hat{C}_3$ symmetric, but which breaks the threefold rotation symmetry spontaneously. The free energy we write down is then obtained by expanding the parent functional around one of the three valleys.

A Landau-Ginzburg free energy consistent with alll the above considerations is given by:
\begin{widetext}
\begin{eqnarray}
F & = & \frac{1}{2} r_\rho |\rho|^2 + U_\rho |\rho|^4
+ \frac{1}{2}r_I |\I |^2 + U_I |\I |^4 
 + \frac{i}{2} r_x  \epsilon_{ij}I^i I^{j*} - U_x \left(i\epsilon_{ij}I^i I^{j*} \right)^2 \nonumber\\
 & & + \lambda_1 \left[ \left(\I \cdot \I \right)\rho^* + \left(\I \cdot \I \right)^*\rho\right]
+ \lambda_2 |\I |^2 |\rho |^2 + \lambda_3 | \I |^2\left( i\epsilon_{ij}I^i I^{j*}  \right) 
+ \lambda_4 |\rho |^2\left( i\epsilon_{ij}I^i I^{j*}  \right)\, ,
\end{eqnarray}
\end{widetext}
where $\I\equiv I_\q$ and $\rho\equiv \rho_{2\q}$. The terms with coefficients $r_x,\lambda_3$ and $\lambda_4$ are not present in the analysis of Ref.~\cite{ZacharKivelsonEmery}. We will now argue that despite the presence of these additional terms, the physical conclusions of Ref.~\cite{ZacharKivelsonEmery} survive. From now on, we will normalize the order parameters such that $U_\rho = U_I = 1$, and require that $U_x<1$ in order for the free energy to be bounded from below. We will also consider the case with $\lambda_2 = \lambda_3 = \lambda_4 = 0$, as these terms only quantitatively change the physics we want to discuss here.

As a first step, we choose a coordinate system such that the order parameters can be written as
\begin{eqnarray}
\I & = & | \I |\left( \cos\alpha\; \hat{\mathbf{e}}_1 + i \sin\alpha\; \hat{\mathbf{e}}_2\right) \\
\rho & = & |\rho| e^{i\theta}\, ,
\end{eqnarray}
where $\hat{\mathbf{e}}_i$ are two orthogonal unit vectors,  and $0\leq \alpha \leq \pi/4$. If $\alpha = 0$ then the IVC order is collinear, whereas if $\alpha = \pi/4$ the Kekul\'e spiral is perfectly circular, which is what we find in HF simulations. For intermediate values of $\alpha$ the Kekul\'e spiral has a non-zero eccentricity $\tan \alpha$. Using the above expressions for the order parameters, the free energy can be written as
\begin{eqnarray}
F & = & \frac{1}{2} r_\rho |\rho|^2 + |\rho|^4 + \frac{1}{2}r_I |\I |^2 + |\I |^4 \nonumber\\
 & & + \frac{1}{2} r_x |\I |^2 \sin 2\alpha - U_x |\I |^4 \sin^2 2\alpha  \nonumber\\
 & & + 2\lambda_1 |\I |^2 |\rho| \cos 2\alpha \cos \theta.
\end{eqnarray}
Next,we minimize $F$ with respect to $\theta$ in the presence of non-zero $|\I |$. From this we find that $\theta$ is either $0$ or $\pi$, with the optimal value being determined by the sign of $\lambda_1$. Note that for $\theta = 0$ or $\pi$, the charge order preserves the $\hat{C}_{2z}$ symmetry (as does the circular Kekul\'e spiral state).

Now, from minimizing $F$ with respect to $\alpha$, it follows that $\alpha = \pi/4$ only if $|\rho|=0$. Minimizing $F$ with respect to $|\rho|$ gives us the reverse implication: if $F$ is minimized for $|\rho|=0$, then this implies that $\alpha = \pi/4$ (and also that $r_\rho > 0$). Combining the above implications, we conclude that there is no charge order if and only if the IKS state is perfectly circular. This interplay between charge order and eccentricity suggests
interesting possibilities to experimentally detect the IKS order. For example, let us consider a situation where the TBG sample is strained in a direction orthogonal to the spiral wavevector $\q$. Via the distortion of the graphene bond hoppings, this will introduce some non-zero eccentricity for the IKS state and thus charge order with half the period of the Kekul\'e spiral. Conversely, if non-zero charge order is induced via an inhomogeneous electrostatic potential with wavevector $2\q$, then the IKS state will respond by changing the amplitude of the Kekul\'e pattern differently in inequivalent AA regions.

Minimizing $F$ exactly with respect to $\alpha$ is cumbersome, requiring us to minimize the expression
\begin{equation}\label{eqmin}
\frac{1}{2}r_x \sin 2\alpha - U_x |\I |^2 \sin^2 2\alpha - 2|\lambda_1| |\rho| \cos 2\alpha.
\end{equation}
Since we are only interested in the parameter regime close to where the circular spiral state is lowest in energy, let us set $2\alpha = \frac{\pi}{2} + \delta$ and expand in powers of $\delta$. Keeping terms to second order in $\delta$, Eq. \eqref{eqmin} becomes
\begin{equation}
\frac{1}{2}r_x \left( 1 - \frac{1}{2}\delta^2\right) - U_x |\I |^2 (1-\delta^2) + 2|\lambda_1| |\rho| \delta.
\end{equation}
Minimizing this expression with respect to $\delta$, we find
\begin{equation}\label{eqminB}
\delta = \frac{|\lambda_1| }{U_x |\I |^2 - r_x/4}|\rho|,
\end{equation}
whence we can write the free energy as
\begin{equation}
F = \frac{1}{2}\left(r_\rho - \frac{4|\lambda_1|^2 |\I |^2}{U_x |\I |^2 - r_x/4} \right) |\rho|^2 + |\rho|^4  + \dots \, ,
\end{equation}
where the dots stand for terms which do not involve the charge density (note that if one keeps higher orders of $\delta$, then the coefficient of $|\rho|^4$ will also receive a correction). It is now clear that there will be a second order phase transition to a phase with non-zero charge order when
\begin{equation}
r_\rho = \frac{4|\lambda_1|^2 |\I |^2}{U_x |\I |^2 - r_x/4}\, .
\end{equation}
As discussed above, in the charge ordered phase the Kekul\'e spiral necessarily becomes an elliptical spiral. It thus follows that by tuning a single parameter it is possible to go from the circular Kekul\'e spiral phase to an elliptical Kekul\'e spiral phase via a second order phase transition. As a direction for future work, it would be interesting to identify experimental knobs that can drive such a transition.

\section{Quenched Disorder}
\label{sec:quenched}
For the KIVC state, which occurs at very small strain, quenched disorder is relatively innocuous. The reason is that this state breaks the physical time-reversal symmetry, while physical impurities in graphene are time-reversal symmetric, which implies that they cannot couple as random-field disorder to the KIVC order. The IKS state, on the other hand,  preserves time-reversal symmetry and is consequently less protected against disorder. For example, because the IKS state has a Kekul\'e pattern on every AA region, graphene-scale bond disorder 
couples to the IVC order as a random field.

Motivated by this observation, we now investigate the effect of bond disorder on the graphene scale in more microscopic detail (we focus on bond disorder for concreteness -- the analysis  for graphene-scale potential disorder is similar). In particular, we consider the following disorder Hamiltonian:
\begin{equation}
H_{\text{dis}} = \sum_{\R}\sum_{j=1,2,3} \delta t(\R + \boldsymbol{\delta}_j/2)(c^\dagger_{\R,A}c_{\R+\boldsymbol{\delta}_j,B} + h.c.)\,,
\end{equation} 
where $\R$ runs over the positions of the A sites, and $\boldsymbol{\delta}_1,\boldsymbol{\delta}_2$ and $\boldsymbol{\delta}_3$ connect each A site to its three neighbouring B sites. In the BM basis, the disorder Hamiltonian becomes
\begin{equation}
H_{\text{dis}} = \frac{1}{A} \sum_{\k \in \text{mBZ}}\sum_{\q} \psi^\dagger_{\k+\q}G_\q(\k) \psi_{\k}\, ,
\end{equation}
where $A$ is the area of the sample, and
\begin{widetext}
\begin{eqnarray}
\left[G_\q(\k)\right]_{(\tau',n'),(\tau,n)} & = & \langle u(\k+\q)_{\tau',n'}|F_{\q}(\k)|u(\k)_{\tau,n}\rangle \\
\left[F_\q(\k) \right]_{(\tau',l',\sigma',\G'),(\tau,l,\sigma,\G)} & = & \delta_{l',l} \delta_{\G,\G'} \left(\begin{matrix} & \delta t(\q_{\tau\tau'}) f(\k+\G + \q_{\tau\tau'}/2) \\  \delta t^*(\q_{\tau\tau'}) f^*(\k+\G+\q_{\tau' \tau}/2)& \end{matrix} \right)_{\sigma',\sigma}
\end{eqnarray}
\end{widetext}
Here, $\tau,l,\sigma$ respectively denote valley, layer and sublattice, $|u(\k)_{\tau,n}\rangle$ is the periodic part of the BM Bloch states, $\G$ are moir\'e reciprocal lattice vectors, and $\K$ is the position of the Dirac point corresponding to the $\tau=+$ valley in the graphene Brillouin zone. We have also defined $\q_{\tau' \tau} = \q + (\tau'-\tau)\K$ and $f(\k) = \sum_{j=1,2,3} e^{i\boldsymbol{\delta}_j\cdot \k}$. Note that at this point we have performed an exact unitary transformation, so $n$ and $n'$ run over all the BM bands, and not only over the active bands.

Our HF simulations indicate that all IVC order is carried by the active bands, indicating the subspace relevant to studying the order parameter. Let us therefore perform a Schrieffer-Wolff transformation to obtain the effect of $H_{\text{dis}}$ on this subspace. We define $G^{ij}_\q(\k) \equiv P^iG_\q(\k)P^j$, where $P^0$ ($P^1$) is the projector onto the active (remote) bands. Assuming that the Fermi energy lies within the active bands, the Schrieffer-Wolff transformed disorder Hamiltonian (up to second order) can be written as follows:
\begin{widetext}
\begin{align}\label{SW}
H_{\text{dis}} & = \frac{1}{A} \sum_{\k,\q} \psi^\dagger_{\k+\q}G^{00}_{\q}(\k)\psi_\k \\
& - \frac{1}{2A^2} \sum_{\k,\q,\q'}\sum_{n,n',\tilde{m}}\psi^\dagger_{n',\k+\q+\q'}\left[G^{01}_{\q'}(\k+\q)\right]_{n'\tilde{m}}\left(\frac{1}{E_{\tilde{m},\k+\q}-E_{n,\k}} + \frac{1}{E_{\tilde{m},\k+\q}-E_{n',\k+\q+\q'}} \right)\left[G^{10}_\q(\k)\right]_{\tilde{m}n}\psi_{n,\k}\, , \nonumber
\end{align}
\end{widetext}
where the sum over $\tilde{m}$ runs over the remote bands, and $E_{n,\k}$ are the energies of the BM bands.
In what follows, we are only interested in the part of $H_{\text{dis}}$ which couples as a random field to the IKS state, i.e. the part of $H_{\text{dis}}$ which is off-diagonal in the valley indices. 

Let us first consider the first-order term in $H_{\text{dis}}$. An important observation is that the flat band wavefunctions vary slowly on the moir\'e scale. Specifically, the BM eigenvectors satisfy
\begin{equation}
\sum_{l,\sigma} \big| u(\k)_{(\tau,n)}^{l,\sigma,\G}\big|^2 \approx 0 \text{ for } |\G| \gtrsim 2|\G_1|\, ,
\end{equation}
if $n$ is one of the two flat bands. Here, $\G_1$ a basis vector of the moir\'e reciprocal lattice. This implies that $||G^{00}_\q(\k)||\approx 0$ if $|\q|\gtrsim 4|\G_1|$. From this we conclude that the short-wavelength disorder does not couple directly to the IKS order parameter. Furthermore, long-wavelength disorder, which does couple to the IKS state via the first order term, gets suppressed by a factor $\theta^2$. Physically, this is because the density of electrons occupying the central bands is very small.

\begin{figure}
    \centering
    \includegraphics[scale=0.6]{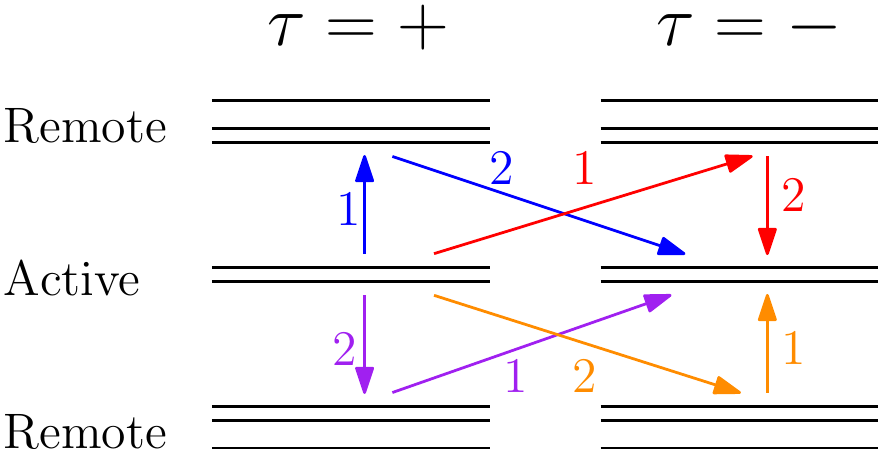}
    \caption{Virtual processes in second order perturbation theory which hop an electron from valley $\tau = +$ to valley $\tau = -$ due to quenched random bond disorder.}
    \label{fig:SW}
\end{figure}

Because the first order term in the Schrieffer-Wolff transformed disorder Hamiltonian is inoperative for short-wavelength bond disorder, let us consider the second order term next. The relevant second order processes which hop an electron from valley $\tau=+$ to valley $\tau = -$ are shown schematically in Fig. \ref{fig:SW}. In the figure, the numbers next to each arrow indicate the order in which the virtual hopping processes take place. There are two processes which involve an intermediate electron in the remote conduction bands, and two processes which involve an intermediate hole in the remote valence bands. Note that in Eq. \eqref{SW}, the energy denominators do \emph{not} have absolute value signs. This means that the energy denominators involving remote conduction band energies will be positive, while energy denominators involving remote valence band energies will be negative. This difference in sign comes from the fact that the virtual electron processes generate terms in the second order Schrieffer-Wolff Hamiltonian of the conventional form $\psi^\dagger\psi$, while the virtual hole processes produce terms of the form $\psi \psi^\dagger$. The difference in sign thus comes from the fermion anti-commutation relations.

From Eq. \eqref{SW} we can estimate the magnitude of the second order term to be
\begin{equation}
E_{\text{dis}}^{|\q|>4|\G_1|}\sim \theta^4 t_{\text{dis}}^2  \langle \Delta^{-1}\rangle\, ,
\end{equation}
where $t_{\text{dis}}$ is the typical energy scale of the bond disorder, and $\langle \Delta^{-1}\rangle$ is the inverse gap between the active bands and the remote bands averaged over the mini-Brillouin zone. The energy gain of pinning to short-wavelength bond disorder is thus reduced by a factor of $\theta^2 t_{\text{dis}}\langle \Delta^{-1}\rangle$ compared to the energy gain of pinning to long-wavelength disorder. As a result, for physically relevant disorder strengths we expect the Imry-Ma domains associated with short-wavelength bond disorder to be much larger than the size of the AA regions. For this reason, we ignore short-wavelength disorder and focus only on long-wavelength or moir\'e-scale random-field disorder. We discuss the effect of moir\'e-scale quenched disorder on the finite-temperature physics in the following section.

\section{Finite-temperature phase transitions}
\label{sec:thermodyn}

The IKS order not only breaks the valley-charge conservation symmetry, but also superlattice translation and three-fold rotation symmetries. It is important to distinguish from the outset the different ways in which the translation and rotation symmetries are broken. In particular, because the IKS states are invariant under a combination of superlattice translation and valley $U(1)$ rotation, only local operators with a non-zero valley charge can detect the translational symmetry breaking. A corollary of this observation is that the translational symmetry-breaking order is replaced by quasi-long range order at non-zero temperature, and is completely lost once vortices of the IVC order proliferate. The rotational symmetry breaking, on the other hand, can be detected by operators which have zero valley charge. This means that the nematic order (ignoring the explicit symmetry-breaking effects of strain), can persist as true long-range order at non-zero temperatures. 

Let us now consider the different possible ways for the IKS order to disappear at finite temperature. We first note that the destruction of IKS order is likely to be driven by the proliferation of fluctuating defects in the order parameter, and not by the unbinding of excitons. The reason is that the binding energy of the condensed inter-valley coherent excitons is expected to be of the order of the Coulomb scale ($\sim 20$ meV), whereas the IKS stiffness extracted from our numerical Hartree-Fock calculations is significantly smaller ($\sim 0.4$ meV). 

Taking the above considerations into account, there are three different possible scenarios for the IKS order to disappear. In the first scenario, a Berezinskii-Kosterlitz-Thouless (BKT) transition occurs at a temperature $T_{\text{BKT}}$, at which the IKS angle disorders via the unbinding of vortex-antivortex pairs. The nematic order, however, persists above $T_{\text{BKT}}$ and disappears at a higher temperature $T_N$\footnote{The small but non-zero strain present in TBG will turn this second phase transition at $T_N$ into a crossover where the nematic order parameter decreases and saturates at a small but non-zero value with increasing temperature.}. In the second scenario, the algebraic IKS order and the nematic order disappear simultaneously via a single phase transition, which most likely is first order. In the third scenario, the nematic order disappears inside the region of the phase diagram with algebraic IKS order, such that $T_N < T_{\text{BKT}}$. In App. \ref{app:thermal}, we show that all these different scenarios can indeed be realized for the special case of commensurate IVC spiral states. In the commensurate case, we find that the finite-temperature physics of IKS states is closely related to that of frustrated or generalized XY models, which also harbor both quasi-long range order and discrete symmetry breaking \cite{LeeNegeleLandau,LeeGrinstein,LeeWu,GranatoKosterlitz,GranatoKosterlitzNightengale,Henley,ChandraColemanLarkin}. Exactly which of the three scenarios is realized depends on the ratio $\rho_s/\sigma_{\text{DW}}$, where $\rho_s$ is the IVC stiffness, and $\sigma_{\text{DW}}$ is the domain wall tension between two different ground states related by $\hat{C}_{3}$. The physical picture which arises in our study of the commensurate models also suggests that the third scenario with $T_N < T_{\text{BKT}}$ is excluded in the incommensurate case relevant for TBG. But regardless of which of the three scenarios occurs experimentally, we expect all transition temperatures to be of the order $T_{\text{BKT}} \lesssim \pi \rho_s/2 \sim 7$ K.

In our discussion of the finite-temperature phase diagram so far we have ignored moir\'e-scale quenched random-field disorder. Because TBG is a two-dimensional material, reintroducing quenched disorder will, strictly speaking, destroy all ordered phases (breaking both discrete and continuous symmetries). However, if the moir\'e-scale disorder is sufficiently weak the finite-temperature phase transitions we discussed for the disorder-free case should still leave detectable imprints on measurable quantities. In particular, for the case of discrete symmetry breaking, it is known that the size of the Imry-Ma domains is exponentially large in $(\sigma_{\text{DW}}/h)^2$, where $\sigma_{\text{DW}}$ again is the domain-wall tension in the ordered state and $h$ is the RMS random-field disorder strength \cite{Binder,SeppalaAvala}. For weak disorder, the Imry-Ma domains will thus be much larger than the relatively small TBG devices currently being used in experiment. In this case, nematicity will survive the presence of quenched disorder as `vestigial' order of the IVC spiral state, similarly to what was discussed in Ref.~\cite{NieKivelson} for 3D models of charge density wave order in the context of the cuprate superconductors. 

\section{Experiments}
\label{sec:experiments}

We have demonstrated that IKS order is ubiquitous at nonzero integer fillings in the presence of small amounts of heterostrain, largely independent of substrate or fine details of twist angle. We now argue that this fact, combined with the absence of IKS order at charge neutrality in favour of a gapless nematic semimetal,  provides a unified explanation of several recent experiments. Evidently, this picture requires us to posit that small heterostrain is  inveitably present in typical experimental samples; this is  however not  an unreasonable assumption, particularly given the  modest heterostrain needed to stabilize IKS order (which is comparable to experimentally observed strains \cite{Kerelsky2019,Choi2019,Xie2019stm}). Before proceeding, we note that in Ref.~\cite{Wagner_unpub} we extend the Hartree-Fock analysis and investigate the effects of heterostrain upon doping away from integer fillings and including IKS order. By considering the chemical potential variations and the Fermi surface degeneracies, we obtain results in the strained regime that are consistent with the experimentally measured compressibility traces (`cascade' transitions) and Landau fans~\cite{Sharpe2019,Serlin2019,Lu2019,Cao2018mott,Cao2018sc,Yankowitz2019,Park_2021,Stepanov2020,AndreiChern,IlaniCascade,WongCascade,KangBernevigVafek,saito2021isospin,rozen2021entropic,Pierce2021,Tomarken2019,yu2021correlated}.

At even integer fillings, the IKS state can be distinguished from the KIVC state by only probing the spin physics. In particular, in a small magnetic field the KIVC state at $\nu = \pm 2$ has a local spin moment of $\lesssim 2\mu_B$ per moi\'re unit cell\footnote{Adding a ferromagnetic valley-Hunds coupling term to the Hamiltonian would induce a magnetic moment of $2 \mu_B$ per unit cell for the KIVC state at $\nu = \pm 2$, even in the absence of a magnetic field. If, as was argued in Refs. \cite{KhalafLedwith,LakeSenthil}, the valley-Hunds coupling is anti-ferromagnetic as a result of e.g. phonon scattering \cite{ChatterjeeBultinck}, then a small magnetic field will cause the spins in the two valleys to cant, producing a magnetic moment per unit cell which is smaller than $2 \mu_B$.}, whereas the IKS state at $\nu = \pm 2$ has a vanishing local moment. An immediate prediction that follows from this observation is that in samples with negligible strain, which would have a strong KIVC gap at neutrality (assuming there is also negligible hBN alignment), the insulators at $\nu = \pm 2$ should have a non-zero local spin moment, whereas strained samples with semi-metallic behaviour at neutrality should have no (spin or orbital) magnetic moment at $\nu = \pm 2$. By applying a small strain to an initially unstrained sample, one should therefore observe a strong first order transition associated with an abrupt disappearance of the local moment as one enters the IKS phase. If the strain in experiment can both be slowly increased and decreased, hysteretic behaviour should be observed for the local spin moment around the KIVC--IKS transition. We note that Ref.~\cite{Yankowitz2019} indeed found evidence for spin-unpolarized insulators at $\nu = \pm 2$ in a sample which is semi-metallic at neutrality, which is consistent with the IKS/nematic SM scenario under the assumption that their samples are heterostrained at the $\sim0.1-0.2\%$ level.

At $\nu = \pm 3$, the IKS insulators also have a smaller local magnetic moment than the QAH insulators which occur in the absence of strain. As both states are spin polarized in a small field, this difference is now due to the large orbital moment of the QAH insulators, which is absent in the IKS insulator. However, the easiest way to distinguish the IKS state from the QAH state is via the transverse or Hall resistance $R_{xy}$. This quantity is zero in the IKS state as dictated by the spinless time-reversal symmetry, but takes on a quantized non-zero value in the QAH state. In Refs.~\cite{Yankowitz2019,BernevigEfetov2020}, insulating states were observed at $\nu = +3$ which show Landau fans in magnetotransport measurements that are consistent with a zero Chern number. Given the semi-metallic behaviour at charge neutrality, we thus expect the samples of Ref.~\cite{Yankowitz2019,BernevigEfetov2020} to be strained and therefore the insulators at $\nu = + 3$ to have IKS order.

As discussed in Sec.~\ref{sec:thermodyn} and Appendix \ref{app:thermal}, the nematic order of the IKS state survives at finite temperature. We therefore predict that all insulators at $\nu = \pm 2, \pm3$ should show strong interaction-induced nematicity, much stronger than what one would naively expect from the small strain present in the sample. This prediction  actually fits perfectly with the experimental observations of Ref.~\cite{Cao2020nematicity}, where nematicity was observed in the superconducting dome between $\nu = -3$ and $\nu = -2$. Indeed, unless the insulators at integer fillings are separated from the superconducting dome by a strong first order transition, one would generally expect the insulators and the superconductors to either both be isotropic or both be nematic.  In Ref.~\cite{Wagner_unpub}, we show that IKS order persists for a finite range of doping around the integers and survives to temperatures\footnote{This refers to the mean-field transition temperature $T_{\textrm{MF}}\sim50$~K which is related to exciton binding energies. As discussed in Sec.~\ref{sec:thermodyn}, there is a lower energy scale $T_\text{BKT}\sim 7$~K controlled by the stiffness, above which the phase of the IKS is disordered (though likely still nematic).} much greater than the experimental $T_c$, suggesting that the superconducting dome could indeed inherit physics from the IKS. Furthermore, the ideal ordering wavevector $\bm{q}$ (which is naturally soft as illustrated in Fig.~\ref{fig:phase_nu2}d) changes continuously as a function of density~\cite{Wagner_unpub}, analogous to the evolution of the nematic axis of the superconductor~\cite{Cao2020nematicity}. Such behavior is harder to rationalize for other rotation symmetry-breaking parent states such as stripes. So even without making any assumptions about the nature of the superconducting state, we can interpret the observations of Ref.~\cite{Cao2020nematicity} as indirect evidence for the IKS state.

While the above evidence is reasonable, it remains to a degree circumstantial. A more definitive diagnostic for IKS order is possible in principle, by detecting the Kekul\'e pattern at the AA regions directly using STM/STS. Kekul\'e order in mono-layer graphene induced by mobile adatoms (or substrate vacancies) \cite{AltshulerKekule,FalkoAdatom} has been measured in Ref.~\cite{GutierrezSTM}, whereas Kekul\'e order induced by large (isotropic) strain \cite{MarianettiStrain,SorellaStrain} has been observed in Ref.~\cite{EomKooSTM}. However, only a very small fraction of the total number of electrons, i.e. those occupying the central bands, participate in the IKS order we find at nonzero integer fillings in TBG. As a result, the signal coming from the graphene-scale $\sqrt{3}\times \sqrt{3}$ Kekul\'e pattern in the IKS state will be significantly smaller than that seen in the above-mentioned monolayer experiments, and hence could lie below the present-day experimental resolution. (However, the STM studies of Ref.~\cite{LiHe,liu2021visualizing} were able to detect a Kekul\'e distortion in monolayer graphene in a high magnetic field at densities that are comparable to those of TBG. The theory here is based on the approximate $SU(4)$ symmetry of the zeroth Landau level, which is close in spirit to the $U(4)\times U(4)$ limit of TBG, but the anisotropies and associated mechanisms are different~\cite{Kharitonov_2012}.) In Sec.~\ref{sec:LG}, we discussed how charge order could be induced in the IKS state by changing the eccentricity of the Kekul\'e spiral. This charge order might be easier to detect experimentally than the Kekul\'e pattern, for example by using high resolution scanning single-electron transistors made of a carbon nanotube \cite{IlaniSET1,IlaniSET2}. 

\section{Discussion}
\label{sec:discussion}

By combining the results obtained in this work at non-zero integer fillings with the findings of Ref. \cite{Parker2020} at neutrality, we conclude that by adding a small amount of uniaxial heterostrain to the BM Hamiltonian one obtains from self-consistent Hartree-Fock a global picture of the TBG phase diagram that is consistent with most or even all experimental observations at integer fillings. In particular, at neutrality moderately strained TBG is semi-metallic, at $\nu = \pm 1$ it is metallic but with a significantly lower carrier density than the non-interacting BM model due to strong IKS order, and at $\nu = \pm 2, \pm 3$ it becomes an IKS insulator. On top of this, we argued that several other properties of the IKS states (e.g. spin polarization, Chern number, nematicity) fit nicely with many of the experimental observations. The estimated temperatures at which the insulating IKS states should appear ($\sim 7$ K) also agree remarkably well with experiment. Furthermore as we explain in Ref.~\cite{Wagner_unpub}, the doped descendants of the finite-strain integer orders examined here exhibit Landau fans and compressibility signatures that are consistent with those measured experimentally.

Our results open up several interesting directions for future work. 
For example, a generalized Pomeranchuk effect has been observed in TBG \cite{Pomeranchuk1,Pomeranchuk2}, which causes high-entropy insulators to win over metallic states with increasing temperature. There is at present no theory to fit the IKS state into this Pomeranchuk scenario. For the uniform QHFM states, several groups have recently calculated the collective mode spectrum \cite{Kwan2020,Khalaf2020,TBGV,KangVafekRG}. In this work, we have not attempted to do this for the IKS states, although it would be helpful to obtain a more complete picture of the low-energy physics at the different integer fillings. Arguably the most interesting direction for future work is to investigate the relation between the IKS states and superconductivity. For example, the KIVC state has recently been argued to be a natural parent state for superconductivity in Refs. \cite{SkyrmionSC,Chatterjee2020}. A future direction would be to adapt this mechanism to the IKS state, or investigate whether local KIVC correlations are strong enough to allow the same mechanism to be operative.
It would also be interesting to consider if a similar approach to coupling superconductivity via topological terms in Ref.~\cite{Christos_2020} could be generalized to the `dominant lobe' physics of the IKS. 
Refs. \cite{Stepanov2020,VafekLi} have subjected TBG to different amounts of Coulomb screening, and found that the strengths of insulating and superconducting orders seem to anti-correlate. Any satisfactory theory of superconductivity should be able to explain this anti-correlation.

\begin{acknowledgments} 
We thank Ashvin Vishwanath and Yahui Zhang for interesting discussions on the subtle nature of the translation symmetry breaking order in the IKS state.    NB is supported by a senior postdoctoral research fellowship of the Flanders Research Foundation (FWO).  We acknowledge support from the European Research Council under the European Union Horizon 2020 Research and Innovation Programme, Grant Agreement No. 804213-TMCS and  from EPSRC Grant EP/S020527/1. Statement of compliance with EPSRC policy framework
on research data: This publication is theoretical work
that does not require supporting research data.
\end{acknowledgments}

\bibliographystyle{unsrt}
\bibliography{bib}

\begin{widetext}

\clearpage
\newpage

\begin{appendix}
\counterwithin{figure}{section} 
\section{Strained BM Model with Non-local Tunneling}\label{secapp:continuummodel}

In this section we discuss the single-particle continuum model in the presence of strain and non-local tunneling
\begin{equation}
    \hat{H}_\textrm{SP}=\hat H_\textrm{BM}+\hat H_\textrm{NLT}.
\end{equation}
This does not include the interaction-induced contribution $\hat{H}_\textrm{DC}$ which is described in Appendix~\ref{appsec:HFsubtraction}.

\subsection{Lattice Vectors}
The conventions for the the basis direct lattice vectors (LVs) and reciprocal lattice vectors (RLVs) for the unstrained TBG moir\'e pattern (where layers 1 and 2 are rotated counterclockwise by $\theta/2$ and $-\theta/2$ respectively) and the graphene monolayer are
\begin{gather}
    \bm{a}_1 =\frac{4\pi}{3k_\theta}(\frac{\sqrt{3}}{2},\frac{1}{2}),\quad\bm{a}_2 =\frac{4\pi}{3k_\theta}(0,1),\quad \bm{G}_1=\sqrt{3}k_\theta(1,0) ,\quad\bm{G}_2=\sqrt{3}k_\theta(-\frac{1}{2},\frac{\sqrt{3}}{2})\\
    \bm{a}_{G1}=\sqrt{3}a(\frac{1}{2},\frac{\sqrt{3}}{2})
    ,\quad \bm{a}_{G2}=\sqrt{3}a(\frac{1}{2},-\frac{\sqrt{3}}{2})
    ,\quad\bm{G}_{G1}=\frac{4\pi}{3a}(\frac{\sqrt{3}}{2},\frac{1}{2}) ,\quad\bm{G}_{G2}=\frac{4\pi}{3a}(\frac{\sqrt{3}}{2},-\frac{1}{2}),\quad \bm{K}_D=\frac{4\pi}{3\sqrt{3}a}(1,0)  
\end{gather}
where $k_\theta=2k_D\sin\frac{\theta}{2}$, and $a$ is the graphene C-C bond length. The intracell coordinates for the $A$ and $B$ atoms of graphene are $\bm{\tau}_A=a(\frac{\sqrt{3}}{2},\frac{1}{2})$ and $\bm{\tau}_B=a(\frac{\sqrt{3}}{2},\frac{1}{2})$. $\bm{K}_D$ is the valley-$K$ (i.e. $\tau=+$) Dirac point for untwisted graphene. The Dirac point in valley $K'$ ($\tau=-$) is found by time reversal.

We now consider how the above change in the presence of strain. Following \cite{Parker2020} and \cite{Bi2019}, we consider a layer-dependent transformation $M_l$ such that the graphene LVs $\bm{r}_l$ and RLVs $\bm{g}_l$ on layer $l$ become
\begin{gather}
    \bm{r}_l=M^T_l \bm{r},\quad \bm{g}_l=M^{-1}_l\bm{g}\\
    M_l\simeq1+\mathcal{E}_l^T\\
    \mathcal{E}\simeq R(\theta)-1+S(\epsilon,\varphi)
    \simeq\begin{pmatrix}
    \epsilon_{xx} & \epsilon_{xy}-\theta\\
    \epsilon_{xy}+\theta & \epsilon_{yy}
    \end{pmatrix}\\
    S(\epsilon,\varphi)=R^{-1}(\varphi)\begin{pmatrix}-\epsilon&0\\0&\nu\epsilon\end{pmatrix}R(\varphi)
\end{gather}
with $\nu=0.16$ the Poisson ratio. We are interested in heterostrain, so the parameters of the layers are related as
\begin{equation}
    \theta_1=-\theta_2=\frac{\theta}{2},\quad \varphi_1=\varphi_2=\varphi,\quad \epsilon_1=-\epsilon_2=\frac{\epsilon}{2}
\end{equation}
In analogy with the pure twist angle case, the new moir\'e RLVs are, in terms of the monolayer RLVs,
\begin{gather}
    \bm{G}_1=(M^{-1}_1-M^{-1}_2)(\bm{G}_{G2}-\bm{G}_{G1}),\quad \bm{G}_2=(M^{-1}_1-M^{-1}_2)\bm{G}_{G1}
\end{gather}

\subsection{Intralayer (Kinetic) Term}

The strained and rotated intralayer kinetic term in valley $\tau$ is
\begin{gather}
    \bra{\bm{k},l}H_\textrm{BM}\ket{\bm{k}',l}=\hbar v_F\left[M_l(\bm{k}-\tau \bm{A}_l)-\bm{K}_\tau\right]\cdot
    \begin{pmatrix}
    \tau\sigma_x\\
    -\sigma_y
    \end{pmatrix}\delta_{\bm{k}\bm{k}'}\\
    \bm{A}=\frac{\beta}{2a}(\epsilon_{xx}-\epsilon_{yy},-2\epsilon_{xy})
\end{gather}
where $\beta=3.14$, and $a$ is the C-C bond length. Note that $\bm{k}$ above is measured with respect to the global momentum origin, and $\bm{K}_\tau=\tau \bm{K}_D$ is the original Dirac momentum in valley $\tau$. 

Alignment of the TBG sample with the hBN substrate leads to a staggered sublattice potential on one or both of the graphene layers. This can be incorporated into our model by projecting the $\sigma_z$ operator onto the active bands of interest~\cite{Kwan2020}, with a coefficient $\Delta$ of order $10~\text{meV}$.

\subsection{Local Interlayer Hopping}
The usual Bistritzer-Macdonald (BM)~\cite{Bistritzer2011} interlayer hopping term for valley $+$ is
\begin{gather}
\bra{\bm{k},1}H_\textrm{BM}\ket{\bm{k}',2} = 
T_1\delta_{\bm{k}-\bm{k}',\bm{0}} + T_2\delta_{\bm{k}-\bm{k}',\bm{G}_1+\bm{G}_2} + T_3\delta_{\bm{k}-\bm{k}',\bm{G}_2}\\
T_1 = \begin{pmatrix}w_{AA}&w_{AB}\\w_{AB}&w_{AA}\end{pmatrix}\\
T_2 = \begin{pmatrix}w_{AA}&w_{AB}e^{i\phi}\\w_{AB}e^{-i\phi}&w_{AA}\end{pmatrix}\\
T_3 = \begin{pmatrix}w_{AA}&w_{AB}e^{-i\phi}\\w_{AB}e^{i\phi}&w_{AA}\end{pmatrix}\\
\phi=\frac{2\pi}{3}
\end{gather}
where the matrices act in sublattice space. The equations in valley $K'$ can be found by time-reversal. Since $\bm{k}$ in both layers is measured with respect to a common basepoint, the hopping term is moir\'e-periodic. Unless otherwise stated, we use $w_{AA}=82.5\,\text{meV}$, $w_{AB}=110\,\text{meV}$ and $\theta=1.08^\circ$.

\subsection{Non-local Tunneling}
In the absence of strain, the non-local tunneling term (linear order in momentum) for valley $+$ is~\cite{Fang2019}
\begin{gather}\label{eqnapp:NLT}
    \bra{\bm{k},1}H_\textrm{NLT}\ket{\bm{k}',2} = 
    \tilde{T}_1(\bm{k},\bm{k}')\delta_{\bm{k}-\bm{k}',\bm{0}} + \tilde{T}_2(\bm{k},\bm{k}')\delta_{\bm{k}-\bm{k}',\bm{G}_1+\bm{G}_2} + \tilde{T}_3(\bm{k},\bm{k}')\delta_{\bm{k}-\bm{k}',\bm{G}_2}\\
\tilde{T}_1(\bm{k},\bm{k}')=-\frac{1}{2}\bigg[
\begin{pmatrix}
\lambda_1 & \lambda_2\\
\lambda_3 & \lambda_1
\end{pmatrix}(p_x-ip_y)
+
\begin{pmatrix}
\lambda_1 & \lambda_3\\
\lambda_2 & \lambda_1
\end{pmatrix}(p_x+ip_y)
\bigg]\\
\tilde{T}_2(\bm{k},\bm{k}')=-\frac{1}{2}\bigg[
\begin{pmatrix}
\lambda_1e^{i\phi} & \lambda_2e^{-i\phi}\\
\lambda_3 & \lambda_1e^{i\phi}
\end{pmatrix}(p_x-ip_y)
+
\begin{pmatrix}
\lambda_1e^{-i\phi} & \lambda_3\\
\lambda_2e^{i\phi} & \lambda_1e^{-i\phi}
\end{pmatrix}(p_x+ip_y)
\bigg]\\
\tilde{T}_3(\bm{k},\bm{k}')=-\frac{1}{2}\bigg[
\begin{pmatrix}
\lambda_1e^{-i\phi} & \lambda_2e^{i\phi}\\
\lambda_3 & \lambda_1e^{-i\phi}
\end{pmatrix}(p_x-ip_y)
+
\begin{pmatrix}
\lambda_1e^{i\phi} & \lambda_3\\
\lambda_2e^{-i\phi} & \lambda_1e^{i\phi}
\end{pmatrix}(p_x+ip_y)
\bigg]
\end{gather}
which respects all the symmetries of the original BM Hamiltonian. Note that the $\bm{p}$ defined above is simply the sum of the two momenta of $\ket{\bm{k},1},\ket{\bm{k}',2}$, but measured with respect to the respective $\bm{K}^l_+=M^{-1}_l \bm{K}_+$ of the two layers. Here $\bm{K}^l_+$ is the new position of the Dirac point in each monolayer in the absence of interlayer coupling and the gauge potential $\bm{A}$. Hence without strain, $\bm{K}^l_+$ is simply the Dirac point rotated by $\pm\theta/2$. The above form of NLT can be derived from Eq.~21 in Ref.~\cite{Fang2019} by rotating their system by $\pi/2$ clockwise and interchanging the layers (since they consider twisting layer 1 by $\theta/2$ in a clockwise fashion).
For calculations that include non-local tunneling, we use $\lambda_1=90~\text{meV}$\r{A}, $\lambda_2=180~\text{meV}$\r{A} and $\lambda_3=0$.

In the next section, we show that to first order in strain/non-local tunneling, the above form holds even in the presence of heterostrain.

\subsection{Combining Strain and Nonlocal Tunneling}
The most natural way to generalize Eqn~\eqref{eqnapp:NLT} in the presence of strain, is to consider the linear momentum argument $\bm{p}$. In the absence of strain, we have
\begin{equation}
    \bm{p}=(\bm{k}-R(\theta/2)\bm{K}_+)+(\bm{k'}-R(-\theta/2)\bm{K}_+)
\end{equation}
where $\bm{k},\bm{k'}$ are measured with respect to the global momentum origin. In the presence of strain, the dominant change should be to consider the transformed Dirac point location in each layer
\begin{equation}
    \bm{p}=(\bm{k}-M^{-1}_1\bm{K}_+-\bm{A}_1)+(\bm{k'}-M^{-1}_2\bm{K}_+-\bm{A}_2)).
\end{equation}
However $\bm{A}_1=-\bm{A}_2$ for heterostrain. Hence the form of the non-local tunneling term is unaffected to $O(\lambda \epsilon)$.

\section{Interactions and Form Factors}
Eigenstates of the single-particle continuum Hamiltonian in Appendix~\ref{secapp:continuummodel} are parameterized as 
\begin{equation}
    \ket{\psi_{\tau,a}(\bm{k})}=\sum_{\bm{G}f}c_{\bm{G}\tau a f}(\bm{k})\ket{\bm{k}+\bm{G},\tau,f}
\end{equation}
where $a$ is a band label, $f$ is a composite index for layer and sublattice, $\bm{G}$ is a RLV, and $\bm{k}$ is a momentum that lives in the moir\'e Brillouin zone (mBZ). The ket on the RHS is a plane wave state with momentum $\bm{k}+\bm{G}$ measured with respect to an arbitrary basepoint which maps onto $\Gamma_\text{M}$ in valley $\tau$.

Define the continuum model form factors~\cite{Liu2021}
\begin{equation}
\lambda_{\tau,a,b}(\bm{k};\bm{q},\bm{G})=\braket{u_{\tau,a}(\bm{k})|u_{\tau,b}(\bm{k}+\bm{q}+\bm{G})}
\end{equation}
where $\bm{k},\bm{q}$ are mBZ momenta. The Bloch functions are defined to be periodic in the mBZ, so the cell-periodic $u$'s are \emph{not} periodic in its momentum argument. In particular, we have
\begin{equation}
    \ket{u_{\tau,a}(\bm{k}+\bm{G})}=e^{-i\bm{G}\hat{\bm{r}}}\ket{u_{\tau,a}(\bm{k})}.
\end{equation}
In terms of these, the interaction Hamiltonian is
\begin{align}
    \hat{H}_{\text{int}}=\frac{1}{2A}\sum_{\stackrel{ss'\tau\tau'}{abcd}}\sum_{\bm{k}^\alpha\bm{k}^\beta\bm{q}\bm{G}}&\lambda_{\tau,a,b}(\bm{k}^\alpha;\bm{q},\bm{G})\lambda^*_{\tau',d,c}(\bm{k}^\beta;\bm{q},\bm{G})V(\bm{q}+\bm{G})\\
    &\times c^\dagger_{\tau as}(\bm{k}^\alpha)c^\dagger_{\tau' cs'}(\bm{k}^\beta+\bm{q})c_{\tau' ds'}(\bm{k}^\beta)c_{\tau bs}(\bm{k}^\alpha+\bm{q}),
\end{align}
where $A$ is the area of the system, and the $c^\dagger$ are continuum model band creation operators. Intervalley scattering terms which are suppressed at small twist angles are neglected~\cite{Bultinck2020}.
We consider a dual-gate screened interaction $\big[\frac{e^2}{2\epsilon_0\epsilon_r q}\tanh{qd}\big]$ with screening length $d=25~\text{nm}$ and relative permittivity $\epsilon_r=10$ unless otherwise stated.

\section{Hartree-Fock and Subtraction Term}\label{appsec:HFsubtraction}

\subsection{Hartree-Fock Generalities}\label{secapp:HFgeneralities}
Consider a Hamiltonian in a general orthonormal basis $c^\dagger_\alpha$
\begin{equation}
   \hat{H}=h_{\alpha\beta}c^\dagger_\alpha c_\beta + \frac{1}{2}V_{\alpha\beta\gamma\delta}c^\dagger_\alpha c^\dagger_\beta c_\delta c_\gamma.
\end{equation}

A Slater determinant is described by its projector $P_{\alpha\beta}=\langle c^\dagger_\alpha c_\beta \rangle$. We may want to define interactions with respect to a reference projector $P^{0}$---by this we mean that the state $P^0$ should have no explicit interaction energy. This motivates the following Hartree-Fock decomposition, and expression for the total energy, for an arbitrary $P$
\begin{gather}
    \hat{H}^{\text{HF}}[P]=h_{\alpha\beta}c^\dagger_\alpha c_\beta+(P_{\gamma\delta}-P^0_{\gamma\delta})(V_{\alpha\gamma\beta\delta}-V_{\alpha\gamma\delta\beta})c^\dagger_\alpha c_\beta\\
    E[P]=h_{\alpha\beta}P_{\alpha\beta}+\frac{1}{2}(P_{\gamma\delta}-P^0_{\gamma\delta})(P_{\alpha\beta}-P^0_{\alpha\beta})(V_{\alpha\gamma\beta\delta}-V_{\alpha\gamma\delta\beta})
\end{gather}
where we have neglected a constant term in the HF Hamiltonian. 

Now imagine that some subset of orbitals have a frozen correlation matrix. In other words, $P$ decomposes into two orthogonal contributions---the \emph{active} part which we will call $P$ again, and the \emph{frozen} part which we will call $P^f$. Furthermore we will distinguish indices $\alpha,\beta$ that live solely in the active subspace, and indices $x,y$ that run over all orbitals. The active space HF Hamiltonian and the energy is then (ignoring constant terms that depend only on $P^f$ and $P^0$) 
\begin{gather}
    \hat{H}^{\text{HF}}[P]=h_{\alpha\beta}c^\dagger_\alpha c_\beta+P_{\gamma\delta}(V_{\alpha\gamma\beta\delta}-V_{\alpha\gamma\delta\beta})c^\dagger_\alpha c_\beta+(P^f_{xy}-P^0_{xy})(V_{\alpha x\beta y}-V_{\alpha x\delta y})c^\dagger_\alpha c_\beta\\
    E[P]=h_{\alpha\beta}P_{\alpha\beta}+\frac{1}{2}P_{\gamma\delta}P_{\alpha\beta}(V_{\alpha\gamma\beta\delta}-V_{\alpha\gamma\delta\beta})
    +(P^f_{xy}-P^0_{xy})P_{\alpha\beta}(V_{\alpha x\beta y}-V_{\alpha x\delta y}).
\end{gather}

Defining $F=P^f-P^0$, this means that we can define an effective single-particle subtraction term that augments the existing single-particle term
\begin{equation}
    H^{DC}_{\alpha\beta}=F_{xy}(V_{\alpha x\beta y}-V_{\alpha x\delta y})
\end{equation}
(DC stands for double-counting, but note that the above also includes the frozen contributions.)

\subsection{Twisted Bilayer Graphene}
We now translate the expression for $\hat{H}^{DC}$ above to the situation for TBG. 
The active subspace consists of the two central bands (for each spin and valley) and an additional number of remote bands. Therefore the frozen correlation matrix $P^f$ involves filling all bands below charge neutrality that are not in the active subspace. For the reference projector, we pick $P^0$ to be the density of the two decoupled graphene sheets at charge neutrality~\cite{XieSub,Bultinck2020} (i.e.~with all interlayer terms switched off).
Hence the matrix $F=P^0-P^f$ is diagonal in mBZ momentum, valley, and spin. Furthermore, for band indices $x,y$ far from the central bands, $F_{x,y}$ can be approximated as 0 since the Dirac cone kinetic energies far exceed the interlayer tunneling scale $\sim 100~\text{meV}$. Typically we keep around 50 bands per spin/valley for $F_{x,y}$---bands that are too far from charge neutrality will be inaccurate anyways since they approach the plane wave cutoff used to diagonalize the continuum model. The DC term is then
\begin{gather}
   \hat{H}^\text{DC}=\sum_{\bm{k},\tau a b s}\mathscr{H}^\text{DC}_{\tau, ab}(\bm{k})c^\dagger_{\bm{k}\tau a s}c^{\phantom{\dagger}}_{\bm{k}\tau b s}\\
   \mathscr{H}^\text{DC}_{\tau, ab}(\bm{k})=2D_{ab}(\bm{k},\tau)-E_{ab}(\bm{k},\tau)\\
   D_{ab}(\bm{k},\tau)=\sum_{\bm{G}''}V(\bm{G}'')\bigg[\sum_{\bm{G}f}c^*_{\bm{G}\tau af}(\bm{k})c_{\bm{G}+\bm{G}'',\tau bf}(\bm{k})\bigg]\bigg[\sum_{\bm{k}'\tau''f'\bm{G}'}F_{\bm{G}'f';\bm{G}'-\bm{G}'',f'}(\bm{k}',\tau'')\bigg]\\
    E_{ab}(\bm{k},\tau)=\sum_{\bm{G}\bm{G}'ff'}c^*_{\bm{G}\tau af}(\bm{k})c_{\bm{G}',\tau b f'}(\bm{k})\bigg[\sum_{\bm{k}'\bm{G}''}F_{\bm{G}'-\bm{G}'',f';\bm{G}-\bm{G}'',f}(\bm{k}',\tau)V(\bm{G''}+\bm{k}-\bm{k}')\bigg]
\end{gather}
where the factor of 2 in the direct term is due to summation over spin ($F$, which is expressed above in the plane wave basis, is independent of spin). 

From the perspective of the active subspace, the total Hamiltonian is then
\begin{equation}
    \hat{H}_\textrm{active}=\hat H_\textrm{BM}+\hat H_\textrm{NLT}+\hat{H}_\textrm{DC}+\hat{H}_\textrm{int}
\end{equation}
where all terms are restricted to the active bands only.

\subsection{General Translation Symmetry Breaking}
We wish to consider general translation symmetry-breaking (TSB) in both directions. 
We assume that the system is of size $N_1\times N_2$, and that the  periodicity along the two axes are $d_1$ and $d_2$ or factors thereof. So for example $d_1=1$ is no symmetry breaking, and $d_1=N_1$ allows for full TSB along the $1$-axis. $d_1,d_2$ must be factors of $N_1,N_2$ respectively. For a given choice of $d_1,d_2$, any mBZ momentum can be decomposed as $\bm{k}=(\tilde{\bm{k}},\bm{m}^k)$, where $\tilde{\bm{k}}$ is a conserved momentum within the reduced mBZ (rBZ), and $\bm{m}^k$ are `miniband' indices that label the rBZ sector. The allowed values (in fractional mBZ coordinates) are
\begin{gather}
    k_1=\tilde{k}_1+m^k_1\times\frac{1}{d_1}\\
    \tilde{k}_1=0,1/N_1,\ldots,(d_1-1)/d_1\\\
    m^k_1=0,1,\ldots,d_1-1,
\end{gather}
and analogously for the $2$-axis. For a given $d_1,d_2$, the HF projector is allowed to mix different minibands, but must be diagonal in rBZ momentum.

\section{Additional HF Results}\label{secapp:additional}

Fig.~\ref{figapp:extra_phase}a presents phase diagrams where the number of active bands has been increased to six per spin/valley (note that the system size has been decreased to $9\times 9$). NLT has not been included, so the phases at negative fillings are similar. Compared with the central-bands-only results, the phase boundaries move around slightly. The IKS at $\nu=\pm1$ is not fully-formed since the extra bands renormalize the central bands further.

Fig.~\ref{figapp:extra_phase}b presents phase diagrams (central bands only) without NLT, showing the approximate PHS between positive and negative fillings. The sublattice potential is applied to one layer only which does not satisfy PHS, but the phase diagrams throughout the explored parameter regime are still very similar.

Fig.~\ref{figapp:nematic} presents a phase diagram at $\nu=+3$ at a different set of parameters that stabilizes an intervening nematic phase between the QAH and IKS at zero substrate. The nematic preserves the combination $\hat{C}_{2z}\hat{\mathcal{T}}$, thereby leaving the Dirac points intact. This gapless state is also fully spin and valley polarized. In fact for the figures in the main text, the nematic is lower in energy than the QAH above a certain value of strain that is, however, larger than the critical strain that leads to the IKS.

Fig.~\ref{figapp:band_coherences} plots the IVC of the $\nu=-3$ IKS where six bands per spin/valley are kept. The coherence is broken down into components corresponding to each BM band. It can be seen that the majority of the symmetry-breaking is occuring in the central band subspace. The nearest remote valence band has non-negligible contributions at $\Gamma_M$, which is not surprising since this is where the remote bands come closest to the central bands (and affect quantitative numbers such as band gaps).

Fig.~\ref{figapp:tauz_fillings} plots the momentum-resolved valley polarization for the IKS at different fillings. Note that the spins have been rotated so that only the spin component containing IVC is shown. The IVC is similar across the fillings except at $\nu=-1$, where the semimetallic nature of this state is revealed by the regions of increased population (and neutral valley polarization) near $\Gamma_M$.

Fig.~\ref{figapp:PH_bandgaps} shows the effect on NLT on the band gaps as a function of strain and filling. For all insulating states, the band gap at positive filling is greater than its counterpart at negative filling, which is consistent with experiments observing more robust insulators on the electron side of charge neutrality.

Fig.~\ref{figapp:parent_energies} compares the HF band structures for different translation symmetric states at $\nu=-2$, at a choice of parameters where the lowest energy HF state is the IKS. While the quantitative details depend on the state, the main qualitative features (large Hartree peak near $\Gamma_M$ and low-energy lobe at some point along the $k_x$ axis) are consistent. In particular, the predicted ideal IKS wavevector $\bm{q}_0$ (which connects the low-energy minimum of valley $K'$ to the high-energy peak of valley $K$) would be around the same for all cases.

Fig.~\ref{figapp:varphi_q0} illustrates the validity of the wavevector selection mechanism outlined in the main text for different strain angles. The qualitative structure of the IKS energy as a function of $\bm{q}$ is reflected in the interaction-renormalized band structure.

Fig.~\ref{figapp:q0_prediction} considers how well the SM band structure can predict the minimum energy wavevector $\bm{q}_0$ of the IKS, for different strain angles. The predicted $\bm{q}_0$ is obtained by simply connecting the minimum energy point in valley $K'$ to the maximum energy point in valley $K$. As shown in the figure, both Hartree and Fock contributions are required to obtain a reasonably good prediction of $\bm{q}_0$, demonstrating that both are crucial for wavevector selection. Note that the predictions for $\varphi=15^\circ,20^\circ$ are particularly poor---this is because the low-energy region of the SM band structure is quite extended in momentum space (see Fig.~\ref{figapp:varphi_q0}), so such a simplistic estimate cannot reliably pick out the correct wavevector. In such regimes, the choice of translation symmetric state is also expected to change the estimates (see Fig.~\ref{figapp:parent_energies}).

Fig.~\ref{figapp:IKS_bandstructure} shows the bandstructure of the IKS across the mBZ. Note that there are four bands excluding spin, since the valleys are generically coupled. The valleys have been shifted by $\pm\bm{q}/2$, so e.g. $\bm{k}$ refers to the momentum coupling states at $\bm{k}+\bm{q}/2$ in valley $K$ and $\bm{k}-\bm{q}/2$ in valley $K'$.

Fig.~\ref{figapp:nu_2_stiffnesses} shows the dispersion of the IKS along two directions. The ideal $\bm{q}_0$ is already an appreciable fraction of the mBZ at the onset of IKS order. Assuming an energy of the form $E=\frac{\rho}{2}\int \mathrm{d}A (\nabla\theta)^2$, where $\mathrm{d}A$ is an area element and $\theta$ is the IVC angle, we extract the stiffnesses $\rho_i$ and the offsets of the minimum of the dispersion $q_{0i}$ along the two directions for different values of strain. We do so by fitting a parabola to the dispersions and extracting $\rho_i$ and $q_{0i}$ according to the formula
\begin{equation}\label{eqapp:stiffness}
    \frac{\Delta E}{N_1N_2}=\frac{(2\pi)^2}{\sqrt{3}}\rho_i\bigg(\frac{q_i-q_{0i}}{|\mathbf{G}_i|}\bigg)^2.
\end{equation}

Fig.~\ref{figapp:energy_breakdown} shows the energy difference between the translationally invariant state and the IKS at a given wavevector broken down into the different energy contributions. The IKS pays an exchange penalty compared to the translationally invariant state, however it has favourable direct and kinetic energy. 

Fig.~\ref{figapp:trial1} shows the energy of the IKS trial state at $\nu=-1$ compared to both the optimal HF IKS state at that filling and the translationally invariant state. The trial state construction at $\nu=-1$ works less well than at $\nu=-2$.

Fig.~\ref{figapp:chiral} shows the phase diagram in strain-$w_{AA}$ space. The translationally invariant ferromagnetic states are favoured as one tends to the chiral limit $w_{AA}=0$.

In Fig.~\ref{figapp:anneal} we show the energy of the IKS HF state compared to the translationally invariant state. Even at zero strain, one can find the IKS solution, albeit it is a metastable state that is higher in energy than the translationally invariant ground state by about 0.7meV. 

In Fig.~\ref{figapp:allphasesaverage}, we show the substrate-strain phase diagrams for a different interaction subtraction scheme (see Section~\ref{secapp:HFgeneralities}). Here we use the `average' scheme $P^0=\frac{1}{2}I$ (in the central band subspace), which matches the scheme used in Ref.~\cite{TBGIV} in the limit of zero strain. Hence this choice retains the approximate particle-hole symmetry (we neglect NLT). The phase diagrams are similar to those in Fig.~\ref{fig:all_phases}.  Note that the gaps of the QAH state at $\nu=3$ are significantly smaller than those of the graphene scheme used in the main text. For our choice of twist angle and interlayer tunneling, the bare BM bands are very narrow. Hence with NLT, the state at zero strain and substrate can sometimes differ from the ones shown in Fig.~\ref{figapp:allphasesaverage}. However within the resolution of the phase diagram, the states at any finite strain/substrate revert to the expected phases.

In Fig.~\ref{figapp:allphasesCN}, we show the substrate-strain phase diagrams for the CN subtraction scheme proposed in Ref.~\cite{Liu2021}, where $P^0$ is constructed by filling the lower BM band. Hence this choice retains the approximate particle-hole symmetry (in the absence of NLT). One quantitative difference with the graphene scheme is that the strain scales for transitions to the IKS are significantly lower (almost two orders of magnitude). The properties of the phases themselves, and their relative positions, are largely unchanged. Note that the gaps of the QAH state at $\nu=\pm3$ are significantly smaller than those of the graphene scheme used in the main text. Also at $\nu=-3$, the state at zero strain and substrate is no longer fully valley polarized, and possesses some IVC.

Given the significantly smaller strain scales of the CN scheme, we briefly comment on why this may be happening and why we believe this scheme is not an appropriate choice (at least for finite strains). We note that the physical properties of the subtraction projectors $P^0$ for the graphene and `average' schemes are largely independent of strain. The only dependence arises from the changing Hilbert subspace of the central bands. Owing to the large gap to the remote bands at zero strain, this effect is weak. On the other hand, the CN projector $P^0$ closely tracks the evolution of the BM bands as a function of strain. For example, $P^0$ changes dramatically depending on the positions of the Dirac points. This likely explains the contrasting behavior to the graphene and `average' schemes. Further evidence against the CN scheme is the heightened fragility of the $\nu=+3$ QAH phase against tiny amounts of strain in Fig.~\ref{figapp:allphasesCN}, even in the presence of large substrate coupling. This is inconsistent with experiments which have observed a robust zero-field QAH phase at this filling with substrate alignment. We anticipate that  fluctuations beyond HF will not significantly change this property of the CN scheme, for reasons that we now sketch. Since the QAH is a translationally-invariant ferromagnet (similar to QHFM states), relaxation of the single Slater-determinant constraint is unlikely to substantially improve its energetics. On the other hand, the IKS has a soft ordering wavevector $\bm{q}$ and its projector has significant variations across the mBZ, suggesting it may benefit comparatively more from beyond-HF corrections. However we caution that this is a qualitative argument; it would be interesting to verify, e.g. via DMRG, whether the CN scheme indeed persistently produces phase structures that are challenging to reconcile with experimental inputs, such as the absence of the QAH phase for reasonable values of strain and substrate potential.

In Fig.~\ref{figapp:FMC}, we plot the form factor $\lambda$ for momentum transfer $-\bm{G}_1$ (for $\bm{G}=0$ the result is trivial due to Bloch function orthonormality). This is relevant for examining the flat-metric condition $\lambda_{\tau,a,b}(\bm{k};\bm{0},\bm{G})=\xi(\bm{G})\delta_{a,b}$ used in part of the analysis in Refs.~\cite{TBGIV,TBGV,TBGVI}. For the our parameters, the flat-metric condition is already moderately violated at zero strain for this shell of reciprocal lattice momenta. Strain shuffles the form factors around, but does not significantly change the scale of mBZ variation.

\clearpage

\begin{figure}[h!]
    \includegraphics[width=0.65\linewidth]{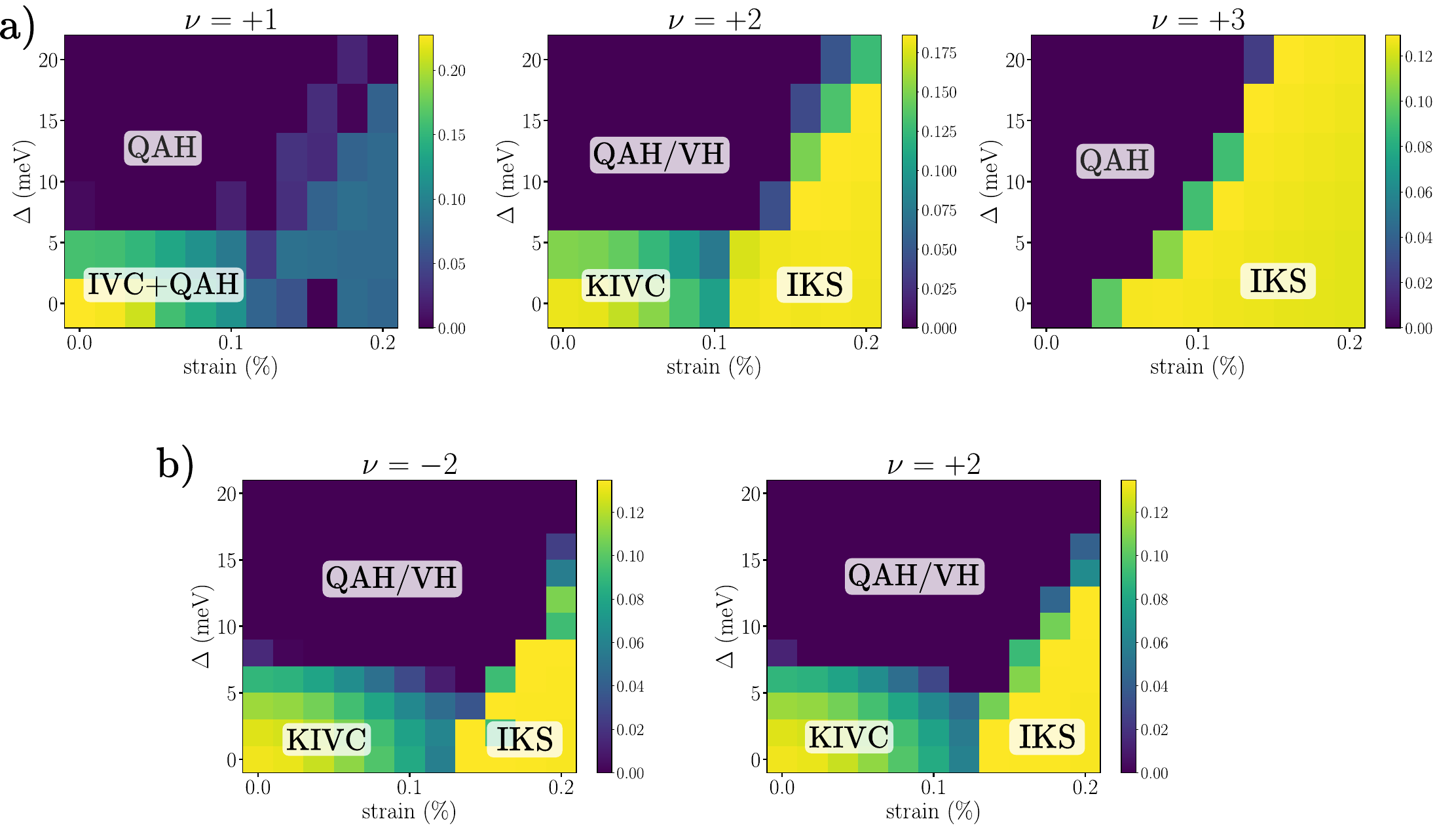}
    \caption{a) Strain-substrate phase diagrams with 6 active bands per spin/valley. System size is $9\times9$ and NLT is not included. b) Strain-substrate phase diagrams with 2 active bands per spin/valley. System size is $12\times12$ and NLT is not included. For all plots, allowed translation symmetry-breaking is period-tripling along $\bm{G}_1$, and strain is applied along $x$ axis.}\label{figapp:extra_phase}
\end{figure}

\begin{figure}[h!]
    \includegraphics[width=0.43\linewidth]{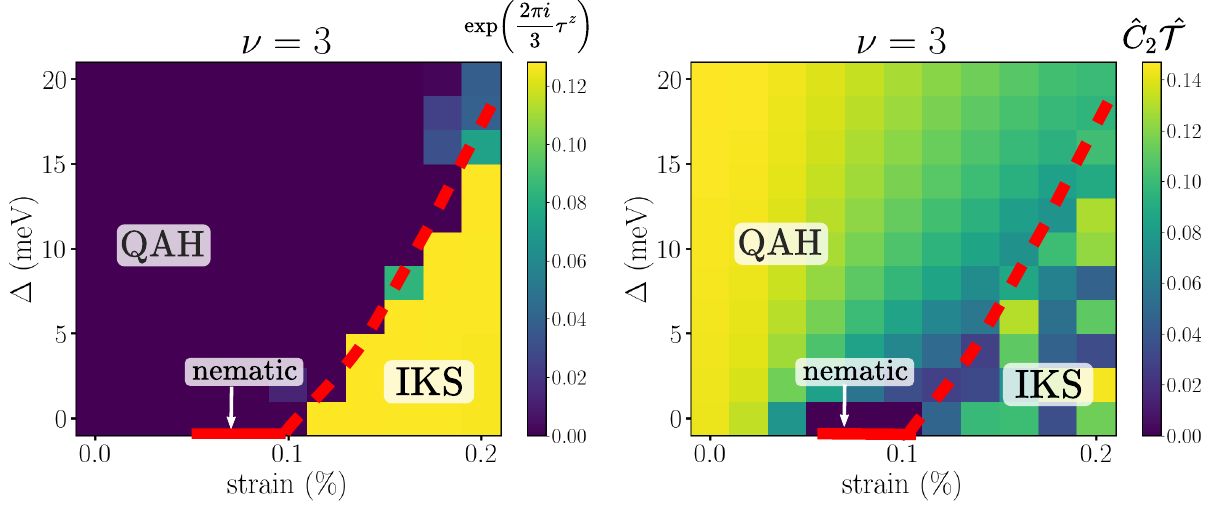}
    \caption{Strain-substrate phase diagram at $\nu=+3$ with different choice of system parameters that leads to a nematic phase: $w_{AA}=85\,\text{meV}$, screening length $d=25\,\text{nm}$, $\theta=1.05^\circ$, and stronger NLT $\lambda_2=2\lambda_1=2\lambda_3=0.36\,\text{eV\r{A}}$. Right figure shows $\hat{C}_{2z}\hat{\mathcal{T}}$-breaking, defined as the Frobenius norm of the difference of the density matrix after acting with $\hat{C}_{2z}\hat{\mathcal{T}}$. The value in the IKS is random because $\hat{C}_{2z}\hat{\mathcal{T}}$ does not commute with $U(1)_V$ rotation (which is broken).}\label{figapp:nematic}
\end{figure}

\begin{figure}[h!]
    \includegraphics[width=0.6\linewidth]{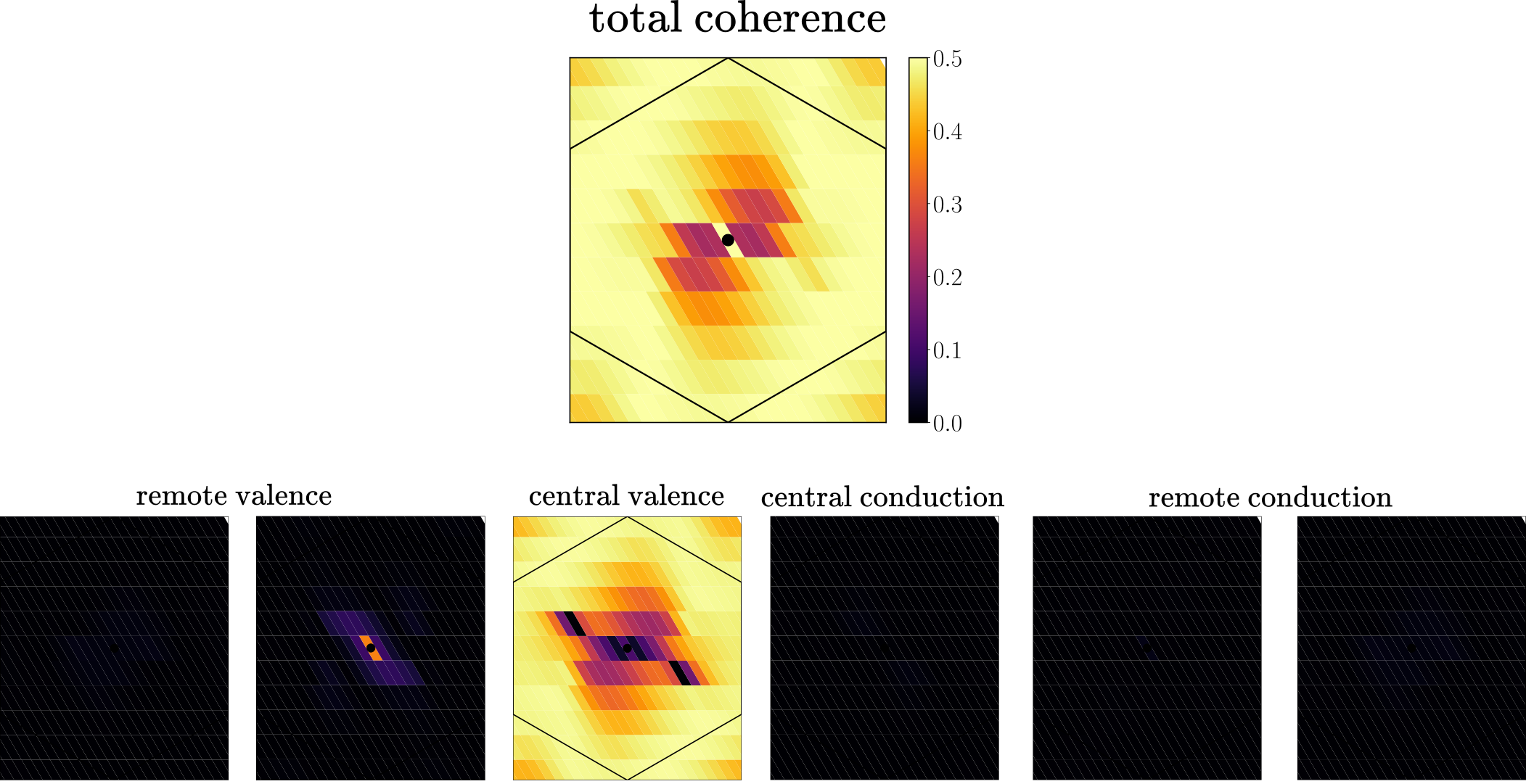}
    \caption{Frobenius norm of the IVC $\sim|\langle c^\dagger_{K,\bm{k}+\frac{\bm{q}}{2}}c_{K',\bm{k}-\frac{\bm{q}}{2}}\rangle|$ at wavector $\bm{q}=\bm{G}_1/3$ of the $\nu=-3$ IKS. System size is $24\times 8$, strain is $0.2\%$, substrate is $\Delta=0$, NLT is not included, and six bands per spin/valley are kept. Top shows the total IVC, bottom row shows the IVC filtered into BM band-diagonal components, starting from the lowest active valence to the highest active conduction band.}
    \label{figapp:band_coherences}
\end{figure}

\begin{figure}[h!]
    \includegraphics[width=0.58\linewidth]{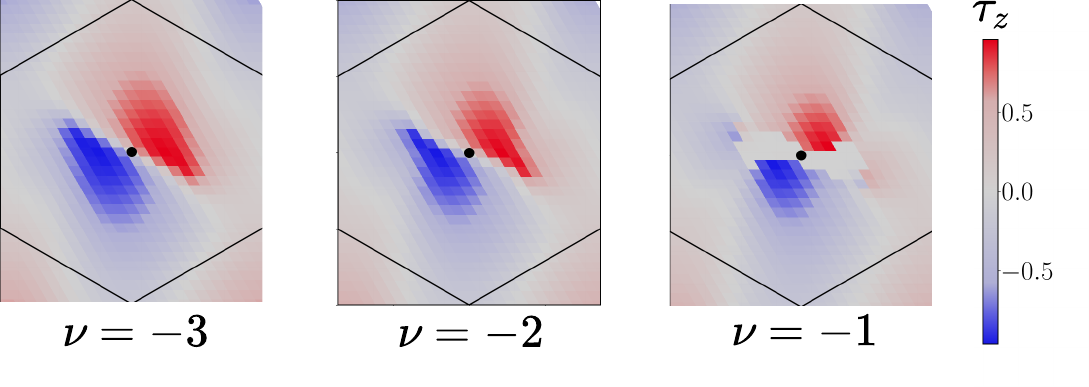}
    \caption{Expectation value of $\tau_z$ of the IKS across the mBZ for different fillings (spins have been polarized in a way such that only the spin component that includes the spiral order is shown). The momentum has been given a valley-dependent boost of $\pm\bm{q}/2$, where $\bm{q}=\bm{G}_1/3$ and $\varphi=0^\circ$. System size is $24\times24$, strain is $0.2\%$, substrate is $\Delta=0$ and NLT is included.}
    \label{figapp:tauz_fillings}
\end{figure}

\begin{figure}[h!]
    \includegraphics[width=0.55\linewidth]{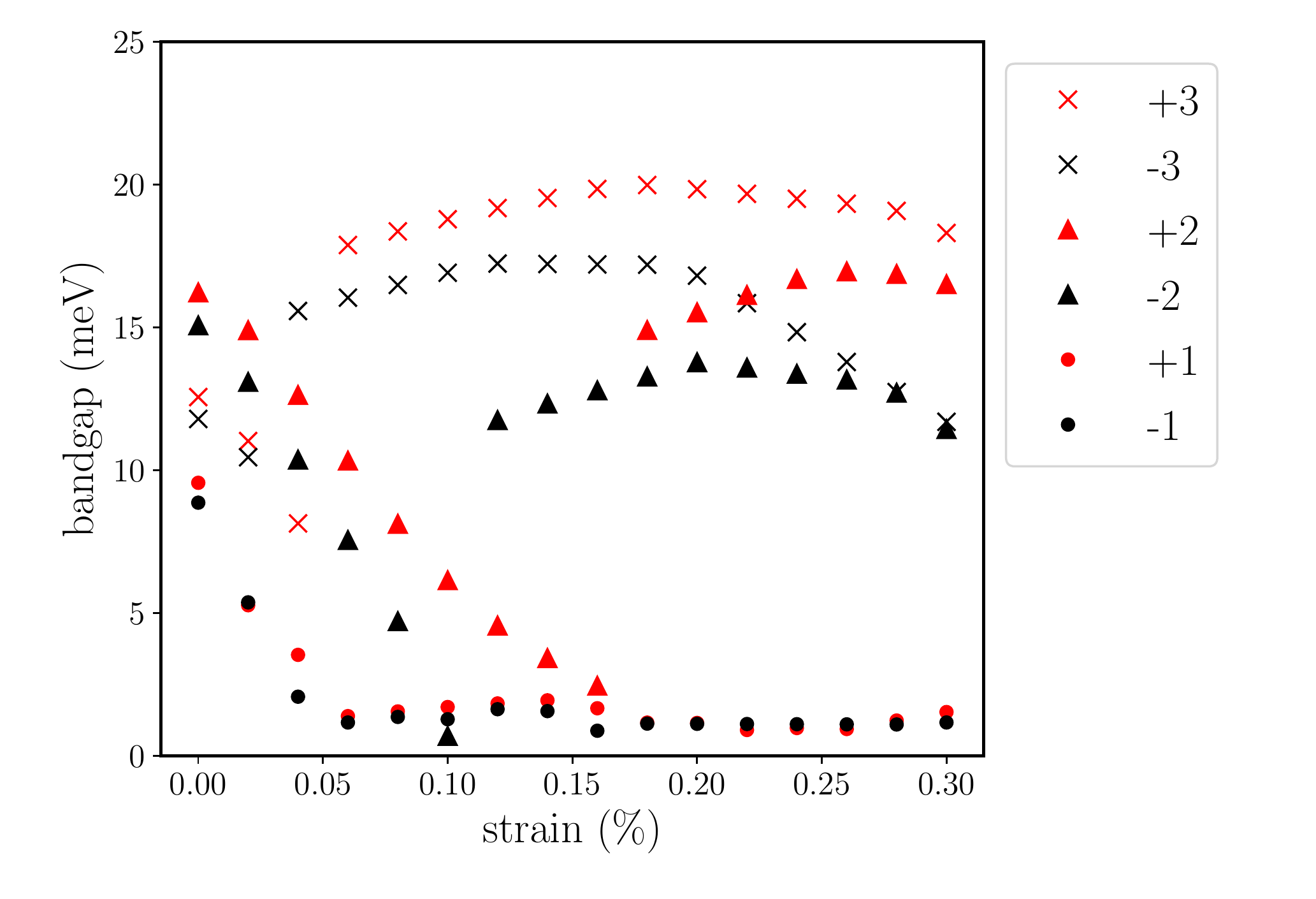}
    \caption{Band gaps along the strain axis for the strain-substrate phase diagrams in Figs.~\ref{fig:phase_nu2},\ref{fig:phase_nu3}, showing the PHS-breaking introduced by NLT.}
    \label{figapp:PH_bandgaps}
\end{figure}

\begin{figure}[h!]
    \includegraphics[width=0.95\linewidth]{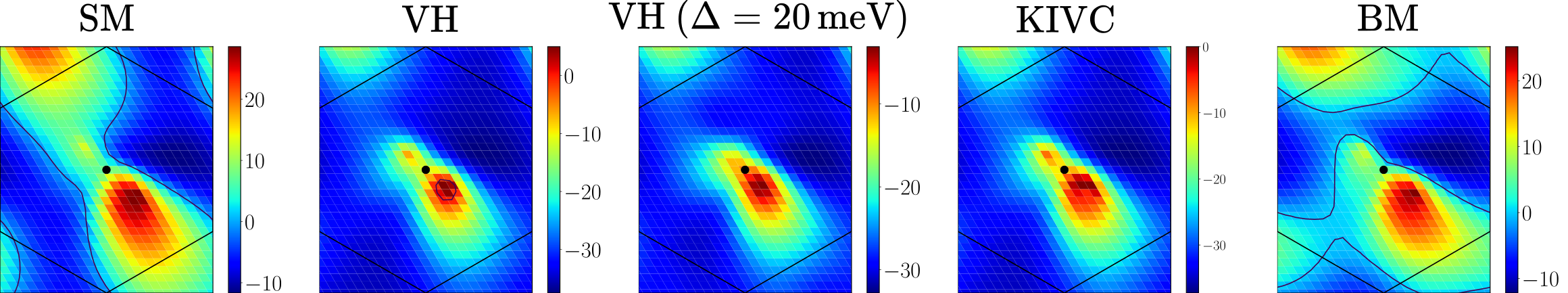}
    \caption{Valley-$K$ HF bandstructures for various translation symmetric states at $\nu=-2$ for parameters in IKS phase (strain=0.2$\%$, $\varphi=0^\circ$, no NLT, no substrate). Lower band is shown. Energies are in meV and measured relative to $E_F$. Fermi surfaces are plotted with dark grey lines. The second VH plot has a 20 meV sublattice splitting applied to generate the density matrix, but not to generate the HF spectrum. For the $U(1)_V$-breaking KIVC, the effective bandstructure is found by taking a linear combination of the valence bands at each $\bm{k}$, weighted by valley polarization. BM consists of filling the non-interacting band structure (without $\hat{H}_{\mathrm{DC}}$) to $\nu=-2$. For all cases, the bandstructure in valley $K'$ is found by taking $\bm{k}\rightarrow -\bm{k}$.}
    \label{figapp:parent_energies}
\end{figure}

\begin{figure}[h!]
    \includegraphics[width=0.85\linewidth]{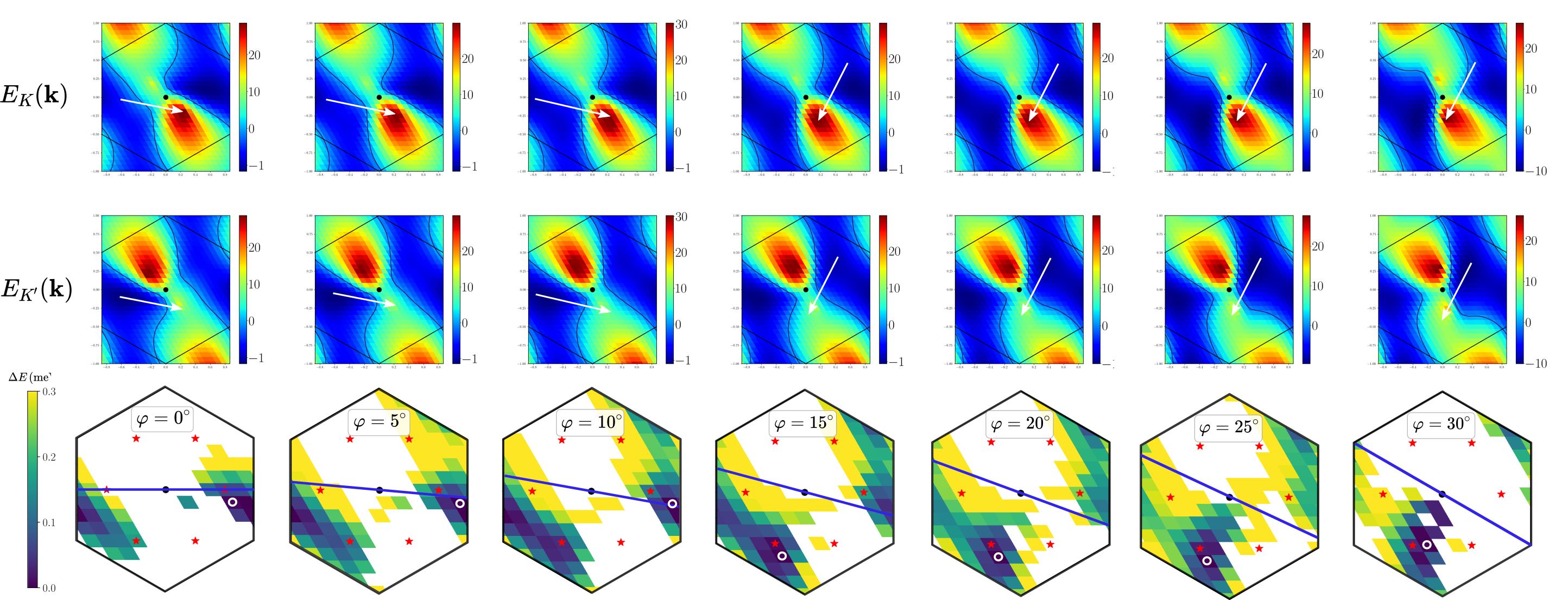}
    \caption{SM bandstructures and IKS $\bm{q}$ energetics at $\nu=-2$. First two rows show HF bandstructures in the two valleys for the SM at $\nu=-2$. White arrows are guides to the eye, showing for each column a suggested $\bm{q}_0$ that connects low-energy features in $E_{K'}$ with high-energy features in $E_K$, and is consistent with the self-consistent HF solution for the IKS.  Bottom row shows IKS energy as a function of $\bm{q}$ (non-IKS states that converged to higher energies were discarded). From left to right, columns denote different strain angles $\varphi=0^\circ,5^\circ,10^\circ,15^\circ,20^\circ,25^\circ,30^\circ$. Strain is $0.2\%$, system size is $24\times24$ for SM and $12\times12$ for IKS. NLT is not included.}
    \label{figapp:varphi_q0}
\end{figure}

\begin{figure}[h!]
    \includegraphics[width=0.65\linewidth]{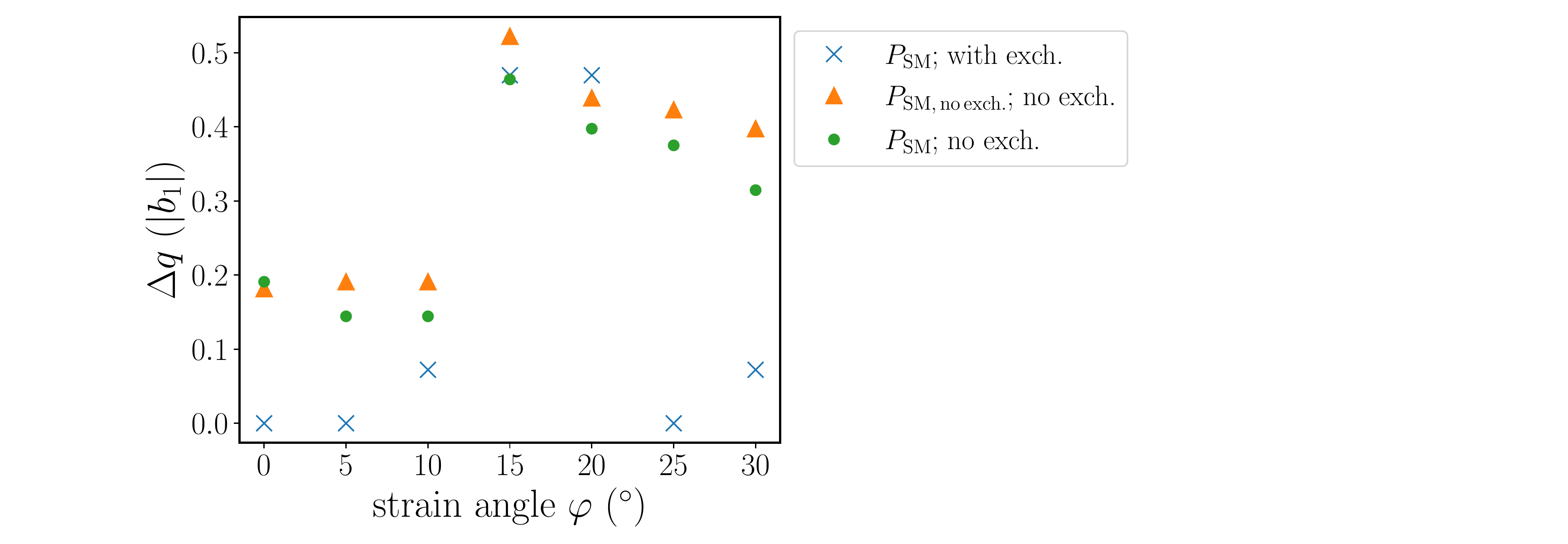}
    \caption{Predicted $\bm{q}_0$ from a SM band structure vs actual ideal IKS wavevector $\bm{q}_0$ (obtained from self-consistent HF, see bottom row of Fig.~\ref{figapp:varphi_q0}) for different strain angles. The predicted $\bm{q}_0$ connects the minimum energy momentum from the lower band in valley $K'$ of the SM, to the maximum energy momentum from the lower band in valley $K$. $\Delta q$ is defined as the Euclidean distance between the predicted and the actual $\bm{q}_0$, normalized to the length of $\bm{G}_1$. Strain is $0.2\%$, filling is $\nu=-2$, NLT is not included, and system size is $24\times24$ ($12\times12$) for the SM (IKS). Blue crosses: Exchange is included, and band structure is from a self-consistent SM. Orange triangles: Exchange is not included, and band structure is from a self-consistent SM. Green circles: Exchange is not included, but the state considered is calculated with exchange included (i.e. same state as blue crosses).}
    \label{figapp:q0_prediction}
\end{figure}

\begin{figure}[h!]
    \includegraphics[width=0.6\linewidth]{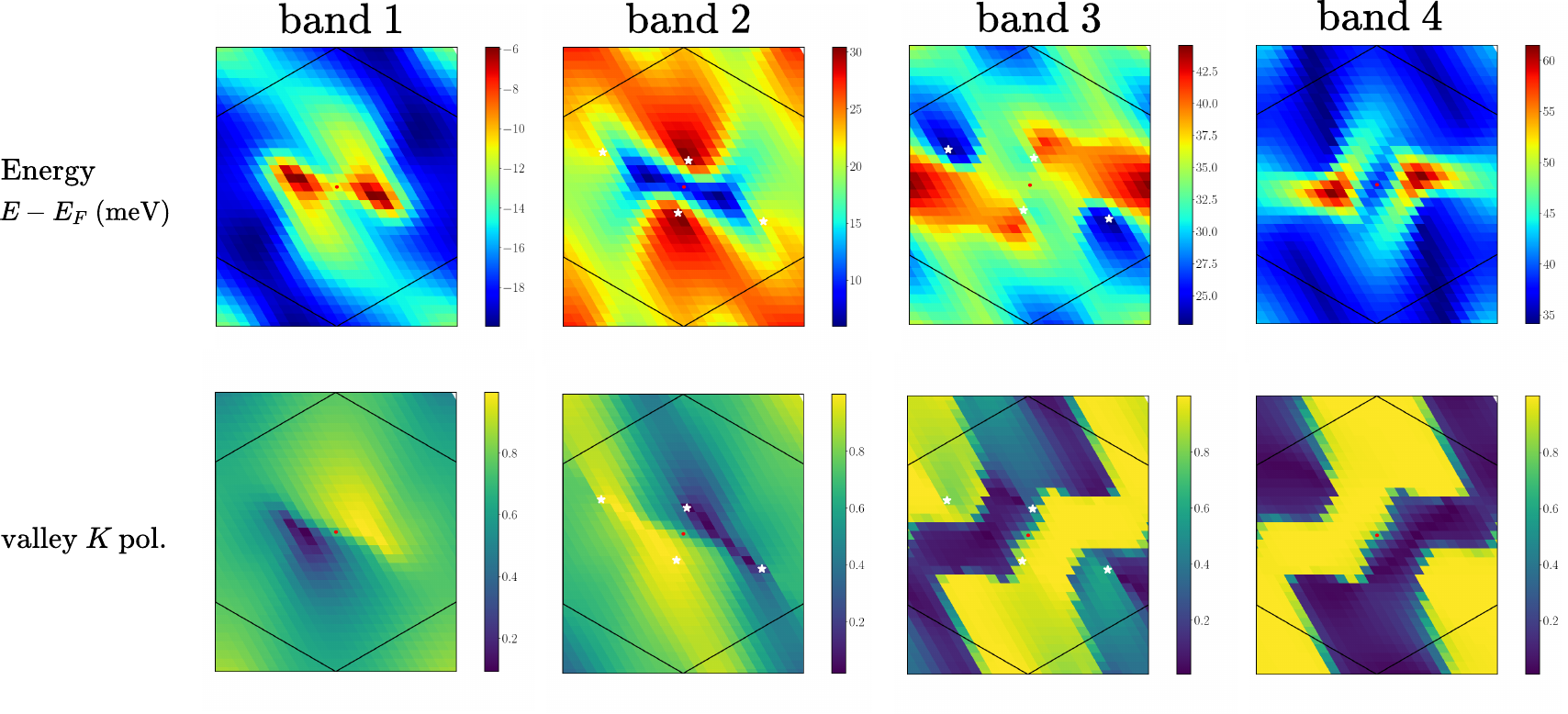}
    \caption{IKS bandstructure at $\nu=-2$, with $0.2\%$ strain, strain angle $\varphi=0^\circ$, system size $24\times24$, and no NLT. All quantities are plotted in the unfolded BZ, with valley $K$ ($K'$) shifted by $-\bm{q}/2$ ($\bm{q}/2$), where $\bm{q}=\bm{G}_1/3$ is the IKS wavevector. Top row shows the HF spectrum for the four bands (since the valleys are coupled, there are four bands excluding spin). Bottom row shows the $K$ valley polarization of each band. White stars denote locations of the Dirac points connecting bands 2 and 3. }\label{figapp:IKS_bandstructure}
\end{figure}

\begin{figure}[h!]
    \includegraphics[width=0.68\linewidth]{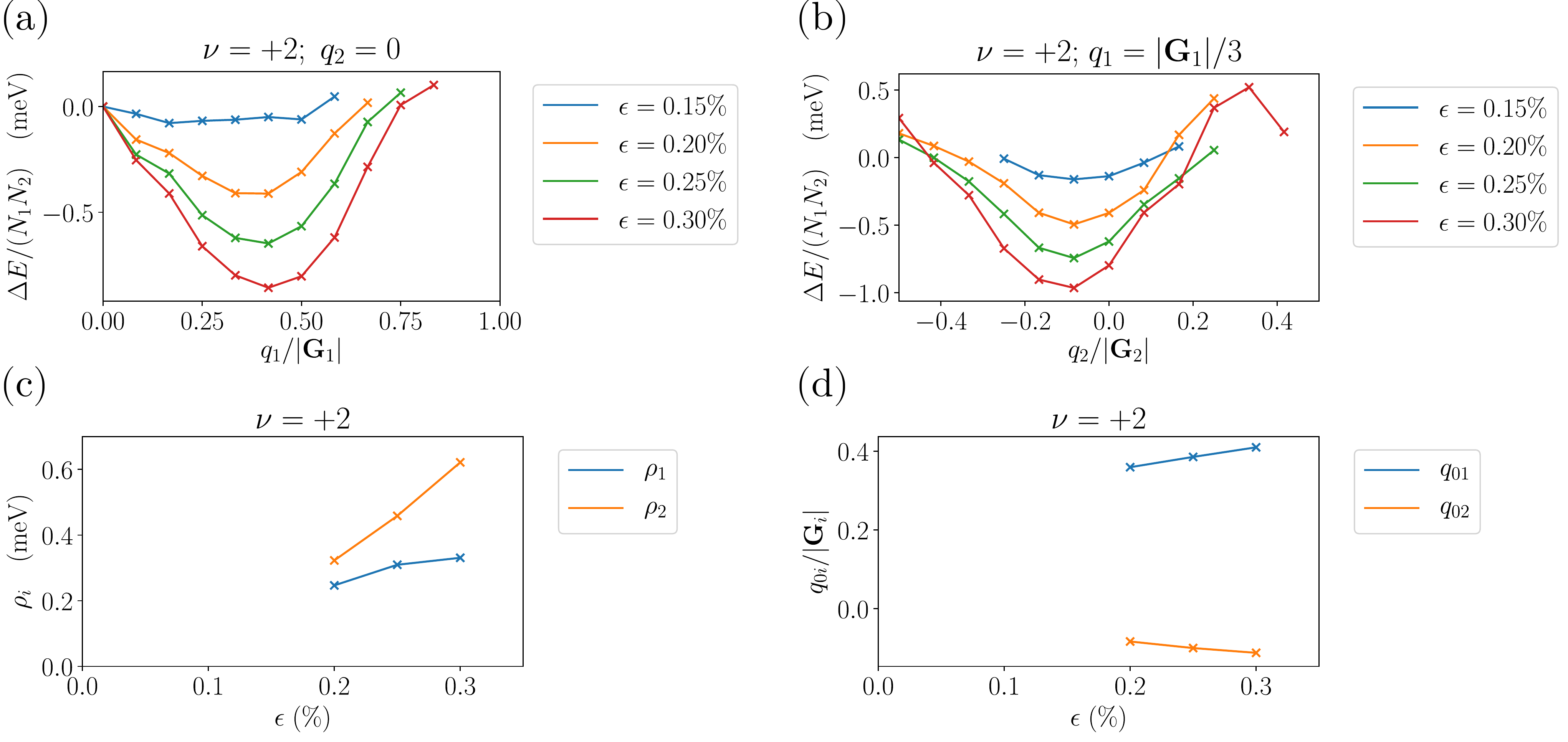}
    \caption{IKS dispersion at $\nu=+2$, for different values of strain, strain angle $\varphi=0^\circ$, system size $12\times12$, and no NLT. (a) Dispersion relation along the $\bm{G}_1$ direction. (b) Dispersion relation along the $\bm{G}_2$ direction for $\bm{q}_1=\bm{G}_1/3$. (c) Extracted stiffnesses $\rho_1$ and $\rho_2$ along the two directions. (d)  Extracted offsets of the minimum of the dispersion $q_1$ and $q_2$ along the two directions.}\label{figapp:nu_2_stiffnesses}
\end{figure}

\begin{figure}[h!]
    \includegraphics[width=0.36\linewidth]{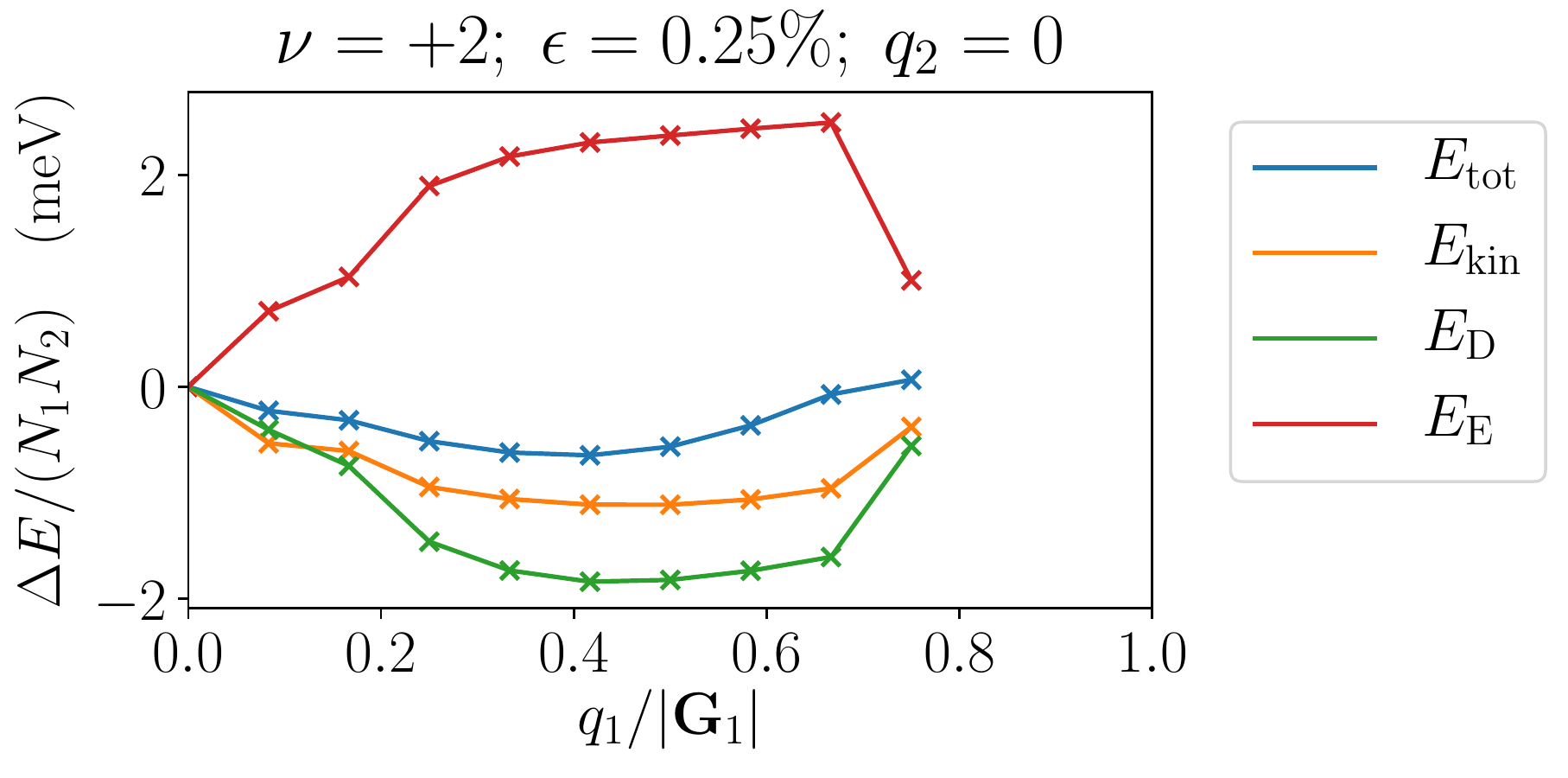}
    \caption{IKS dispersion at $\nu=+2$, for strain $\epsilon=0.25\%$, strain angle $\varphi=0^\circ$, system size $12\times12$, and no NLT. We show the energy difference between the IKS state at a given wavevector and the translationally invariant state broken down into the different energy contributions: Total energy ($E_\textrm{tot}$), kinetic energy ($E_\textrm{kin}$), direct energy ($E_\textrm{D}$) and exchange energy ($E_\textrm{E}$). }\label{figapp:energy_breakdown}
\end{figure}

    \begin{figure}[h!]
    \includegraphics[ width=0.45\linewidth]{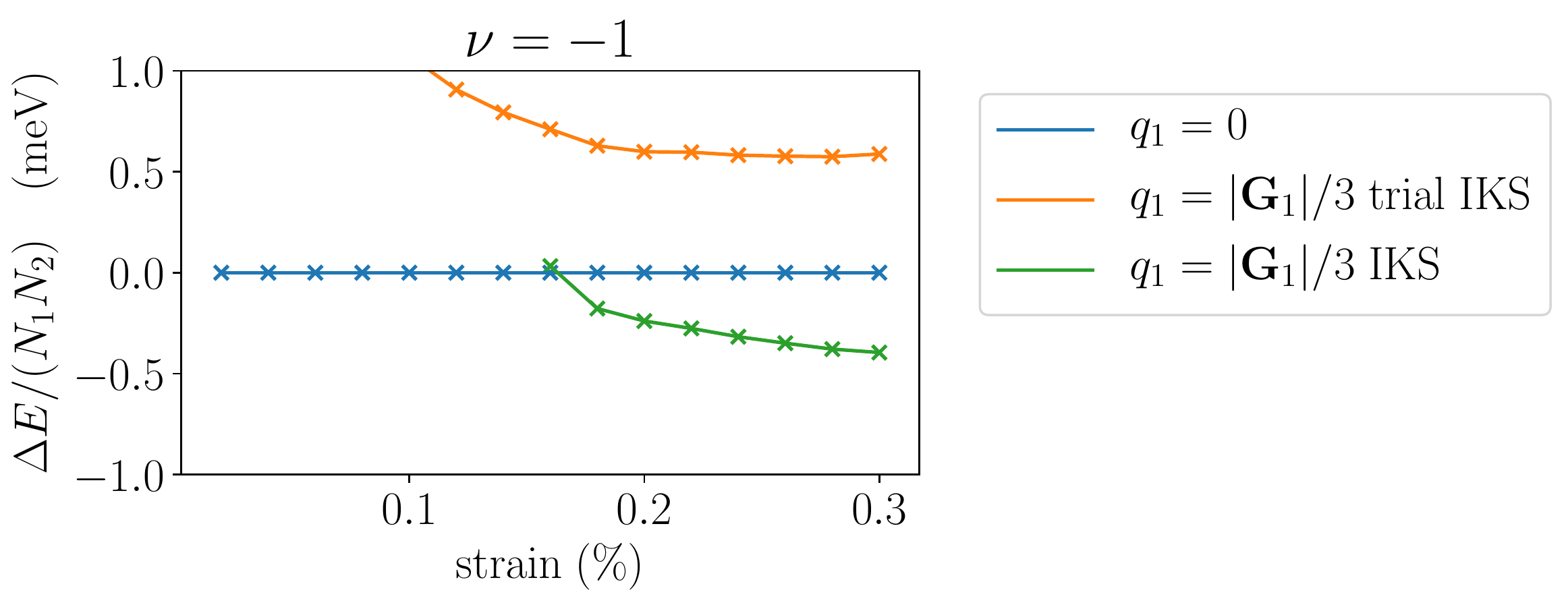}
    \caption{Comparison of the energies of the translationally invariant state, the IKS state and the IKS trial state at $\nu=-1$. The IKS trial state is obtained by adding the IKS HF solution at $\nu=-3$ to the translationally invariant solution at $\nu=-2$. System size is $12\times12$, $\Delta=0\ $ meV and NLT is included.}
    \label{figapp:trial1}
    \end{figure}
    
    \begin{figure}[h!]
    \includegraphics[ width=0.36\linewidth]{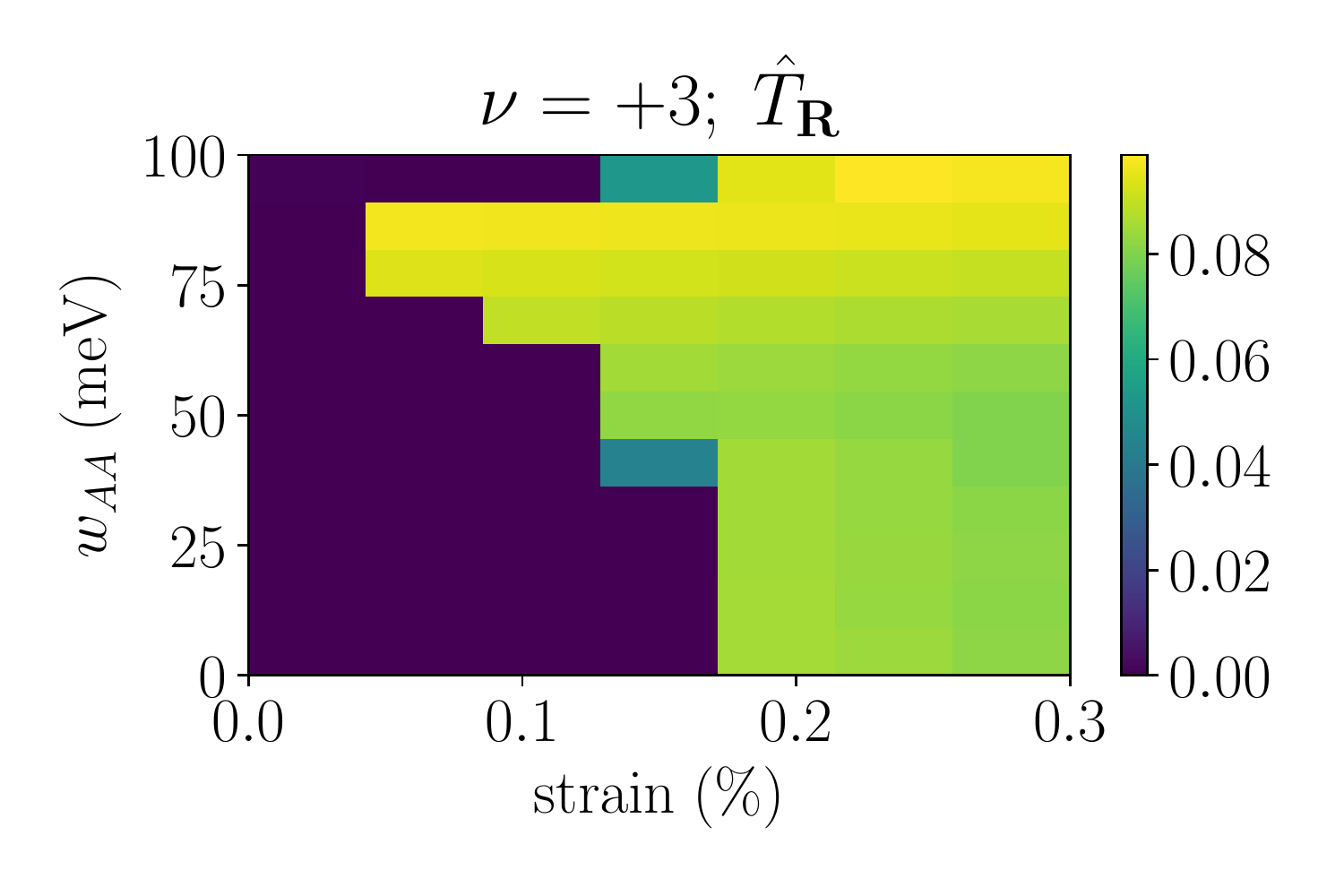}
    \caption{Translational symmetry breaking order parameter of the HF ground state at $\nu=+3$ for different values of strain and $w_{AA}$ (for $w_{AB}=110$meV fixed). System size is $12\times12$, $\Delta=0$meV and NLT is not included.}
    \label{figapp:chiral}
    \end{figure}

    \begin{figure}[h!]
    \includegraphics[ width=0.5\linewidth]{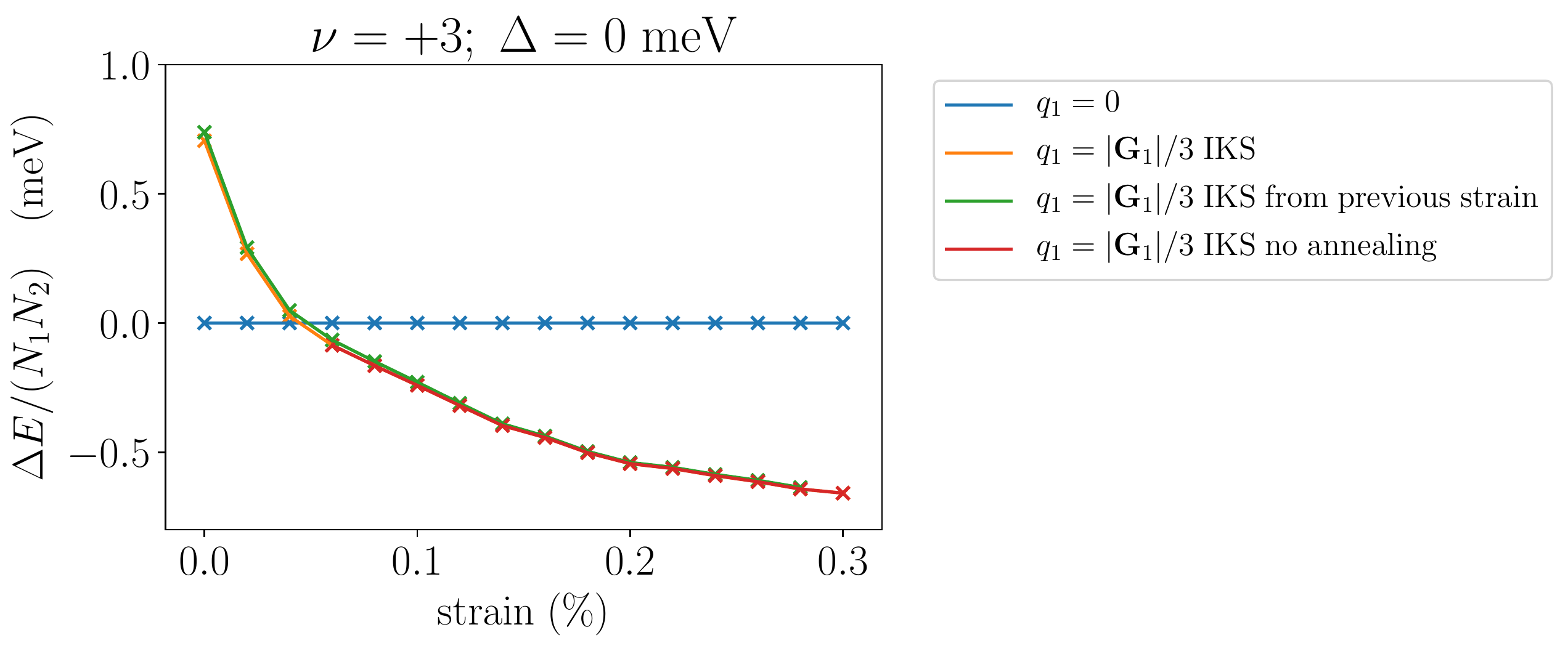}
    \caption{Energy of the $q_1=|\mathbf{G}_1|/3$ IKS state (red) compared to the lowest energy translationally invariant state (blue). In order to find the IKS HF state when it is not the ground state, we use an annealing procedure. We decrease the strain value starting from $0.3\%$ in steps of $0.02\%$ and use the IKS HF state from the previous strain run as an input state. The energy of the input state (green) is close in energy to the lowest energy IKS HF state at that strain value (yellow). System size is $12\times12$, $\Delta=0$meV and NLT is included.}
    \label{figapp:anneal}
    \end{figure}

    \begin{figure}[h!]
    \includegraphics[ width=0.6\linewidth]{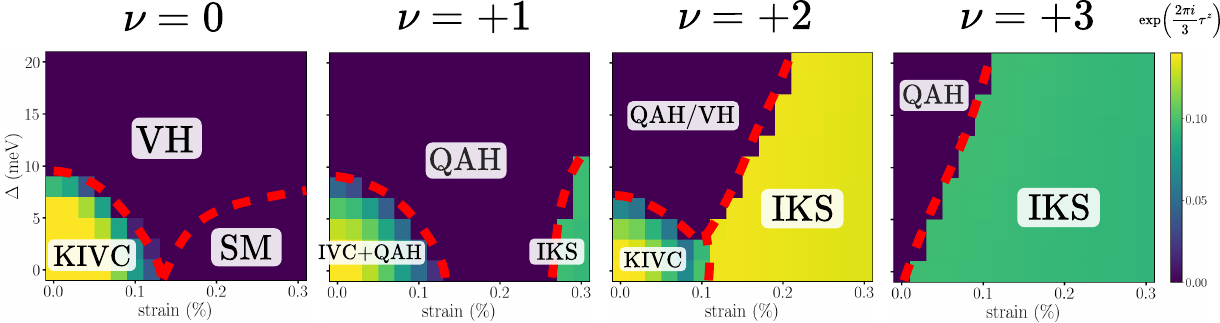}
    \caption{Summary of phases found in self-consistent Hartree-Fock at all non-negative integer fillings $\nu$ for different heterostrains and substrate potentials $\Delta$, without non-local tunneling. Compared to the main text (Fig.~\ref{fig:all_phases}) which used a graphene subtraction scheme, the calculations here used the `average' scheme (see discussion in App.~\ref{secapp:additional}). The axes as are the same as Fig.~\ref{fig:all_phases} in the main text.}
    \label{figapp:allphasesaverage}
    \end{figure}

    \begin{figure}[h!]
    \includegraphics[ width=1\linewidth]{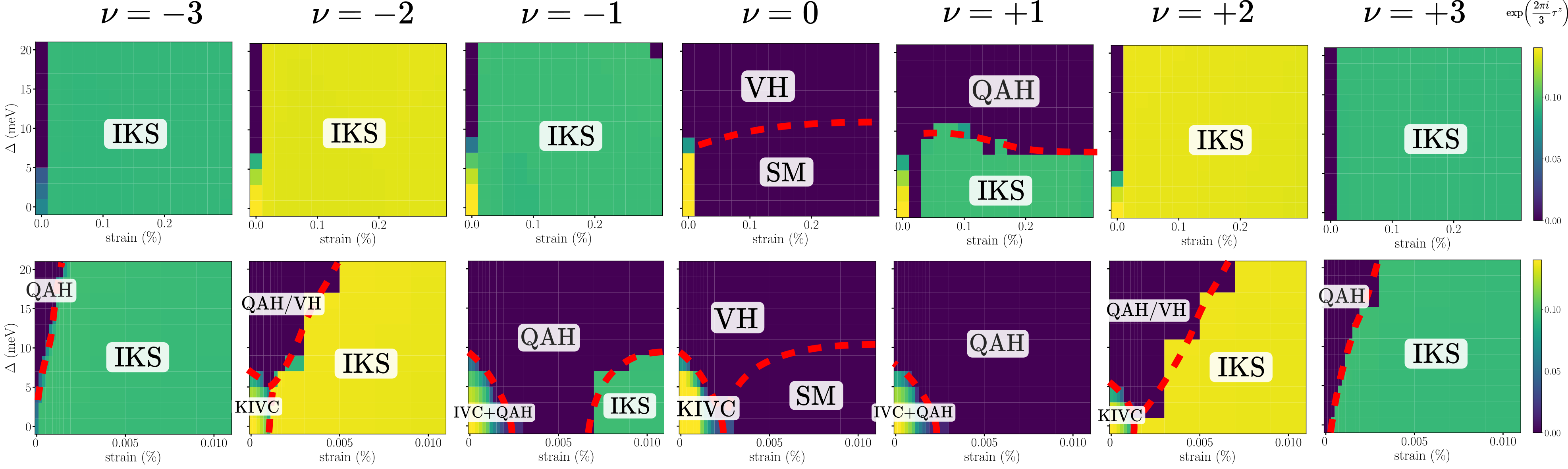}
    \caption{Summary of phases found in self-consistent Hartree-Fock at all integer fillings $\nu$ for different heterostrains and substrate potentials $\Delta$, with non-local tunneling included. Compared to the main text (Fig.~\ref{fig:all_phases}) which used a graphene subtraction scheme, the calculations here used the CN scheme (see discussion in App.~\ref{secapp:additional}). Both rows use the same parameters, except for the different strain scales on the horizontal axes. The top row has the same axes as Fig.~\ref{fig:all_phases} in the main text. The bottom row considers much smaller strains to illustrate the ordering of the competing phases.}
    \label{figapp:allphasesCN}
    \end{figure}

    \begin{figure}[h!]
    \includegraphics[ width=0.6\linewidth]{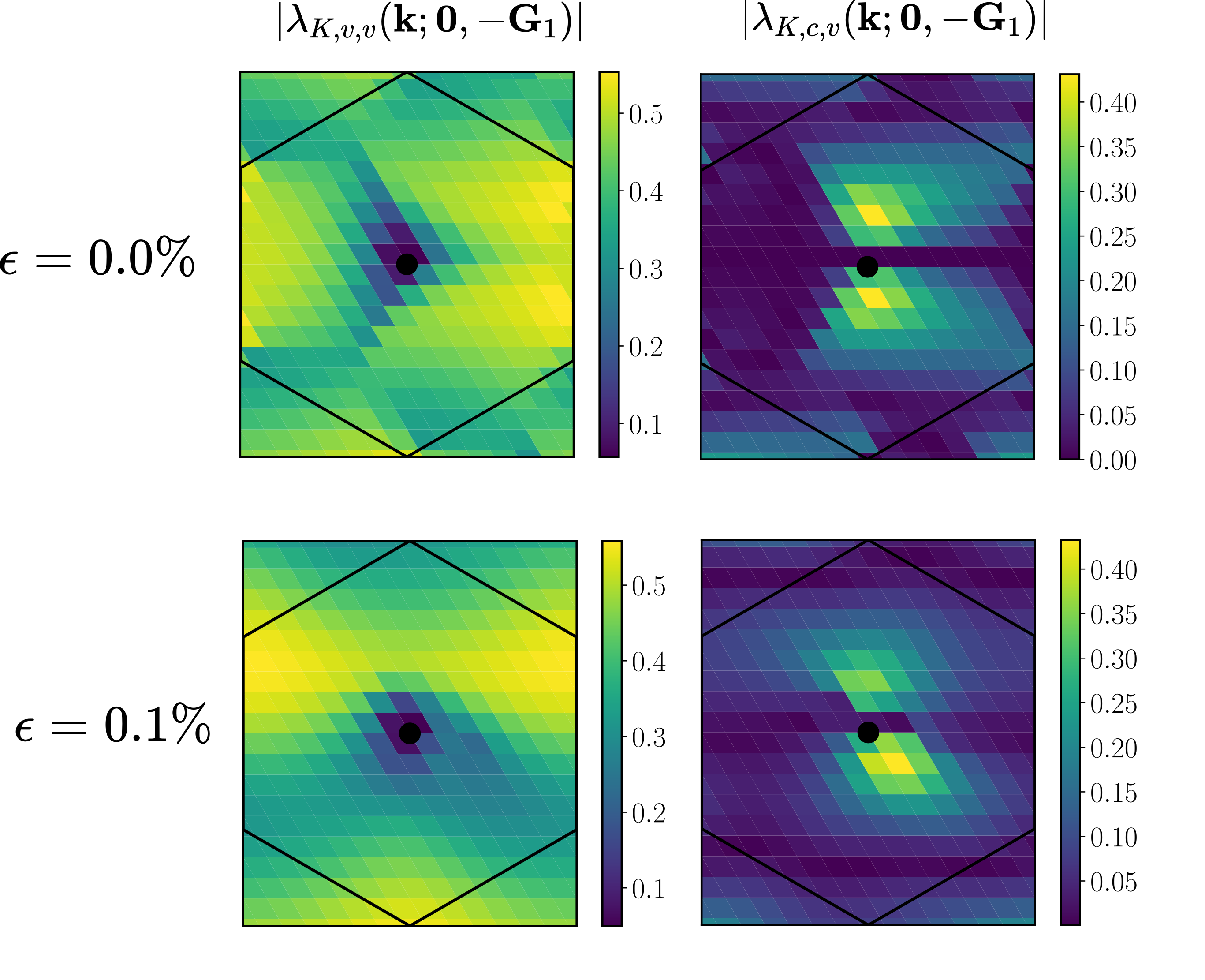}
    \caption{Plots of the magnitude of the form factor for momentum transfer $-\bm{G}_1$ as a function of the mBZ momentum for combinations of valence $v$ and conduction $c$ bands. Under the flat-metric condition, the left column needs to be constant, and the right column should vanish.}
    \label{figapp:FMC}
    \end{figure}

\clearpage

\section{Thermal fluctuations in commensurate IVC spiral states}
\label{app:thermal}

In this appendix we introduce and study statistical mechanics models that allow us to develop an intuition for the effects of thermal fluctuations in IVC spiral states. An important difference, however, between the models studied here and the IVC spiral states in TBG, is that in the statistical mechanics models we take the ordering vector to be commensurate with the lattice whereas our numerical Hartree-Fock results suggest that the optimal ordering vector for the IVC spirals in TBG is in fact incommensurate. 

Before discussing the triangular lattice case relevant for TBG, we will first consider a simpler model on the square lattice. The reason is that we can relate this model to previously studied generalized XY models, and thus obtain a fairly complete understanding of its phase diagram. It also allows us to introduce the essential physical concepts in a concise and clear way. The same physical concepts will then play an equally important role in our model for XY spiral states on the triangular lattice.

\begin{figure}[h]
    \centering
    \includegraphics[scale=0.5]{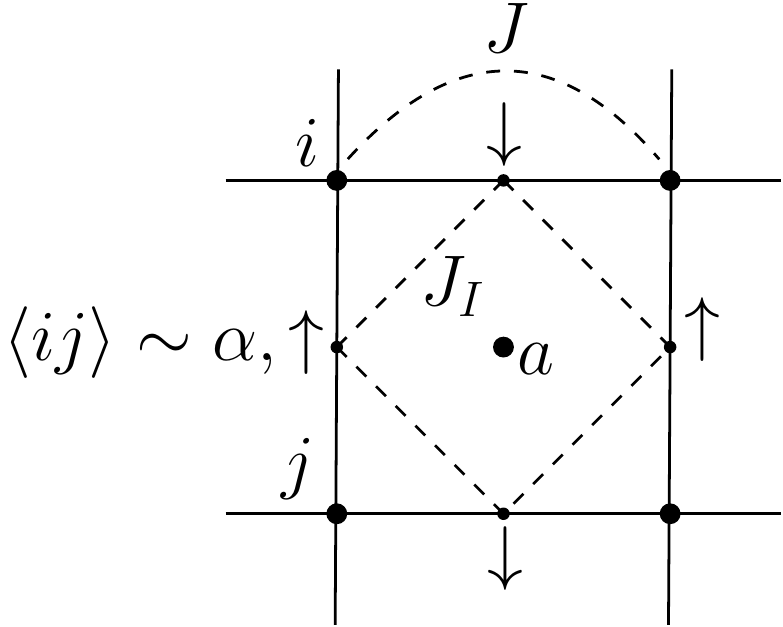}
    \caption{Square lattices and spin couplings used for the coupled Ising-XY model.}
    \label{fig:square}
\end{figure}

\subsection{Coupled Ising-XY model on the square lattice}

The square lattice model has XY spins on the lattice sites, and an Ising spin $s$ on every edge.  This is shown schematically in Fig. \ref{fig:square}. The edges of the square lattice containing the XY spins will be denoted both in the usual way by $\langle ij\rangle$, which labels nearest-neighbour pairs of sites, or as sites $\alpha$ of a $45^\circ$ rotated square lattice,  depending on the context. Later on we will also need to refer to the sites of the dual lattice, which will be labelled as $a,b,c,\dots$.

Having defined the lattice, we now define the Hamiltonian as a sum of two terms:

\begin{eqnarray}
H & = & H_s + H_\theta\\
H_s & = & J_I \sum_{\langle \alpha\beta\rangle} s_\alpha s_\beta \\
H_\theta & = & -J\sum_{\langle ij\rangle} \cos(\theta_i - \theta_j -\frac{\pi}{2}(1+s_{\langle ij\rangle}))
\end{eqnarray}
The first term $H_s$ is simply an anti-ferromagnetic Ising Hamiltonian on the $45^\circ$ rotated dashed square lattice (see Fig. \ref{fig:square}). The second term in the Hamiltonian, $H_\theta$, couples the nearest-neighbour XY spins $\theta_i$ and $\theta_j$ depending on the values of the Ising spin on the edge $\langle ij\rangle$. In particular, if we simply replace the Ising spins by their value in one of the two ground states, then we see that the XY spins will form a stripe pattern with either alternating rows or columns of oppositely oriented spins. 

The physics of the coupled Ising-XY model introduced here is very similar to that of previously studied frustrated or `generalized' XY models, which also exhibit both algebraic order and discrete symmetry breaking \cite{LeeNegeleLandau,LeeGrinstein,LeeWu,GranatoKosterlitz,GranatoKosterlitzNightengale,Henley,ChandraColemanLarkin}. However, to the best of our knowledge, the connection between these models and XY stripe order has not been explored before.

\subsubsection{The solution with completely disordered Ising spins: $J_I =0$}

Let us first study the simplified case where the Ising spins are completely disordered, which is obtained by taking $J_I = 0$. In this limit, only the Hamiltonian $H_\theta$ remains. The effect of the freely fluctuating Ising spins is seen mostly easily using the following dual formulation of the XY model:

\begin{equation}
\beta \tilde{H}^{dual}_\theta = \beta H_h = \frac{\pi}{2K}\sum_{a,\mu} (\Delta_\mu h_a)^2 + i\frac{\pi}{2}(1+s_{a,\mu})\epsilon_{\mu\nu}\Delta_\nu h_{a}\, ,
\end{equation}
where $K = \pi J/T$, and $h_{a}$ is an integer-valued `height' field living on the dual lattice. In the above expression, the index $\mu$ runs over the values $x$ and $y$, $\Delta_\mu$ is a lattice derivative, and we have introduced the notation $s_{a,\mu} = s_{\langle a, a+e_\mu\rangle}$. The sum over the Ising spins can now be done exactly in a straightforward way, and one finds that it enforces the constraint

\begin{equation}
\Delta_\mu h_a = 0 \text{ mod } 2 \, ,
\end{equation}
which implies that the even and odd values for the height field decouple. Focusing on the even sector, we can define a new integer valued field $\tilde{h}$ as $h = 2\tilde{h}$. After a change of variable, we obtain

\begin{equation}
\beta H_{\tilde{h}} = \frac{\pi}{2\tilde{K}} \sum_{a,\mu} (\Delta_\mu \tilde{h}_a)^2\, ,
\end{equation}
with $\tilde{K} = K/4$. Because the BKT transition occurs when $\tilde{K}\sim 2,$ we arrive at the conclusion that the fluctuations of the Ising spins reduce the KT temperature by a factor of four. Contrary to what one might have expected, disordering the Ising variables does not completely disorder the XY spins. This is because the XY spins, even with fluctuating ferro- and anti-ferromagnetic interactions, still want to be collinear. So after disordering the Ising spins, the correlation function of $\cos(\theta)$ will decay exponentially, whereas the correlation function of $\cos(2\theta)$ still decays algebraically. 

\subsubsection{Sketch of the complete phase diagram}

\begin{figure}
    \centering
a)
\includegraphics[scale=0.45]{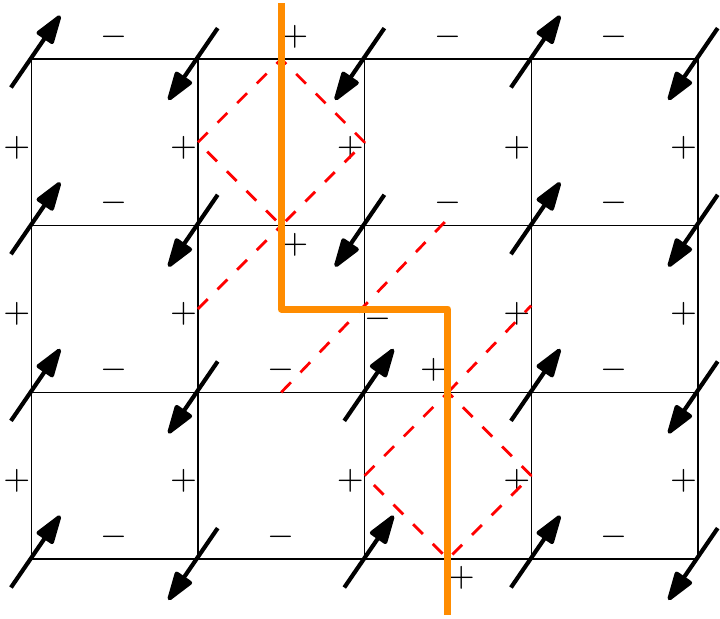}
\hspace{0.5 cm}
b)
\includegraphics[scale=0.45]{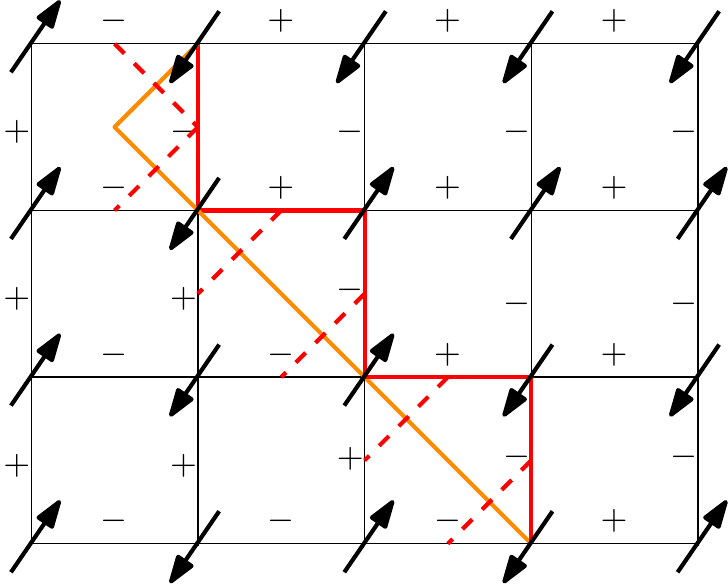}
\caption{Domain walls in the coupled Ising-XY model. (a) $\pi$-domain wall of the XY spins. The orange line denotes the domain wall path, and dashed red lines are Ising terms which contribute to the domain wall energy. (b) Domain wall of the Ising spins. Red full lines are XY coupling terms which contribute to the Ising domain wall energy.}
\label{fig:squareDW}
\end{figure}

Having explored the regime where the Ising spins are completely disordered, let us now attempt to sketch the complete phase diagram of the coupled Ising-XY model. We will fix the $XY$ coupling $J$, which sets the overall energy scale, and ask what happens when we tune $J_I$ and the temperature $T$.  

The case where $J_I = 0$ was solved exactly above, and it was found that there is a `half-KT' transition at a temperature which is four times smaller than the critical temperature for the conventional KT transition. We now consider what happens if we turn on a small Ising coupling $J_I$.  As shown in Fig. \ref{fig:squareDW} (a), this introduces a small domain wall energy of $8J_I$ per unit length for a $\pi$-domain wall of the XY spins. The reason that the domain wall energy for the XY spins is not of order $J$, is that we can lower its energy by flipping all the Ising spins along the domain wall path (see figure). A domain wall for the Ising spins, on the other hand, costs an energy of $4(J+J_I)$ per unit length because it introduces some frustration for the XY spins and is thus much more costly than the XY domain wall in the regime where $J_I \ll J$. As a result, for small $J_I$ we expect that the $\pi$-domain walls of the XY spins will proliferate first at a small temperature $T\sim J_I$. This triggers an Ising phase transition where the $\cos(\theta)$ variable becomes disordered, but $\cos(2\theta)$ retains its algebraic order. In more physical terms, this Ising transition is a deconfinement transition of half vortices, which are connected by strings of $\pi$-domain walls for the XY spins with a corresponding string tension of $8J_I$. At higher temperatures, there will be a half-KT transition characterized by the unbinding of half-vortices and half-antivortices.

If $J_I \gg J$, then the Ising spins remain frozen up to very high temperatures. So with increasing temperature, there will first be a conventional KT transition where vortex-antivortex pairs unbind. Although the XY spins are disordered above the KT temperature, the system still breaks the $C_4$ symmetry. This will be detectable via operators like $\cos(\theta_{(i,j)} - \theta_{(i+1,j)})$ and $\cos(\theta_{(i,j)} - \theta_{(i,j+1)})$, which are invariant under global rotations of the XY spins, but do detect the breaking of lattice rotation symmetry. At a higher temperature $T\sim J_I$, there is an Ising transition which restores the $C_4$ symmetry.

The final remaining question is how the regions of the phase diagram with $J_I\ll J$ and $J_I \gg J$ are connected at intermediate values of $J_I \sim J$. The way in which the half-KT, Ising and conventional KT transition lines meet was studied in previous works \cite{ShiLamacraftFendley,SernaChalkerFendley}. Interestingly, it was found that all these transition lines merge in a single Ising transition line as shown in Fig.~\ref{fig:IsingPHD}(a) (in Refs.~\cite{Foda,BauerTroyerSchoutens,HuijseBauerBerg}, it was argued that the multi-critical points where the Ising and KT lines meet correspond to a CFT with central charge $c=3/2$ and emergent supersymmetry). Physically, the single Ising transition line at $J_I\sim J$ is a simultaneous deconfinement and unbinding transition of the half-vortices. Along this line, the temperature is between the critical temperatures of the half-KT and the conventional KT transitions, so from the moment the half vortices are no longer confined in pairs to form full vortices, they also simultaneously unbind from their anti-vortex partners \cite{ShiLamacraftFendley,SernaChalkerFendley}.

\begin{figure}
    \centering
    a)
    \includegraphics[scale=0.37]{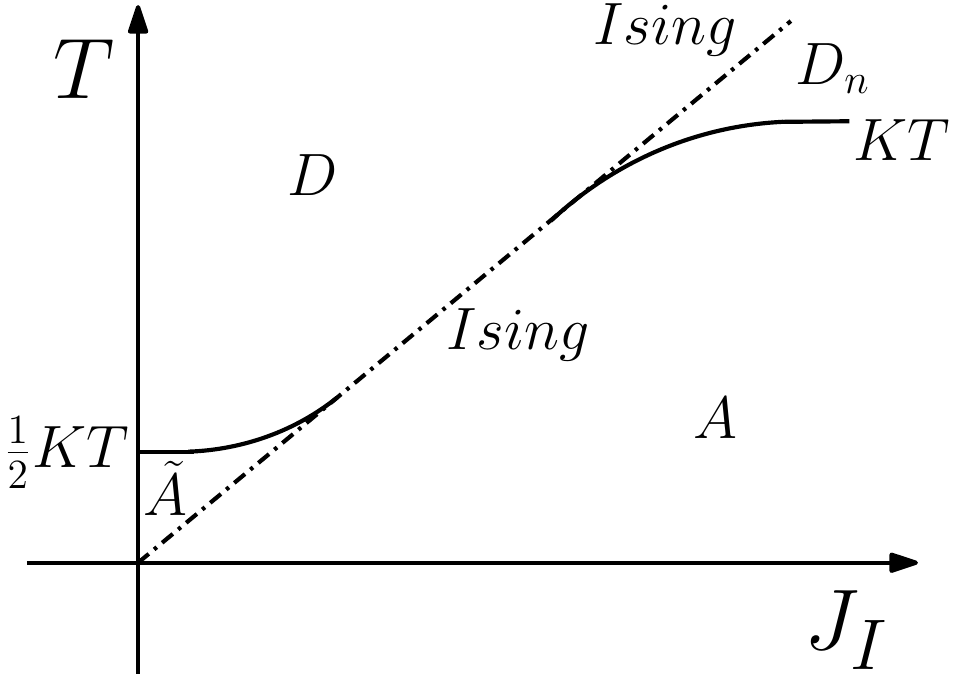}
    \hspace{0.3 cm}
    b)
    \includegraphics[scale = 0.37]{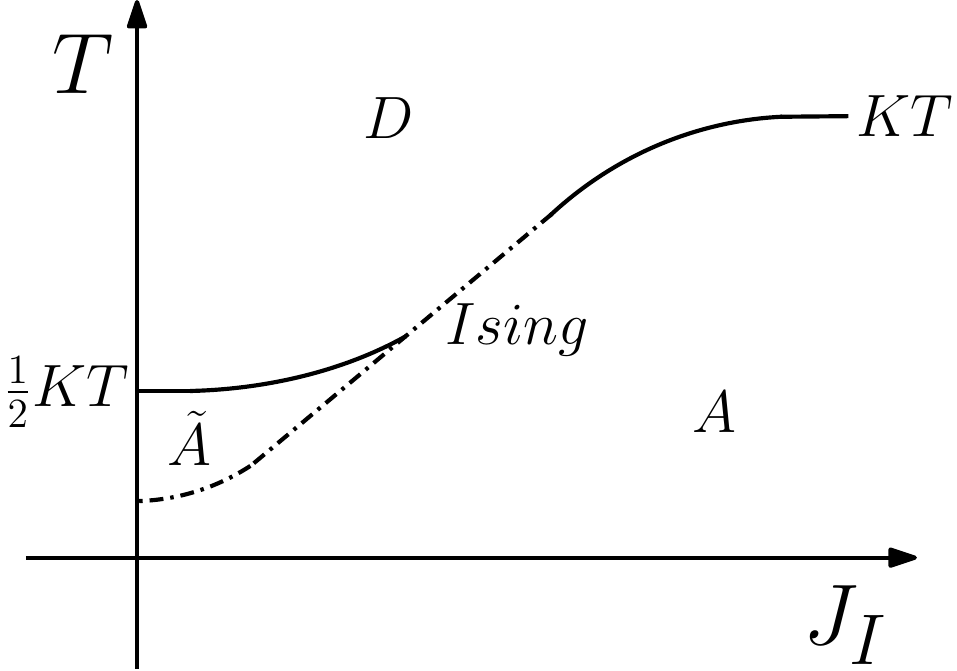}
    \caption{Schematic phase diagram of coupled Ising-XY model on the square lattice, at fixed XY coupling $J$. (a) With $C_4$ symmetry. (b) With $C_4$-breaking staggered magnetic field for the Ising spins. `$D$' is the completely disordered phase, `$D_n$' is the phase with disordered XY spins but with non-zero nematic order, $A$ is the conventional algebraic phase with quasi-long range order, and $\tilde{A}$ is the algebraic phase with deconfined half-vortices and thus exponentially decaying correlations of $\cos n\theta$ with odd $n$.}
    \label{fig:IsingPHD}
\end{figure}

Finally, we should comment on the effects of a small field which explicitly breaks the Ising or $C_4$ symmetry (which would correspond to strain in TBG). In the regime where $J_I\ll J$, the main effect of a non-zero staggered field $\pm h\sum_{i}(s_{i,x} - s_{i,y})$ for the Ising spins is to increase the string tension between the half-vortices from $8J_I$ to $8J_I+2h$. As a result, the Ising transition between the conventional algebraic phase and the algebraic phase with deconfined half-vortices will shift to higher temperatures. In the regime where $J_I \gg J$, the sharp Ising transition will become a crossover. Slightly above the KT transition, the system should still exhibit much stronger nematicity than expected from the small symmetry-breaking field (small strain). The resulting phase diagram with non-zero staggered field is shown in Fig.~\ref{fig:IsingPHD}(b).

\subsection{Triangular lattice model}

\begin{figure}
    \centering
    a)
    \includegraphics[scale=0.4]{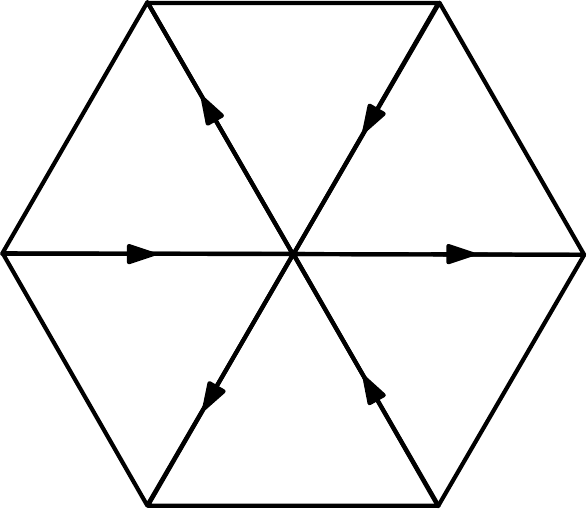}
    \hspace{0.5 cm}
    b)
    \includegraphics[scale=0.3]{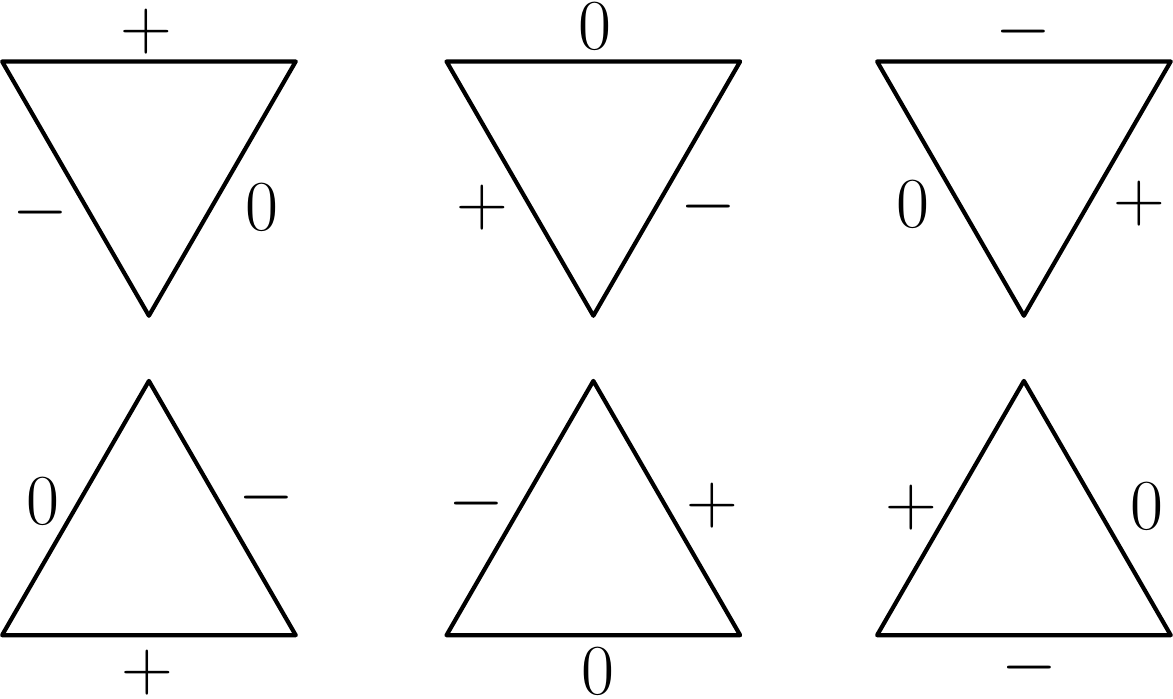}
    \caption{(a) Orientations of the different bonds on the triangular lattice. (b) The lowest energy plaquettes favored by $H_s^t$.}
    \label{fig:triangular}
\end{figure}

The Hamiltonian of the triangular lattice model consists of a sum of two terms:

\begin{eqnarray}
    H^t & = & H_{\theta}^{t} + H_s^t \\
    & = & -J\sum_{\langle ij\rangle} \cos(\Delta\theta_{\langle ij\rangle} - \frac{2\pi}{3}s_{\langle ij\rangle}) + H_s^t \label{eq:period} \,
\end{eqnarray}
where the sum in the first term is over nearest neighbour sites. Similar to the square lattice model, the triangular lattice model has an XY spin on every site. The quantity $\Delta\theta_{ij}$ is defined for nearest neighbour XY spins using the bond orientations shown in Fig. \ref{fig:triangular}(a). In particular, $\Delta \theta_{\langle ij\rangle}$ is equal to either $\theta_i - \theta_j$ or to $\theta_j - \theta_i$, depending on the orientation of the bond. The convention is that if an arrow on a particular bond (see Fig. \ref{fig:triangular}) points from $\theta_i$ to $\theta_j$, then $\Delta \theta_{\langle ij\rangle} = \theta_j - \theta_i$. Otherwise, $\Delta \theta_{\langle ij\rangle} = \theta_i - \theta_j$.

On every edge, the triangular lattice model has an additional degree of freedom $s$, which can take on three different values: $s \in \{-,0,+\}$. The Hamiltonian $H_s^t$ involves only the edge degrees of freedom, and consists of a sum of terms which attribute an energy to every plaquette of the triangular lattice. $H_s^t$ distinguishes between two types of plaquette decorations: the ones shown in Fig.~\ref{fig:triangular}(b) have zero energy, whereas all other plaquette decorations receive an energy penalty $\Delta_s$.

With these definitions, it is straightforward to see that the ground states of $H^t$ consist of uni-directional XY spiral states running along one of the three lattice directions. The spirals have a period of three, as follows from the factor $2\pi/3$ in Eq. \eqref{eq:period}. The Hamiltonian is invariant under three-fold rotations, so the XY spirals spontaneously break this symmetry. Similar to the square lattice model, the $C_{3}$ breaking happens via the edge degrees of freedom $s_{\langle ij\rangle}$. Once the edge spins pick a direction, they induce a uni-directional spiral order for the XY spins via the term $2\pi s_{\langle ij\rangle}/3$ inside the cosine interaction.

We want to point out that $H^t$ is not invariant under $\theta_i \rightarrow -\theta_i$, even if one also simultaneously takes $s_{\langle ij\rangle} \rightarrow - s_{\langle ij\rangle}$. This is because $H_s$ is not invariant under $s_{\langle ij\rangle} \rightarrow - s_{\langle ij\rangle}$, as can be seen from the set of zero energy plaquettes in Fig. \ref{fig:triangular}(b). For our purposes, this does not form a problem because TBG has no symmetry which only changes the sign of the IVC angle. Despite what is suggested at first sight by the use of the bond orientations in Fig. \ref{fig:triangular}(a), $H^t$ does have a $C_{2z}$ symmetry, which acts as $\theta_{\br} \rightarrow -\theta_{-\br}$ and $s_{\br} \rightarrow s_{-\br}$. This is consistent with how $C_{2z}$ acts on the IVC angle in TBG. As pointed out in the main text, the IVC spiral states preserve the $C_{2z}$ symmetry.

\begin{figure}
    \centering
    a)
    \includegraphics[scale=0.3]{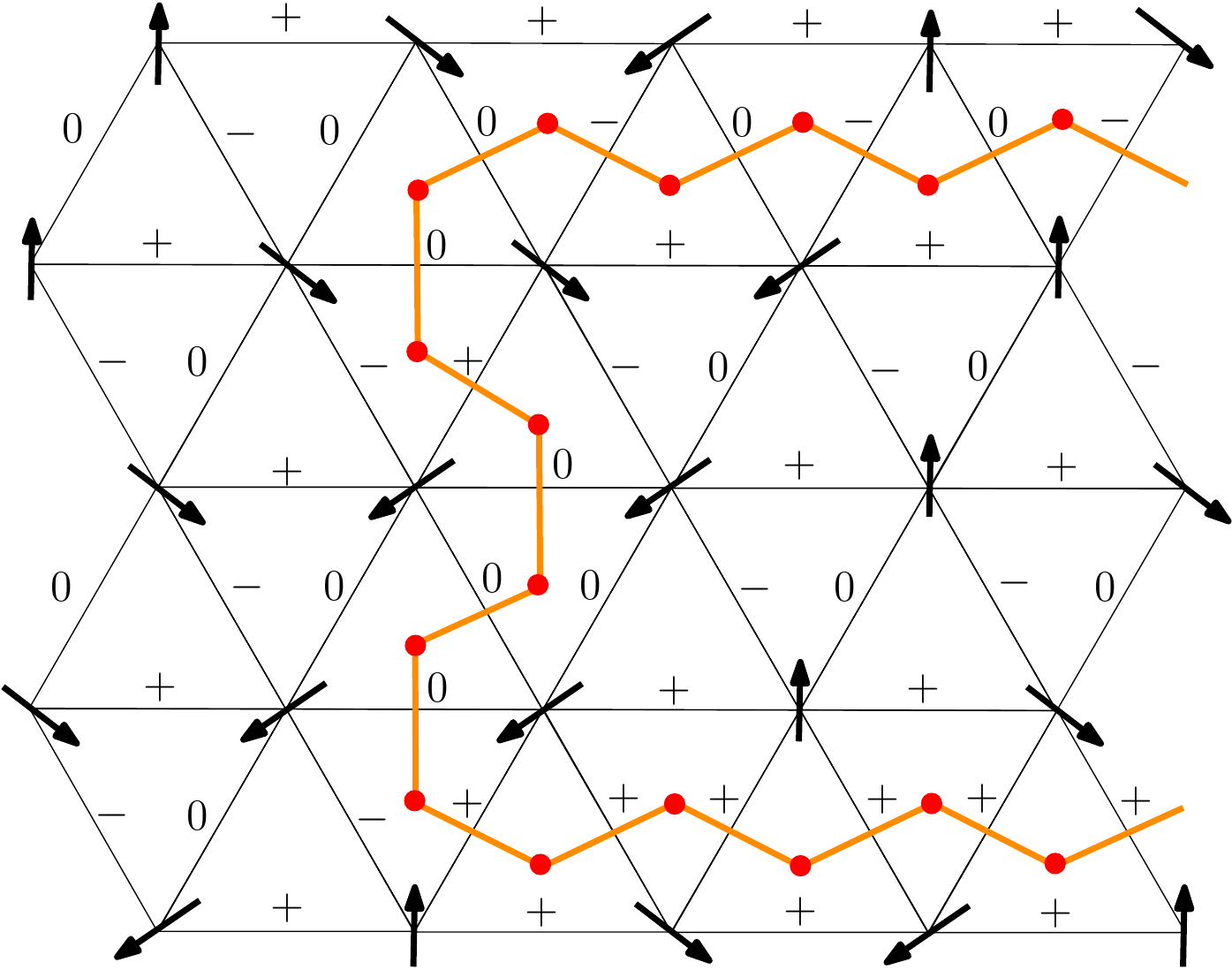}
    \qquad
    b)
    \includegraphics[scale=0.3]{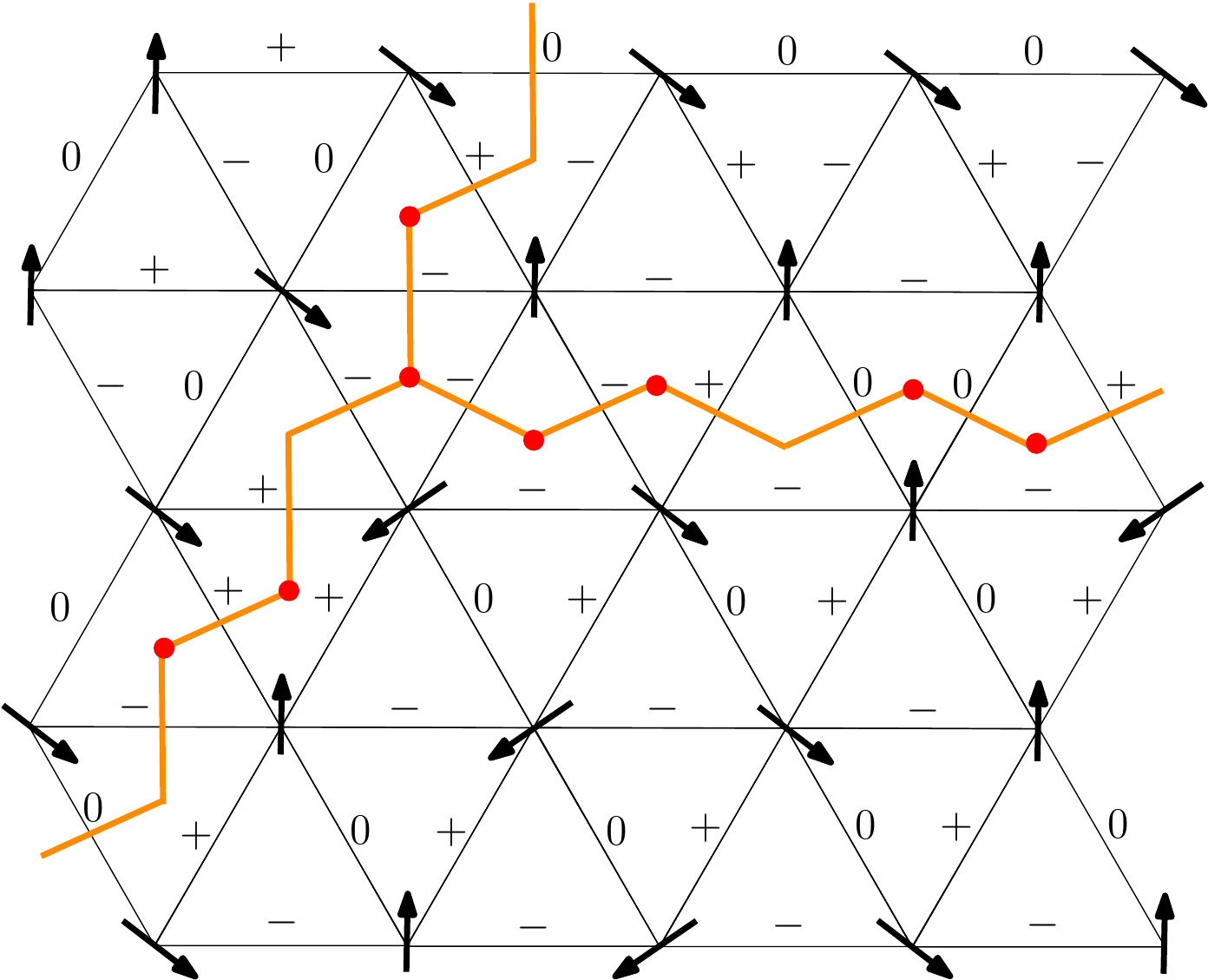}
    \caption{Different domain walls in the XY spiral state on the triangular lattice. (a) Domain wall between two ground states related by translation. (b) Domain walls between different ground states related by $C_3$. Red dots represented terms in $H_s$ which contribute to the domain wall energy.}
    \label{fig:triangularDW}
\end{figure}

Despite the different lattice geometry, and the different period of the XY spirals, the physics of the triangular lattice model is very similar to that of the square lattice model discussed above. For $\Delta_s \ll J$, domain walls between different ground states related by a translation or threefold rotation have an energy on the order of $\Delta_s$ per unit length. This is illustrated in Fig. \ref{fig:triangularDW}. As a result, at a temperature $T_3 \sim \Delta_s$, the nematic order is lost and correlation functions of $\cos n\theta$ decay  exponentially if $n$ is not a multiple of three. The operators $\cos 3n\theta$, on the other hand, retain their algebraic order. The XY spins become completely disordered above a temperature $T_{KT/3} \sim \pi J/18$, which is smaller than the conventional KT temperature by a factor of $9$. For $\Delta_s \gg J$, the nematic order survives to very high temperatures $T_3 \sim \Delta_s$. The XY spins disorder at a lower temperature $T_{KT} \sim \pi J/2$ via a conventional KT transition. 

The remaining question is how the regions of the phase diagram with $\Delta_s \ll J$ and $\Delta_s \gg J$ are connected at intermediate $\Delta_s \sim J$. An interesting possibility suggested by the square lattice model is that there is a single second order transition in the universality class of the $3-$state Potts model which separates the algebraic phase from the completely disordered phase. One reason to consider this possibility is that the physical picture underlying the existence of such a direct second order transition in the square lattice model can be applied to the triangular lattice model as well. In this picture, the direct second order transition would be a simultaneous deconfinement and unbinding transition of fractional vortices which are $1/3$ of a conventional XY vortex. 

We want to point out that similar physics was discussed in a very recent paper which used a generalized XY model to study hexatic-nematic liquid crystals \cite{Lubensky}. In particular, the authors of that work found numerical evidence for continuous transitions in the $3-$state Potts universality class, which in our triangular lattice model would correspond to the nematic transition at $T_3\sim \Delta_s$ for $\Delta_s \gg J$, and the fractional vortex deconfinement transition at $T_3 \sim \Delta_s$ for $\Delta_s \ll J$. In the generalized XY model of Ref.~\cite{Lubensky}, however, all phase transitions seem to merge in a single multi-critical point and there is no evidence for a direct second order transition between the algebraic phase and the completely disordered phase (more results are needed to confirm this). We leave a detailed numerical Monte Carlo study of the triangular lattice model phase diagram, and its connection to the phase diagram of hexatic-nematic liquid crystals, for future work.

\end{appendix}
\end{widetext}

\end{document}